\documentclass[10 pt]{article}
\usepackage[latin1]{inputenc}
\usepackage[english]{babel}
\usepackage{latexsym}
\usepackage{amsmath}
\usepackage{amsfonts}
\usepackage{amssymb}
\usepackage{amsthm}
\ifx\macrosloaded\relax\endinput\else\let\macrosloaded\relax\fi
\newtheorem{thm}{Theorem}[section]

\catcode`\@=11
\newcommand{\eqinsec}{\relax\@addtoreset{equation}{section}}
\renewcommand{\theequation}{\ifx\showlabels\iftrue\the\id\else\thesection.\arabic{equation}\fi}
\catcode`\@=13
\newcounter{supeq}
\newenvironment{subeq}
\stepcounter{equation} 
\def\theequation{\ifx\showlabels\iftrue\the\id\else\thesection.\arabic{equation}\fi}
\newtoks\id
\newcommand{\eqlabel}[1]{\label{#1}\global\id={(#1)}}

\newcommand{\tr}{\mbox{tr}}
\newcommand{\be}{\begin{equation}}
\newcommand{\eeq}{\end{equation}}
\newcommand{\bea}{\begin{eqnarray}}
\newcommand{\eea}{\end{eqnarray}}
\newcommand{\beaa}{\begin{eqnarray*}}
\newcommand{\eeaa}{\end{eqnarray*}}
\newcommand{\bseq}{\begin{subeq}}
\newcommand{\eseq}{\end{subeq}}
\newcommand{\ba}{\begin{array}}
\newcommand{\ea}{\end{array}}
\newcommand{\eql}{\eqlabel}

\def \rectangle#1#2{\hbox{\vrule\vbox to #2
{\hrule\hbox to
#1{\hfil}\vfil\hrule}\vrule}}
\newcommand{\edd}{\end{document}}

\renewcommand{\c}{\cdot}
\newcommand{\NI}{\noindent}

\newcommand{\Lb}{\underline{L}}

\newcommand{\Si}{\Sigma}
\newcommand{\ga}{\gamma}

\newcommand{\dd}{\mbox{${\bf D}$}}

\newcommand{\lap}{\mbox{$\bigtriangleup$}}
\newcommand{\lapp}{\mbox{$\bigtriangleup  \mkern-13mu /$\,}}

\newcommand{\nabb}{\mbox{$\nabla \mkern-13mu /$\,}}

\newcommand{\ddb}{\mbox{$\dd \mkern-13mu /$\,}}

\newcommand{\pr}{\partial}

\newcommand{\hot}{\widehat{\otimes}}

\newtheorem{Le}{Lemma}[section]

\newcommand{\Lie}{\mbox{$\cal L$}}

\newcommand{\lie}{\hat{\Lie}}

\newcommand{\nn}{\nonumber}
\newcommand{\Sch}{{Schwarzschild}}

\newcommand{\chib}{\underline{\chi}}

\newcommand{\de}{\delta}
\newcommand{\De}{\Delta}
\newcommand{\e}{\epsilon}

\newcommand{\chih}{\hat{\chi}}

\newcommand{\chibh}{\underline{\hat{\chi}}}

\newcommand{\und}[1]{\underline{#1}}

\newcommand{\Nb}{\und{N}}

\newcommand{\Cb}{\und{C}}

\newcommand{\ub}{{\und{u}}}
\renewcommand{\c}{\cdot}

\renewcommand{\aa}{\underline{\alpha}}
\newcommand{\bb}{\underline{\beta}}
\renewcommand{\a}{\alpha}
\renewcommand{\b}{\beta}
\newcommand{\dual}{\mbox{}^{\star}\!}

\newcommand{\si}{\sigma}

\newcommand{\ro}{\rho}

\newcommand{\divv}{\mbox{div}\mkern-19mu /\,\,\,\,}

\newcommand{\om}{\omega}
\newcommand{\oom}{\Omega}

\newcommand{\omb}{\underline{\omega}}

\newcommand{\etab}{\underline{\eta}}
\newcommand{\la}{\lambda}

\newcommand{\ep}{\epsilon}

\newcommand{\dddd}{{\bf D} \mkern-13mu /\,}

\newcommand{\QQ}{{\widetilde{\cal Q}}}
\newcommand{\QQb}{\underline{\widetilde{{\cal Q}}}}

\newcommand{\pii}[1]{^{(#1)}\mkern -1.5mu\pi}

\newcommand{\lu}[2]{^{(#2)}\mkern -1.5mu #1}

\newcommand{\acc}{\bar{K}}

\newcommand{\li}[1]{{\lu{{\bf i}}{#1}}}
\newcommand{\lj}[1]{{\lu{{\bf j}}{#1}}}
\newcommand{\lm}[1]{{\lu{{\bf m}}{#1}}}
\newcommand{\lmm}[1]{{\lu{\und{{\bf m}}}{#1}}}
\newcommand{\lnn}[1]{{\lu{{\bf n}}{#1}}}
\newcommand{\lnnn}[1]{{\lu{\und{{\bf n}}}{#1}}}

\newcommand{\ML}{\!\!\!\!\!\!\!\!\!}
\newcommand{\un}{\underline}

\newcommand{\tW}{\tilde{R}}
\def\th{\theta}
\def\ze{\zeta}
\begin{document}
\author{Francesco Nicol\`o \footnote{Dipartimento di Matematica,
Universit\`{a} degli Studi di Roma ``Tor Vergata", Via della Ricerca Scientifica, 00133-Roma, Italy}
\\
\\Universit\`{a} degli Studi di Roma ``TorVergata"}
\title{\LARGE{The peeling in the ``very external region" of non linear perturbations of the Kerr spacetime}}
\date{}
\maketitle
\begin{abstract}

{\NI Let an initial data metric $\overline{g}$ be, outside a ball $B_{R_0}$ centered in the origin, the induced metric on $\Si_0$ of a Kerr spacetime (with a mass $M$ and angular momentum $J$ whose ratio, $J/M$, depends on the size of $R_0$) plus small corrections which decay at spacelike infinity faster than $r^{-3}$; let, in the same region, a symmetric tensor $\overline{k}$ be the second fundamental form of the Kerr spacetime plus small corrections which decay at spacelike infinity faster than $r^{-4}$, let $\overline{g}$ and $\overline{k}$ satisfy the constraint equations. Then, using the previous results of \cite{Ch-Kl:book} and \cite{Kl-Ni:book}, the global existence of the external region of a global spacetime, outside the region of influence of $B_{R_0}$ follows. 

\NI In this external region the various components of the Riemann tensor decay along the outgoing null directions in agreement with the suggestion of the ``Peeling Theorem". }
\end{abstract}
\newpage
\section{Introduction and results}\label{S1}

\NI The problem of the global stability for the Kerr spacetime is a very difficult and open problem. The more difficult issue is that of proving the existence of solutions of the vacuum Einstein equations with initial data ``near to Kerr" in the whole external region up to the event horizon, which is also an unknown of the problem.\footnote{Of course for small perturbations the horizon is expected to stay ``near" to the Kerr horizon for $r_+=m+\sqrt{m^2-a^2}$.} If we consider the existence problem in a part of the external region sufficiently far from the Kerr event horizon, we call hereafter the ``very external region", the result is included,
in the version of Minkowski stability result proved by S.Klainerman and the present author with initial data near the flat ones (see \cite{Kl-Ni:book} and also \cite{Ch-Kl:book}). 
How far the ``very external region" has to be from the event horizon\footnote{More precisely,  the intersection of this region with $\Si_0$ from the ball centered in the origin with radius $r_+=m+\sqrt{m^2-a^2}$.} will depend on the mass, $M$, of the Kerr spacetime, see later for a detailed discussion.

\NI What is known up to now relative to the whole outer region are some relevant uniform boundedness results for solutions to the wave equation in the Kerr spacetime used as a background spacetime, see Dafermos-Rodnianski, \cite{Daf-Rod:Kerr} and references therein.\footnote{See also for the $J=0$ case, \cite{Blue} and references therein.}
\smallskip

\NI If we are interested to the asymptotic behaviour of the Riemann components toward (a portion of)\footnote{As we are concerned to a spacetime region whose initial data are given on $\Si_0$ outside a ball centered in the origin of radius $R_0>>r_+$ this implies that we cannot obtain the asymptotic behaviour for the Riemann tensor on the whole null infinity, but only on a portion of it.} the null infinity, more precisely to the question if  the non linear perturbation of the ``very external region" of the Kerr spacetime satisfy the peeling properties \!\footnote{The decay of the Riemann components satisfying or not the ``Peeling Theorem" have been a problem studied by many authors, see \cite{Kr:log1}, \cite{Kr:log2} and references therein.} one has to observe that also this much easier, but not easy, problem is an open problem.

\NI In \cite{Ch-Kl:book} and in \cite{Kl-Ni:book}, the null asymptotic behaviour of some of the null components of the Riemann tensor, specifically the $\a$ and the $\b$ components, see later for their definitions, is different from the one expected from the ``Peeling Theorem",\cite{Wald}, as the decay proved there is slower. Nevertheless in a subsequent paper S.Klainerman and the present author, \cite{Kl-Ni:peeling}, proved for these components an asymptotic behaviour consistent with the ``Peeling Theorem" under stronger asymptotic conditions for the initial data.  Unfortunately these conditions do not include initial data ``near to Kerr", due to the fact that, as the angular moment $J$ of the Kerr spacetime is different from zero, in the Kerr initial data there are metric components decaying as $r^{-2}$, a decay not sufficient to guarantee the peeling in \cite{Kl-Ni:peeling}.

\NI Therefore the new result we expect to be true and that we would like to prove is the following one:
\smallskip

\NI {\bf Expected result:}\ {\em Assuming the initial data on the hypersurface $\Si_0/B_{R_0}$  near to the Kerr initial data, where $B_{R_0}$ is a ball centered at the origin sufficiently large,  it is possible to prove a global existence result for the ``very external region" such that the null Riemann components decay in agreement with the ``Peeling Theorem".}
\smallskip

\NI In fact we prove a weaker result where the size of the Kerr mass, $M$, depends on how large is $R_0$, namely:
\smallskip

\NI{\bf Proved result:}\ {\em Assuming the initial data on the initial hypersurface $\Si_0/B_{R_0}$,  where 
$B_{R_0}$ is a ball centered at the origin sufficiently large, near to the initial data of a Kerr spacetime whose mass $M$ is (upper) bounded by a quantity depending on the size $R_0$, it is possible to prove a global existence result for the ``very external region" such that the null Riemann components decay in agreement with the ``Peeling Theorem".}
\smallskip

\NI The detailed version of this theorem is given at the end of this section. Its proof can be divided in five steps which we summarize here and will describe in detail in the following sections:
\smallskip

\NI {\bf a)} From the explicit expression of the Kerr metric, in the Boyer-Lindquist coordinates,  $\{t,r,\theta,\phi\}$,\footnote{See the appendix eq.\ref{3.148} for the definitions of the various coefficients.}
\begin{eqnarray}
ds^2 =-\frac{\De-a^2\sin^2\theta}{\Si}dt^2+\frac{\Sigma}{\Delta}dr^2+\Sigma d\theta^2-\frac{4Mar\sin^2\theta}{\Sigma}d\phi dt+R^2\sin^2\theta d\phi^2 \nn
\end{eqnarray}
we can write its restriction to the $t=0$ hypersurface, $\Si_0$, as
\bea
{\bf g}\!&=&\!{\bf g}_{(Sch)}+\left(\frac{a^2\sin^2\theta}{r^2}+O\left(\frac{a^2m}{r^3}\right)+O\left(\frac{a^4}{r^4}\right)\right)\!dr^2
+\left(\frac{a^2\cos^2\theta}{r^2}\right)\!r^2d\theta^2\nn\\
\!&+&\!\left(\frac{a^2}{r^2}+O\left(\frac{a^2m}{r^3}\right)+O\left(\frac{a^4}{r^4}\right)\right)\!r^2\sin^2\theta d\phi^2\ .\eql{Kerrmet}
\eea
where ${\bf g}_{(Sch)}$ denotes the restriction of the Schwarzschild metric on the initial hypersurface:
\[{\bf g}_{(Sch)}=(1-\frac{2M}{r})^{-1}dr^2+r^2(d\theta^2+\sin\theta^2d\phi^2)\ .\] 
It follows that the components of the correction to the ${\bf g}_{(Sch)}$ metric,  $({\bf g}_{(Kerr)}-{\bf g}_{(Sch)})$,  have terms  of order $O({a^2}/{r^2})$ or higher.
Therefore if we consider in the external region initial data which are a ``small modification" of those of a Kerr spacetime with a small parameter $a={J}/{m}\leq \varepsilon_0$ we can write, for the metric on $\Si_0$,
\bea
{\bf g}={\bf g}_{(Kerr)}+\de{{\bf g}}={\bf g}_{(Sch)}+\left[({\bf g}_{(Kerr)}-{\bf g}_{(Sch)})+\de{{\bf g}}\right]\eql{Indata1a}
\eea
and if $\de{{\bf g}}$ is a small correction to the Kerr initial data with a better decay toward spacelike infinity, then it follows that the whole correction to ${\bf g}_{(Sch)}$ satisfies the conditions of the theorem proved in \cite{Kl-Ni:book}.

\NI Observe that the initial decay conditions in the present case are even stronger than those required in \cite{Kl-Ni:book} where the components of the metric restriction on $\Si_0$ have to decay faster than $r^{-\frac{3}{2}}$. This will turn out to be crucial in the sequel.

\NI These considerations allow us to conclude that, outside the dependance region of the sphere $B_{R_0}$, we have a global spacetime satisfying all the properties proved in \cite{Kl-Ni:book} which can be interpreted as a ``small" non linear perturbation of the very external region of a Kerr spacetime with a small angular momentum.

\NI Moreover  expression \ref{Kerrmet} and the fact that in \cite{Kl-Ni:book} the decay required for $\bf g$ is $r^{-\frac{3}{2}}$ allows to use part of the extra decay of the term $O(a^2/r^2)$ to avoid the condition $a=J/M$ small. In fact we can choose $\ga<1/2$ and require
\bea
\frac{a^2}{R_0^{\ga}}\leq\ep_0\ \ \mbox{equivalent to}\ \ \frac{J}{M}\leq \sqrt{R_0^{\ga}\ep_0}\ .\eql{1.3}
\eea
If we consider the Kerr spacetimes satisfying the inequality $a={J}/{M}\leq M$ \footnote{Those Kerr spacetimes which do not have a naked singularity and can therefore be interpreted as representing the spacetime around a black hole with angular momentum $J$.} then condition \ref{1.3} is satisfied when
\bea
M\leq \sqrt{R_0^{\ga}\ep_0}\eql{1.3a}
\eea
which tells us that provided the Kerr mass satisfies inequality \ref{1.3a} no conditions on $J$ is imposed.\footnote{Remark that, unfortunately, this does not allow to choose $R_0\cong 2M$ as to prove the global existence in the external region, see \cite{Kl-Ni:book}, Chapter 3, the following inequality $M/R_0\leq \sqrt{\varepsilon_0}$ has to be satisfied.}
In the sequel we denote $B_{R_0}\equiv B$. 
\smallskip

\NI {\bf b)} From the results in \cite{Kl-Ni:book}, it follows that the decay  in the null directions of the various components of the Riemann tensor\footnote{Components relative to a null frame adapted to the null outgoing and incoming cones which foliate the ``very external region" we will denote ${\cal K}$.} do not follow the ``Peeling theorem". More precisely the components $\a$ and $\b$ decay as $r^{-\frac{7}{2}}$ while we ``expect $r^{-5}$ and $r^{-4}$ respectively.\footnote{In principle some $\log$ powers can be present, see J.A.V.Kroon,\cite {Kr:log2}.}
In a subsequent paper, \cite{Kl-Ni:peeling}, S.Klainerman and the author proved that the decay suggested from the ``Peeling theorem" could be obtained assuming a stronger spacelike decay for the initial data. In particular we proved the following result:
\begin{thm}\label{TT1.2}
Let assume that on $\Si_0/B$ the metric and the second fundamental form have the following asymptotic behaviour
\footnote{Here $f=O_q(r^{-a})$ means that $f$ asymptotically
behaves as $O(r^{-a})$ and its partial derivatives $\partial^kf$, up to order $q$ behave as $O(r^{-a-k})$. Here with $g_{ij}$ we mean the components written in Cartesian coordinates. }
\bea
&&g_{ij}={g_S}_{ij}+O_{q+1}(r^{-(3+\ep)})\nn\\
&&{k}_{ij}=O_{q}(r^{-(4+\ep)})\eql{1.1b}
\eea
where ${g_S}$ denotes the restriction of the
Schwarzschild metric on the initial hypersurface:
\[g_S=(1-\frac{2M}{r})^{-1}dr^2+r^2(d\theta^2+\sin\theta^2d\phi^2)\ .\] 
Let us assume that a smallness condition for the initial data is satisfied.\footnote{The details of the smallness condition are in \cite{Kl-Ni:peeling}. }
Then along the outgoing null hypersurfaces $C(u)$ (of the external region $\cal K$) the following limits hold, with $\ep'<\ep$ and $\ub$ and $u$ the generalization of the Finkelstein variables $u=t-r_*\ \,\ \ u=t+r_*$ in the {\Sch} spacetime:
\bea
&&\lim_{C(u);\ub\rightarrow\infty}r(1+|u|)^{(4+\ep)}\aa=C_0\nn\\
&&\lim_{C(u);\ub\rightarrow\infty}r^2(1+|u|)^{(3+\ep)}\bb=C_0\nn\\
&&\lim_{C(u);\ub\rightarrow\infty}r^3\ro=C_0\nn\\
&&\lim_{C(u);\ub\rightarrow\infty}r^3\si=C_0\eql{peel1}\\
&&\lim_{C(u);\ub\rightarrow\infty}r^4(1+|u|)^{(1+\ep)}\b=C_0\nn\\
&&\sup_{(u,\ub)\in {\cal K}}r^{5}(1+|u|)^{\ep'}|\a|\leq C_0\ .\nn
\eea
\end{thm}
\NI$\aa,\bb,...$ are the null components of the Riemann tensor defined with respect to a null frame adapted to the double null foliation, see \cite{Kl-Ni:book} Chapter 2, we recall here their definition for completeness,
\bea
&&\a(R)(e_a,e_b)=R(e_a,e_4,e_b,e_4)\ ,\ \aa(R)(e_a,e_b)=R(e_a,e_3,e_b,e_3)\ \ \ \ \ \ \ \ \ \nn\\
&&\b(R)(e_a)=\frac{1}{2}R(e_a,e_4,e_3,e_4)\ ,\ \ \bb(R)(e_a)=\frac{1}{2}R(e_a,e_3,e_3,e_4)\ \ \ \ \ \ \eql{11.187aa}\\
&&\ro(R)=\frac{1}{4}R(e_3,e_4,e_3,e_4)\ ,\ \si(R)=\frac{1}{4}\ro(^\star R)=\frac{1}{4}{^\star R}(e_3,e_4,e_3,e_4)\ \ \ \ \ \ \ \ \nn
\eea
where $\{e_a,e_b,e_3,e_4\}$ is a null frame adapted to the foliation.

\NI This result was obtained, basically, in two steps. 
The first one consisted in proving that a family of energy-type norms, $\tilde{\cal Q}$, made with the Bel-Robinson tensor associated to the Riemann tensor $R$, of the same type as those used to prove the global existence of the spacetime in the external region $\cal K$ see \cite{Kl-Ni:book}, but with a different weight in the integrand, were bounded in terms of the same norms relative to the initial data. This allowed to prove that the various null components of the Riemann tensor, besides the decay in $r$ (or $\ub$) already proved, have a decay factor in the in the $|u|$ variable. 
In the second step  it was proved, integrating along the incoming cones, that the extra decay in the $u$ variable can be transformed in an extra decay in the $r$ variable proving the final result.
\smallskip

\NI Unfortunately Theorem \ref{TT1.2} cannot be applied to the present case as the initial conditions of the previous theorem do not include the initial data \ref{Indata1a}, due to the part $({\bf g}_{(Kerr)}-{\bf g}_{(Sch)})$ which decays as $O(r^{-2})$. 

\NI This can be understood observing that the extra weight in the $\tilde{\cal Q}$ norms, required in \cite{Kl-Ni:peeling}, is a factor
$|u|^{({3+\ep})}$ and, as on the initial hypersurface $|u|$ is equivalent to $r$,  it follows that the decay $O(r^{-2})$ of $({\bf g}_{(Kerr)}-{\bf g}_{(Sch)})$ is not sufficient to make the initial data norms $\tilde{\cal Q}|_{\Si_0}$ bounded. To be more explicit let us look at the second term of the $\tilde{\cal Q}$ norm in \cite{Kl-Ni:peeling}, eq.(2.6), on the initial hypersurface,
\beaa
 \int_{\Si_0\cap{\cal K}}|u|^{3+\e}Q(\lie_O {R})(\bar{K},\bar{K},T,T)\ .
\eeaa
Recalling the expression of $Q(\lie_O R)(\bar{K},\bar{K},T,T)$, see \cite{Kl-Ni:book}, Chapter 4,
\beaa 
Q(\lie_O R)(\bar{K},\bar{K},T,T)\!&=&\! \frac 1 8 \ub^4|\a|^2+\frac 1 8u^4|\aa|^2+\frac 1 2(\ub^4+\frac{1}{2}u^2\ub^2)|\b|^2\\
\!&+&\!\frac 1 2 (u^4+4\ub^2u^2+\ub^4)(\ro^2+\si^2)+\frac 1 2(u^4+\frac{1}{2}u^2\ub^2)|\bb|^2
\eeaa
and observing that on $\Si_0\cap{\cal K}$ $u,\ub$ behave like $r$, we have
\beaa
 &&\int_{\Si_0\cap{\cal K}}|u|^{3+\e}Q(\lie_O R)(\bar{K},\bar{K},T,T)\ \cong\nn\\
 &&\int_{R_0}^{\infty}dr r^{2+(3+\ep)+4}\left[|\a|^2+|\aa|^2+|\b|^2+(\ro^2+\si^2)+|\bb|^2\right]\ .
 \eeaa
This integral is convergent if all the null components of $\lie_O R$ decay faster than $r^{-5}$ which implies that the various components of $\bf g$ have to decay faster than $r^{-3}$, while  $({\bf g}_{(Kerr)}-{\bf g}_{(Sch)})$ decays as $r^{-2}$ and in particular $\ro(\lie_OR)$ decays as $r^{-4}$.\footnote{One could expect a decay $r^{-3}$, but the result is better as the part of the $\ro$ component which decays as $r^{-3}$ is spherically symmetric and therefore cancelled by $\lie_O$. Here the fact that $\lie_O$ is a ``modified" Lie derivative, see later on for its definition, is not relevant.}
 \smallskip 

\NI  The idea to overcome this difficulty is based, intuitively, on the fact that the Kerr spacetime is static and $\frac{\partial}{\partial t}$ is a Killing vector field. Therefore if, instead of considering the Riemann components, we consider their time derivatives, they do not depend anymore on the Kerr part of the Riemann tensor, Their initial data can have a good decay and then, reapplying Theorem \ref{TT1.2}, a good decay along the null directions. Once we have this decay by a simple ``time integration" we can obtain the decay of the Riemann tensor.

\NI This argument as described here is not rigorous, but could be directly implemented if we consider a ``linear perturbation" of the (very external) Kerr spacetime.\footnote{This is, basically, part of a paper by G.Caciotta and T.Raparelli, to appear, where it is also discussed in which sense we can define a linear perturbation of the Kerr spacetime.} In the non linear case the proof of its validity has to be redone from scratch, but it turns out that the basic idea is still valid in the following sense:
\smallskip

\NI{\em Let us consider instead of the time derivative of the Riemann tensor its (modified) Lie derivative $\lie_TR$,\footnote{See later for its precise definition.} where $T$ is not anymore a Killing vector, but only ``nearly Killing",\footnote{With ``nearly Killing" we mean that its deformation tensor is small with respect to some Sobolev norms.} then we can define some $\tilde{\cal Q}$ norms relative to $\lie_TR$ with appropriate weights and prove that they are bounded in terms of the corresponding initial data norms.}

\NI To prove that these norms are bounded in terms of the initial data ones, is not obvious even if we are in a given spacetime $\cal K$. The reason is that the ``Error"\footnote{with Error we mean the difference between the norms on the final boundary of a region of $\cal K$ and those defined on (the portion of) the initial hypersurface.} which has to be controlled depends on the $\ro$ component of the $\lie_T$ derivative of the Riemann tensor, $\lie_TR$, which cannot be estimated by the $\tilde{\cal Q}$  norms (they can only provide an estimate of $\ro-\overline{\ro}$, see \cite{Kl-Ni:book}, Chapter 5); on the other side $\ro$ is estimated by an integration along the incoming directions which, to be done, requires the control of the other Riemann null components which are at their turn estimated in terms of the $\tilde{\cal Q}$ norms. This clearly suggests the need of a bootstrap mechanism that we show in detail in the following sections.
\smallskip

\NI {\bf c)} The initial data correction, $\de{\bf g}$ have to be chosen with a decay in $r$ such that all the components of $\lie_TR$ on $\Si_0$ have
 a decay which makes the $\tilde{\cal Q}$ norms on $\Si_0$  finite. Once this is achieved to prove that the same norms in the whole region $\cal K$ are bounded in terms of the initial data ones amounts to prove that the ``Error", which is a spacetime integral describing the difference between the final and initial norms, can be bounded by the same $\tilde{\cal Q}$ norms mutiplied by a factor proportional to $\varepsilon_0{R_0}^{-1}$, where $\varepsilon_0$ is a parameter defining the smallness of the initial data corrections. Formally we have
 \bea
 \tilde{\cal Q}\leq \tilde{\cal Q}|_{\Si_0}+c\frac{\varepsilon_0}{R_0}\tilde{\cal Q}
 \eea 
 implying
 \[\tilde{\cal Q}\leq \frac{1}{1-c\frac{\varepsilon_0}{R_0}}\tilde{\cal Q}|_{\Si_0}\ .\]

\NI The control of these norms allows to prove,  with the same techniques used in \cite{Kl-Ni:book} and in the first part of \cite{Kl-Ni:peeling}, the following decay along the outgoing null directions for the null components of $\lie_T R$,
$\a( \lie_T R),\  \b( \lie_T R),.....$\ ,
\begin{eqnarray}
&&\ML\sup_{\mathcal{K}}r^{\frac{7}{2}}|u|^{\frac{5}{2}+\frac{\e}{2}}|\a(\lie_{T}R)| \leq C_0\ ,
\quad\sup_{\mathcal{K}}r^{\frac{7}{2}}|u|^{\frac{5}{2}+\frac{\e}{2}} |\b(\lie_{T}R)|\leq C_0\nn\\
&&\ML\sup_{\mathcal{K}}r^3|u|^{3+\e'}|\ro(\lie_{T}R)-\overline{\ro(\lie_{T}R)}| \leq C_0\ ,\quad 
\sup r^3|u|^{3+\frac{\e}{2}}|\si(\lie_{T}R)-\overline{\si(\lie_{T}R)}|\leq C_0\nn\\
&&\ML\sup_{\mathcal{K}}r^2|u|^{4+\frac{\e}{2}}\bb(\lie_{T}R)\leq C_0\ ,\quad\sup_{\mathcal{K}}r|u|^{5+\frac{\e}{2}}|\aa(\lie_{T}R)|\leq C_0\ .\eql{LTRdec1a}
\end{eqnarray}
As said before while discussing point {\bf b)}, this is only the first step for getting the right decay. The next one, which is basically a repetition of what has been done in \cite{Kl-Ni:peeling}, consists in transforming  the $|u|$ decay in a $r$ decay, integrating along the (portion) of the incoming cones. As proved in detail later on, Theorem \ref{T4.1}, this provides the following estimates
\begin{eqnarray}
&&\ML\sup_{\mathcal{K}}r^{5}|u|^{1+\frac{\e''}{2}}|\a(\lie_{T}R)| \leq C_1,\quad\sup_{\mathcal{K}}r^{4}|u|^{2+\frac{\e''}{2}} |\b(\lie_TR)|\leq C_1\nn\\
&&\ML\sup_{\mathcal{K}}r^3|u|^{3+\frac{\e'}{2}}|\ro(\lie_{T}R)-\overline{\ro(\lie_{T}R)}| \leq C_1,\quad \sup r^3|u|^{3+\frac{\e'}{2}}|\si(\lie_{T}R)-\overline{\si(\lie_{T}R)}|\leq C_1\nn\\
&&\ML\sup_{\mathcal{K}}r^2|u|^{4+\frac{\e}{2}}\bb(\lie_TR)\leq C_1,\quad\sup_{\mathcal{K}}r|u|^{5+\frac{\e}{2}}|\aa(\lie_TR)|\leq C_1\ \ .\ \eql{LTWdec11a}
\end{eqnarray} 
where 
\[\e''<\e'<\e\ .\]
\smallskip

\NI {\bf d)} Differently from \cite{Kl-Ni:peeling} our goal is not yet obtained. In fact we have a decay in agreement with the peeling only for the null components of $\lie_TR$. To get the decay for the null components of $R$ we have to integrate the previous components along the timelike curves $r=const$. To do it we have first to connect the null components of  $\lie_TR$ to the partial $t$-derivatives of the null components of $R$. 

\NI Observe that, as we will show in detail later on, the decay coming from inequalities \ref{LTWdec11a} seems sufficiently good for integrating in the $t$ variable and getting again uniform decay estimates, but as we will integrate the time derivatives of the various components, $\pr_t\a(e_a,\e_b),\pr_t\b(e_a),....$ we have first to prove that these time derivatives satisfy the same bounds as those in \ref{LTWdec11a}. 

\NI To connect, for instance, $\a(\lie_{T}R)$ to $\pr_t\a(e_a,\e_b)$ one has first observe that $\lie_{T}$ is the modified Lie derivative introduced in \cite{Ch-Kl:book}, see its definition in \cite{Kl-Ni:book}, page 75, def.(3.2.2); therefore one has first to estimate, starting from the proved decay of  $\a(\lie_{T}R)$, the decay satisfied by $\a(\Lie_{T}R)$ and then from it the one satisfied by $\pr_t\a(e_a,\e_b)$. This can be done using the equations which connect these various expressions. To connect, for instance, $\a(\lie_TR)$ with $\a(\Lie_TR)$ we use the definition,
\bea
\Lie_TR= \lie_{T}R+\frac 1 2 {^{(T)}}[R]-\frac 3 8(\tr{^{(T)}}\pi)R\eql{piT1a}
\eea
where ${^{(T)}}\pi$ is the deformation tensor of the vector field $T$. 

\NI From \ref{piT1a} and the explicit expression of ${^{(T)}}\pi$, eq.\ref{5.20}, one realizes that to prove that $\a(\Lie_{T}R)$ and $\pr_t\a(e_a,\e_b)$ satisfy the same bounds  of $\a(\lie_TR)$  requires to control the decay of the connection coefficients of the spacetime $\cal K$ and of the components of the non derived Riemann tensor $R$ which we have, up to now, at our disposal, therefore those provided in \cite{Kl-Ni:book}.

\smallskip

\NI {\bf e)} This is a slightly delicate point. In fact if we assume for the connection coefficients of the spacetime $\cal K$ and of the components of the Riemann tensor $R$ exactly the decay proved in \cite{Kl-Ni:book} it is easy to realize that this will not allow to obtain a decay for $\pr_t\a(e_a,e_b)$ sufficient to integrate and to obtain the expected $r^{-5}$ decay for $\a(R)$, but at most we can prove a decay $O((\log r)r^{-5})$ (Viceversa this would be enough for $\b(R)$, while the other components already satisfy the appropriate ``Peeling decay").

\NI It is  to face this problem that we take advantage from the fact, observed discussing point {\bf a)}, that the corrections to the {\Sch} metric on $\Si_0/B$ decay faster than the minimum (spacelike) decay  required in \cite{Kl-Ni:book}. This fact will allow, as proved in the first part of \cite{Kl-Ni:peeling}, to obtain for the various null components of $R$ (with the exception of $\ro$) and for those connection coefficients or their combinations which are identically zero in Kerr, an extra $|u|^{\frac{1}{2}-\de}$ decay factor with $\de>0$ arbitrarily small.\footnote{It is important to observe that the combination of connection coefficients which appear in the ${^{(T)}}\pi$ components are, in fact, identically zero in Kerr.}

\NI This result allows us to prove that $\a(\Lie_{T}R)$ and $\pr_t\a(e_a,\e_b)$ satisfy the same bounds  as $\a(\lie_TR)$ and that, therefore, we can perform the integration along the $T$ integral curves, obtaining the expected decay. Therefore  we have proved the following theorem:
\begin{thm}\label{spaceK}
Assume that the initial data on the initial hypersurface $\Si_0/B$ (where $B$ is a ball centered at the origin of linear size $R_0$ sufficiently large) are near to the initial data of a Kerr spacetime with angular momentum $J$ satisfying the condition ${J}/{M}\leq M$ and the mass $M$ satisfying, with $\ga>0$ sufficiently small,\footnote{$\ga\leq\frac{1}{4}$ wolud be enough.}
\bea
M\leq \sqrt{R_0^{\ga}\varepsilon_0}\ .\eql{Mcond}
\eea
Denoted with ${\bf g}$ the metric on $\Si_0$ we have  
\bea
{\bf g}={\bf g}_{(Kerr)}+{{\bf \de g}}={\bf g}_{(Sch)}+\left[({\bf g}_{(Kerr)}-{\bf g}_{(Sch)})+\de{{\bf g}}\right]\eql{Indata1}
\eea 
where
\bea
({\bf g}_{(Kerr)}-{\bf g}_{(Sch)})=O\left(\frac{a^2}{r^2}\right)\ .\eql{Indata2}
\eea
Moreover let the remaining correction ${\bf\de g}$ be such that,
\[{\bf\de g}=o_q\!\left(\frac{\varepsilon_0}{r^{-3}}\right)\]
and let $k$ be a symmetric tensor such that
\[k= k_{(Kerr)}+o_{q-1}\!\left(\frac{\varepsilon_0}{r^{-4}}\right)\ , \]
with $q\leq 6$.\footnote{A function $f$ is $o_q(r^{-k})$ if $\partial_lf=o(r^{-k-l})$ for $l=0,1,..,q$.} 

\NI Under these assumptions it is possible to prove a global existence result for the ``very external region", $\cal K$, endowed with a double null canonical foliation with $g$ and $k$ the restrictions to $\Si_0$ of the Lorentzian metric $\bf g$ and  of the second fundamental form associated to the vector field
\[T=\oom(e_3+e_4)\]
where $\e_3,e_4$ are the null vector fields adapted to the double canonical foliation and $\oom$ is defined through the relation
\[{\bf g}(L,\Lb)=-(2\oom)^{-2}\ ,\]
where $L,\Lb$ are the null geodesic vector field associated to the double canonical foliation.
Finally all the null Riemann components decay in agreement with the ``Peeling Theorem"\ .
\end{thm}
\NI The result proved in this theorem tell us, in broad terms, that perturbing the (external) initial data of a slow rotating  Kerr spacetime the ``peeling behaviour" of the various Riemann components is preserved. One can argue that the allowed ``perturbation" is very mild, see the assumptions on $\de{\bf g}$ in the theorem, but the previous results in \cite{Kl-Ni:peeling} suggest, although do not prove, that a lower decaying correction $\de{\bf g}$ to Kerr would destroy the peeling, at least for $\a(R)$. If for instance we assume 
an $r^{-2}$ decay all the strategy based on the fact that the $r^{-2}$ decaying part of the metric does not depend on $t$ would fail. Even with a decay $r^{-\ga}$ with  $\ga\in [2, 3]$ the present proof could not be completed and the \cite{Kl-Ni:peeling} results would again suggest a peeling violation.
\smallskip

\NI{\bf Remark:} {\em  As done for $J$ the smallness of the correction $\de g$ can be relaxed if we increase (not in a linear way) the value of $R_0$. 

\NI In the sequel, just for notational simplification we will consider the situation with $R_0$ given and sufficiently large and $a=J/M\leq \sqrt{\varepsilon_0}$\ .}
\medskip

\ \ \ \

\ \ \ \

\ \ \ \

\NI {\bf Acknowledgments:} {\em The author is deeply indebted to Sergiu Klainerman for pointing to him the importance  of considering the Lie derivative, with respect to  the ``time" vector field $T$, of the Riemann tensor to obtain more detailed estimates for the various components of the Riemann tensor. Besides he is also indebted for many illuminating discussions he had with him about this subject and many related ones. He also thanks G.Caciotta for pointing to him some delicate aspects of the proof to be clarified.
He also wants to state strongly that the present result is deeply based on the previous works \cite{Kl-Ni:book}, \cite{Kl-Ni:peeling} and on the original seminal work by D.Christodoulou and S.Klainerman, \cite{Ch-Kl:book}.
Therefore nothing has been ``gracefully \footnote{A very improper adjective.} acknowledged", but all the due credits have been explicitly given at the best of the author's knowledge.}
\newpage

\section{Proof of the result}\label{S.2}
\subsection{The very external region  $\cal K$}
The main difference between the result we obtain and the {``Expected result"} is that in the  present proof we assume the ratio of the Kerr angular momentum with the mass, $J/M$, small.\footnote{As we said before all the proof can be performed under the more general condition \ref{Mcond}.} This allows us to treat the initial data difference between Kerr and {\Sch} as a perturbation of the initial data. Even in this case to get the global existence in the external region is not at all easy, but it is a result already at our disposal, see \cite{Kl-Ni:book}. If  we do not require $J/M$ small,\footnote{More precisely if we do not require $J/M$ depending on $R_0$.} together with proving the asymptotic decay we have to prove the global existence of spacetimes which in the ``very external region" are non linear perturbations around a $(J,M)$-Kerr spacetime and we cannot fully use the results of \cite{Kl-Ni:book}. The main difficulty is  that the rotation generators vector fields cannot be considered anymore approximately ``Killing" and this makes difficult to prove the boundeness of some of the ${\cal Q}$ norms.\footnote{ More specifically the ${\cal Q}(\lie_OR)$ norms, see \cite{Kl-Ni:book} Chapter 3 for their definitions or later on .}
\smallskip

\NI Assuming, therefore, $J/M$ small,
\[J=Ma\leq M\varepsilon_0^{\ \!\!\frac{1}{2}}\ ,\]
we can apply the results in \cite{Kl-Ni:book}. Moreover, as it will turn out to be necessary later on, we can use the fact that the decay of the correction to the {\Sch} metric on $\Si_0$ is faster than the decay required there  and that, therefore, the Riemann tensor has an extra factor  ``decay" in the $|u|$ variable. This implies that also the  decay behaviour of some of the connection coefficients is improved by the same factor. Even this result has not to be proved here as it is already contained in the first part of \cite{Kl-Ni:peeling}. Therefore we only recall this result:
\begin{thm}\label{Ktheorem2}
Assume the initial data $\{{\overline g}, {\overline k}\}$ on the initial hypersurface $\Si_0/B$ have the following asymptotic behaviour
\bea
&&{\overline g}-g_{Kerr}=o_4(r^{-3})\nn\\
&&{\overline k}-k_{Kerr}=o_3(r^{-4})\eql{Indatatheorem2}
\eea
and a smallness condition ${\cal J}\leq \varepsilon_0$ is given as in \cite{Kl-Ni:book}, then there exists a complete spacetime ${\cal K}$\footnote{It is the very external region outside the influence region of the initial ball $B$.} with a double null canonical foliations such that its Riemann components behave in the following way, with $C_1=O(\varepsilon_0)$,
\bea
&&\sup_{\mathcal{K}_2}r^{\frac{7}{2}}|u|^{\frac{1}{2}-\de}|\a(R)|\leq C_1\ \ ,
\quad\sup_{\mathcal{K}_2}r^{\frac{7}{2}}|u|^{\frac{1}{2}-\de} |\b(R)|\leq C_1\nn\\
&&\sup_{\mathcal{K}_2}r^3|\ro(R)| \leq C_1\ \ ,\quad \sup_{\mathcal{K}_2} r^3|u|^{1-\de}|\si(R)|\leq C_1\eql{LTRdec2}\\
&& \sup_{\mathcal{K}_2}r^2|u|^{2-\de}\bb(R)\leq C_1\ \ ,\quad\sup_{\mathcal{K}_2}r|u|^{3-\de}|\aa(R)|\leq C_1\ .\nn
\eea
with $\de>0$ and arbitrarily small. Moreover all the connection coefficients (and their combinations) which are identically zero in the Kerr spacetime acquire the extra decay factor $|u|^{\frac{1}{2}-\de}$. In particular the following inequalities hold
\bea
&&\sup_{\mathcal{K}}r^{2}|u|^{1-\de}|\tr\chi+\tr\chib|\leq C_1\nn\\
&&\sup_{\mathcal{K}}r^{2}|u|^{1-\de}|\om+\omb|\leq C_1\ .\eql{Otheorem2}
\eea
\NI Moreover all the norms of the connection coefficients, $\tilde{\cal O}$,  are proportional to $\varepsilon_0$. 
\end{thm}
\NI{\bf Proof:} 
With the exception of \ref{Otheorem2} whose proof is in the appendix, the proof of this result is the content of part I of Theorem 1.2, of Theorem 2.2 and of the definition given there of the $\tilde{\cal O}$ norms, see Appendix A.2, of \cite{Kl-Ni:peeling}.  Nevertheless in the appendix we give a simplified proof of this result which does not use the results proved in \cite{Kl-Ni:peeling}.
\medskip

\NI{\bf Remarks:} {\em
\smallskip

\NI a) It is important to remark, see subsection \ref{S.S2.8}, that the estimates \ref{Otheorem2} for $\om+\omb$ and $\tr\chi+\tr\chib$ are valid in the whole spacetime, but the estimates on $\Si_0$ of the same quantities are stronger due to the decay assumptions for the metric and the second fundamental form $k$. These stronger initial data assumptions are needed to have bounded $\tilde{\cal Q}$ norms on $\Si_0$.
\smallskip

\NI b) As it will be clear in the following sections and specifically in Subsection \ref{S.s2.7}, Theorem \ref{T5.1}, the result we prove requires all the estimates which are valid in spacetime ${\cal K}$; nevertheless a weaker result  can be obtained using only the estimates proved in \cite{Kl-Ni:book}. In this case, see Theorem \ref{T5.1(K1)}, we obtain for the $\a$ component a decay with an extra logarithmic term in agreement with some results discussed in J.A.V.Kroon works, see \cite{Kr:log1}, \cite{Kr:log2} and references therein.}

\subsection{The boundness of the $\tilde{\cal Q}$ norms}
Once the $\cal K$ spacetimes are defined together with the asymptotic behaviour of the various Riemann components and connection coefficients, we can start proving the better asymptotic behaviour for the $\a$ and $\b$ components of the Riemann tensor. As explained at point ${\bf b)}$ of the introduction we cannot apply the result of \cite{Kl-Ni:peeling} in a direct way and we have to look first for a better asymptotic behaviour of the $\lie_TR$ null components. This is done defining some Bel-Robinson norms similar to those introduced in \cite{Kl-Ni:peeling}, see equations (2.6),...,(2.11), but relative to $\lie_TR$ and with a different weight factor, $|u|^{5+\ep}$ instead of  $|u|^{3+\ep}$. We define, therefore, the following $\tilde{\cal Q}$ norms:
\begin{eqnarray}\label{norme1}
&& \QQ(u,\ub) = \QQ_1(u,\ub)+\QQ_2(u,\ub)+\sum_{i=2}^{q-2}{\QQ}_{(q)}(u,\ub)\nn\\
&& \QQb(u,\ub) = \QQb_1(u,\ub)+\QQb_2(u,\ub)+\sum_{i=2}^{q-2}{\QQb}_{(q)}(u,\ub)\	,
\end{eqnarray}
where, denoting ${\cal K}$ the ``very external region" whose existence has already been proved and the regions $V(u,\ub)\subset {\cal K}$, with $S(u,\ub)=C(u)\cap\Cb(\ub)$,\footnote{Recall that in ${\cal K}$ a double null canonical foliation exists made by outgoing and incoming ``cones" $\{C(u)\},\{\Cb(\ub)\}$.}
\[V(u,\ub)=J^{(-)}(S(u,\ub))\cap{\cal K}\ ,\]
\begin{eqnarray}\label{norms1}
\QQ_1(u,\ub) &\equiv & \int_{C(u)\cap V(u,\ub)}|u|^{5+\e}Q(\lie_T
{\tW})(\bar{K},\bar{K},\bar{K},e_4)\nonumber\\
& &+\int_{C(u)\cap{V(u,\ub)}}|u|^{5+\e}Q(\lie_O
\tW)(\bar{K},\bar{K},T,e_4)\nonumber\\
\QQ_2(u,\ub) &\equiv &\int_{C(u)\cap{V(u,\ub)}}|u|^{5+\e}Q(\lie_O\lie_T
\tW)(\bar{K},\bar{K},\bar{K},e_4)\nonumber\\
& &+\int_{C(u)\cap{V(u,\ub)}}|u|^{5+\e}Q(\lie^2_O\tW)(\bar{K},\bar{K},T,e_4)\label{Q_2}\\
& & +\int_{C(u)\cap{V(u,\ub)}}|u|^{5+\e}Q(\lie_S\lie_T\tW)(\bar{K},\bar{K},\bar{K},e_4)\nn
\end{eqnarray}
\bea
\widetilde{\cal Q}_{(q)}(u,\ub)&\equiv&\sum_{i=2}^{q-2}\left\{\int_{C(u)\cap
V(u,\ub)}|u|^{5+\e}Q(\lie_{T}\lie_{O}^{i}\tW)(\acc,\acc,\acc,e_4)\right.\nn\\
&&\left.+\int_{C(u)\cap V(u,\ub)}|u|^{5+\e}Q(\lie_{O}^{i+1}\tW)(\acc,\acc,T,e_4)\right.\eql{2.2}\\
&&\left.+\int_{C(u)\cap V(u,\ub)}|u|^{5+\e}Q(\lie_{S}\lie_{T}\lie_O^{i-1}\tW)(\acc,\acc,\acc,e_4)\right\}\nn
\eea
\begin{eqnarray}\label{norms2}
\QQb_1(u,\ub)& \equiv &\int_{\un{C}(\ub)\cap{V(u,\ub)}}|u|^{5+\e}Q(\lie_T{\tilde R})(\bar{K},\bar{K},\bar{K},e_3)\nonumber\\
& &+\int_{\un{C}(\ub)\cap{V(u,\ub)}}|u|^{5+\e}Q(\lie_O
{\tilde R})(\bar{K},\bar{K},T,e_3)\ \ ,\nonumber
\end{eqnarray}
\begin{eqnarray}
\QQb_2(u,\ub) &\equiv &\int_{\un{C}(\ub)\cap{V(u,\ub)}}|u|^{5+\e}Q(\lie_O\lie_T{\tilde R})(\bar{K},\bar{K},\bar{K},e_3)\nonumber\\
& &+\int_{\un{C}(u)\cap{V(u,\ub)}}|u|^{5+\e}Q(\lie^2_O{\tilde R})(\bar{K},\bar{K},T,e_3)\label{Qb_2}\\
& & +\int_{\un{C}(\ub)\cap{V(u,\ub)}}|u|^{5+\e}Q(\lie_S\lie_T{\tilde R})(\bar{K},\bar{K},\bar{K},e_3)\nonumber
\end{eqnarray}
\bea
{\QQb}_{(q)}(u,\ub)&\equiv&\sum_{i=2}^{q-2}\left\{\int_{C(\la)\cap
V(\la,\nu)}|u|^{5+\e}Q(\lie_{T}\lie_{O}^{i}\tW)(\acc,\acc,\acc,e_3)\right.\nn\\
&&\left.+\int_{C(u)\cap V(u,\ub)}|u|^{5+\e}Q(\lie_{O}^{i+1}\tW)(\acc,\acc,T_0,e_3)\right.\eql{2.4}\\
&&\left.+\int_{C(u)\cap V(u,\ub)}|u|^{5+\e}Q(\lie_{S}\lie_{T_0}\lie_O^{i-1}\tW)(\acc,\acc,\acc,e_3)\right\}\ .\nn
\eea
and
\begin{eqnarray}
\QQ_1({\Si_0}) &\equiv &
\int_{\Si_0\cap{V(u,\ub)}}|u|^{5+\e}Q(\lie_T {\tilde R})(\bar{K},\bar{K},\bar{K},T)\nonumber\\
& & +\int_{\Si_0\cap{V(u,\ub)}}|u|^{5+\e}Q(\lie_O{\tilde R})(\bar{K},\bar{K},T,T)\\
\QQ_2({\Si_0}) &\equiv & \int_{\Si_0\cap{V(u,\ub)}}|u|^{5+\e}Q(\lie_O\lie_T {\tilde R})(\bar{K},\bar{K},\bar{K},T)\nonumber\\
& & +\int_{\Si_0\cap{V(u,\ub)}}|u|^{5+\e}Q(\lie_O^2{\tilde R})(\bar{K},\bar{K},T,T)\nonumber\\
& &+ \int_{\Si_0\cap{V(u,\ub)}}|u|^{5+\e}Q(\lie_S\lie_T{\tilde R})(\bar{K},\bar{K},\bar{K},T)\ \ .
\end{eqnarray}
\bea
\sum_{i=2}^{q-2}\widetilde{\cal Q}_{(q)}({\Si_0})&\equiv&\sum_{i=2}^{q-2}\left\{\int_{\Si_0\cap{V(u,\ub)}}|u|^{5+\e}Q(\lie_{T}\lie_{O}^{i}\tW)(\acc,\acc,\acc,T)\right.\nn\\
&&\left.+\int_{\Si_0\cap{V(u,\ub)}}|\la|^{5+\e}Q(\lie_{O}^{i+1}\tW)(\acc,\acc,T,T)\right.\eql{2.2a}\\
&&\left.+\int_{\Si_0\cap{V(u,\ub)}}|\la|^{5+\e}Q(\lie_{S}\lie_{T}\lie_O^{i-1}\tW)(\acc,\acc,\acc,T)\right\}\nn
\eea
We also introduce the following quantity: 
\begin{equation}
\QQ_{\cal{K}}\equiv\sup_{\{u,\ub|S(u,\ub)\subseteq\cal{K}\}}\{\QQ(u,\ub)+\QQb(u,\ub)\}\ .
\end{equation}
The initial conditions which define the spacetime ${\cal K}$ have been chosen in such a way that the $\QQ, \QQb$ norms are bounded on $\Si_0$ (due to the fact that we are considering ${\tilde R}=\lie_TR$ instead of $R$). This is discussed in greater detail in subsection \ref{S.S2.8} of the appendix.
\smallskip

\NI As discussed in the introduction, see {\bf b)}, to fulfill the expectation that, everywhere in $\cal K$, the ``Kerr part is ``basically" subtracted in $\lie_TR$, we have to prove that these norms are bounded everywhere, that is that $\QQ_{\cal{K}}$ is bounded. To prove it, proceeding in a similar way as in \cite{Kl-Ni:book}, we define a spacetime integral $\widehat{\cal E}(u,\ub)$, we call the ``Error",
\bea
{\widehat{\cal E}(u,\ub)}:=\left(\QQ(u,\ub)+\QQb(u,\ub)\right)-\left(\QQ({\Si_0}\cap V(u,\ub))+\QQb({\Si_0}\cap V(u,\ub))\right)\ \ \ \ \ 
\eea
and prove that the ``Error" can be bounded by 
\bea
{\widehat{\cal E}(u,\ub)}\leq \varepsilon_0 f(R_0)\!\!\left(\QQ(u,\ub)+\QQb(u,\ub)\right)
\!+\!\!\sup_{\Si_0\cap{V(u,\ub)}}\!\!\big(|r^3|u|^{3-\de}\overline{\ro({\tilde R})}|^2+|r^3|u|^{3-\de}\overline{\si({\tilde R})}|^2\big)\ \ \ 
\eea
with $f$ a function decaying in $R_0$.
\smallskip

\NI The boundedness of  $\QQ_{\cal{K}}$, as we said before, allows us to control the $L^2(C)$ and $L^2(\Cb)$ weighted norms for the null components of ${\tilde R}=\lie_TR$ and their derivatives up to fourth order, with the exception of $\ro({\tilde R})$ and $\si({\tilde R})$, and, using them, to obtain the $\sup$ norms and the decay of the same quantities.\footnote{This is done in any detail in Chapter 5 of \cite{Kl-Ni:book} for the derivatives of $R$ up to second order, the higher derivatives are easier and are estimated exactly in the same way.} The terms $\ro({\tilde R})$ and $\si({\tilde R})$ have to be treated differently. 
In fact only $\ro-{\overline\ro}, \si-{\overline\si}$ can be estimated by the $\tilde{\cal Q}$ norm associated to  $\lie_O{\tilde R}$,
via Poincar\`e Lemma. 

\NI The norms $\sup_{{\cal K}}|r^3|u|^{3-\de}(\overline{\ro({\tilde R})},\overline{\si({\tilde R})})|$ are estimated directly, using the evolution equation for $\ro$ and $\si$ along the incoming cones, in terms of the corresponding  initial data norm.\footnote{Of course we have also to prove that the initial data is finite and small, we discuss it later on.}

\NI Here the need of a bootstrap mechanism shows up: $\sup_{{\cal K}}|r^3|u|^{3-\de}(\overline{\ro({\tilde R})},\overline{\si({\tilde R})})|$ are estimated, using the transport equations along the incoming cones, in terms of the initial data $\sup_{\Si_0\cap{\cal K}}|r^3|u|^{3-\de}(\overline{\ro({\tilde R})},\overline{\si({\tilde R})})|$ and of an integral part which depends on the remaining null Riemann components. On the other side these remaining components are estimated in terms of the ${\cal Q}({\tilde R})$ norms which at their turn, to prove they are bounded, require a control of the $(\ro({\tilde R}),\si({\tilde R}))$ components. 

\subsection{The bootstrap assumption and the ``Error" estimate.}

The strategy to estimate the ``Error terms" is, basically, a much simpler version of  the one used in \cite{Kl-Ni:book}. The reason of the simplification is that in \cite{Kl-Ni:book} the estimate of the error terms required some bootstrap assumptions on the $\cal R$ norms of the Riemann components and on the $\cal O$ norms of the connection coefficients, see the discussion in Chapter 4. This was due to the fact that the control of the $\cal Q$ norm was an intermediate step to prove global existence. Here the global existence of the spacetime $\cal K$, is already proved, and we need a simpler bootstrap assumption only to estimate $(\ro(\lie_TR),\si(\lie_TR))$. The various steps are the following:
\smallskip

\NI a) We know that due to the initial conditions on $\Si_0$ the following estimate holds, with ${\tilde R}=\lie_TR$:
\begin{Le}\label{L2.1a}
Under the initial data assumptions on $\overline g$ and $\overline k$ given in Theorem \ref{Ktheorem2} we have the following estimates, 
\bea
\sup_{\Si_0\cap{V(u,\ub)}}|r^{6-\de}\overline{\ro({\tilde R})}|\leq c\varepsilon_0\ \ ,\ \ \sup_{\Si_0\cap{V(u,\ub)}}|r^{7-\de}\overline{\si({\tilde R})}|\leq c\varepsilon_0\ .
\eea
\end{Le}
\NI{\bf Proof:} The proof of Lemma \ref{L2.1a} is in the appendix.
\smallskip

\NI Therefore we can define a spacetime region ${\overline V}\equiv V({\overline u}=-R_0,\ {\overline \ub}=R_0+\de u)\subset {\cal K}$ where, by continuity,
\bea
\sup_{(u,\ub)\in\overline V}|r^3|u|^{3-\de}(\overline{\ro({\tilde R})},\overline{\si({\tilde R})})|^2\leq \varepsilon_1{\ \!\!\!^2}\ .\eql{ass2}
\eea
with $\varepsilon_1=c\varepsilon_0>\varepsilon_0$, but sufficiently small. In this region it is easy to prove, using also \ref{ass2}, that,
with an appropriate constant $c$,\footnote{We do not report the proof here as it is just a particular case of the proof of point iii).}
\bea
\sup_{(u,\ub)\in {\overline V}}\{\QQ(u,\ub)+\QQb(u,\ub)\}
\leq c\varepsilon_0(\varepsilon_0+\varepsilon_1)\ .\eql{ass3}
\eea
b) This ``local" result allows to state a bootstrap mechanism:
\medskip

\NI i) Define $J^{(-)}(S(u',\ub'))\equiv{\cal K}'\subset {\cal K}$ as the largest possible region where
\bea
&&\ML\sup_{(u,\ub)\in{\cal K}'}|r^3|u|^{3-\de}(\overline{\ro({\tilde R})},\overline{\si({\tilde R})})|^2\leq \varepsilon_1{\ \!\!\!^2}
\eql{ass3a}
\eea

\NI ii) Prove that using \ref{ass3a} the error in the region ${\cal K}'$ satisfies the following bound
\bea
&&\ML\sup_{(u,\ub)\in{\cal K}'}\widehat{\cal E}(u,\ub)\eql{ass4}\\
&&\ML\leq \varepsilon_0f(R_0)\!\!\left(\sup_{(u,\ub)\in{\cal K}'}\left(\QQ(u,\ub)+\QQb(u,\ub)\right)
+\sup_{(u,\ub)\in{\cal K}'}\big(|r^3|u|^{3-\de}\overline{\ro({\tilde R})}|^2+|r^3|u|^{3-\de}\overline{\si({\tilde R})}|^2\big)\right)\nn
\eea
where $f(R_0)$  goes to zero as $R_0\rightarrow\infty$ .
\smallskip

 \NI iii) This implies that, for an appropriate constant $c$
\bea
&&\ML\sup_{(u,\ub)\in{\cal K}'}\left(\QQ(u,\ub)+\QQb(u,\ub)\right)\eql{ass5}\\
&&\ML\leq \frac{1}{1-c\varepsilon_0}\left[\left(\QQ({\Si_0}\cap{\cal K}')+\QQb({\Si_0}\cap{\cal K}')\right)
+c\varepsilon_0\big(|r^3|u|^{3-\de}\overline{\ro({\tilde R})}|^2+|r^3|u|^{3-\de}\overline{\si({\tilde R})}|^2\big)\right]\ \ \ \nn
\eea

\NI iv) The estimate of  $\left(\QQ(u,\ub)+\QQb(u,\ub)\right)$ in \ref{ass5} allows to prove, using the transport equation for $|r^{3-\frac{2}{p}}|u|^{3-\de}(\overline{\ro({\tilde R})},\overline{\si({\tilde R})})|_{p,S}$ (and the analogous one for $\nabb\ro$), the following estimate 
\bea
&&\ML\ML\sup_{(u,\ub)\in{\cal K}'}|r^3|u|^{3-\de}(\overline{\ro({\tilde R})},\overline{\si({\tilde R})})|^2\leq \sup_{\Si_0\cap{{\cal K}'}}|r^3|u|^{3-\de}
(\overline{\ro({\tilde R})},\overline{\si({\tilde R})})|^2\eql{ass5a}\\
&&\ML+c\left[\sup_{(u,\ub)\in{\cal K}'}\left(\QQ(u,\ub)+\QQb(u,\ub)\right)
+\varepsilon_0{\ \!\!^2}\!\!\sup_{(u,\ub)\in{\cal K}'}\big(|r^3|u|^{3-\de}\overline{\ro({\tilde R})}|^2+|r^3|u|^{3-\de}\overline{\si({\tilde R})}|^2\big)\right]\nn\\
&&\ML\leq \sup_{\Si_0\cap{{\cal K}'}}|r^3|u|^{3-\de}\overline{\ro({\tilde R})}|^2\nn\\
&&\ML+c\left[\left(\QQ({\Si_0}\cap{\cal K}')+\QQb({\Si_0}\cap{\cal K}')\right)
+(\varepsilon_0+\varepsilon_0{\ \!\!^2})\!\!\sup_{(u,\ub)\in{\cal K}'}\big(|r^3|u|^{3-\de}\overline{\ro({\tilde R})}|^2+|r^3|u|^{3-\de}\overline{\si({\tilde R})}|^2\big)\right]\nn
\eea
which implies, choosing $\varepsilon_0$ sufficiently small
\bea
&&\sup_{(u,\ub)\in{\cal K}'}|r^3|u|^{3-\de}(\overline{\ro({\tilde R})},\overline{\si({\tilde R})})|^2\\
&&\leq \frac{c}{1-c\varepsilon_0}
\left(\sup_{\Si_0\cap{{\cal K}'}}|r^3|u|^{3-\de}(\overline{\ro({\tilde R})},\overline{\si({\tilde R})})|^2
+\left(\QQ({\Si_0}\cap{\cal K}')+\QQb({\Si_0}\cap{\cal K}')\right)\right)\nn\\
&&\leq \frac{c}{1-c\varepsilon_0}
\left(\varepsilon_0{\ \!\!^2}+\left(\QQ({\Si_0}\cap{\cal K}')+\QQb({\Si_0}\cap{\cal K}')\right)\right)\leq \frac{c\varepsilon_0{\ \!\!^2}}{1-c\varepsilon_0}\ .\nn
\eea

\NI v) Choosing $\varepsilon_0$ sufficiently small the previous inequality implies that
\bea
\sup_{(u,\ub)\in{\cal K}'}|r^3|u|^{3-\de}(\overline{\ro({\tilde R})},\overline{\si({\tilde R})})|^2\leq \frac{\varepsilon_1{\ \!\!^2}}{2}\eql{ass7}
\eea
which contradicts the definition of ${\cal K}'$ unless it coincides with the whole spacetime ${\cal K}$. We have, therefore proved the estimate
\bea
\sup_{(u,\ub)\in{\cal K}}|r^3|u|^{3-\de}(\overline{\ro({\tilde R})},\overline{\si({\tilde R})})|^2\leq {\varepsilon_1{\ \!\!^2}}\ .\eql{ass7aass7}
\eea

\NI vi) Using these estimates and the assumption made on $\varepsilon_1$ we can compute the error in the whole region ${\cal K}$ proving that
\bea
&&\ML\sup_{(u,\ub)\in{\cal K}}\widehat{\cal E}(u,\ub)\eql{ass8}\\
&&\ML\leq c\varepsilon_0\!\left(\sup_{(u,\ub)\in{\cal K}}\left(\QQ(u,\ub)+\QQb(u,\ub)\right)
+\sup_{(u,\ub)\in{\cal K}}\big(|r^3|u|^{3-\de}\overline{\ro({\tilde R})}|^2+|r^3|u|^{3-\de}\overline{\si({\tilde R})}|^2\big)\right)\ .\nn
\eea
which, at its turn, implies that, for an appropriate constant $c$, repeating the result at point iv),
\bea
&&\ML\sup_{(u,\ub)\in{\cal K}}\left(\QQ(u,\ub)+\QQb(u,\ub)\right)\eql{ass9}\\
&&\ML\leq \frac{1}{1-c\varepsilon_0}\left(\QQ({\Si_0}\cap{\cal K})+\QQb({\Si_0}\cap{\cal K})
+{\varepsilon_0}\sup_{(u,\ub)\in{\cal K}}\big(|r^3|u|^{3-\de}\overline{\ro({\tilde R})}|^2+|r^3|u|^{3-\de}\overline{\si({\tilde R})}|^2\big)\right)\nn\\
&&\ML\leq c\varepsilon_0(\varepsilon_0+\varepsilon_1{\ \!\!^2})\nn
\eea

\NI vii) The control of these ${\tilde{\cal Q}}$ norms and of $|r^3|u|^{3-\de}(\overline{\ro({\tilde R})},\overline{\si({\tilde R})})|^2$ allows to control the $\sup$ norms of all the components of $\lie_TR$ with some weights (not yet the final ones).

\NI If we had used only the estimates proved in \cite{Kl-Ni:book} then in all the previous steps the exponent $3\!-\!\de$ would be substituted by $\frac{5}{2}$. 
\smallskip

\NI{\bf Remarks:} {\em 
\smallskip

\NI 1) The estimates \ref{ass7} which are proved in the sequel could be improved, with some extra work, for the $\si$ component due to the fact that its estimate on $\Si_0$ is a better one. The result we prove here is obtained looking at the transport equation for $\ro$ and $\si$ instead of $\overline{\ro}$ and $\overline{\si}$ and does not allow to distinguish between $\ro$ and $\si$.
\smallskip

\NI 2) In the previous argument  that $\varepsilon_1$ be small is not really needed. In fact, as said before, we do not need it as the spacetime $\cal K$ is already given.
\smallskip

\NI 3) Here and in the whole work we denote with $c$ many different constants, which, assumed $\varepsilon_0$ small, do not depend on $\varepsilon_0$ or $\varepsilon_1$.
\smallskip

\NI 4) In this argument and more in general in the whole work the fact the $R_0$ has to be large is used, as in $\cite{Kl-Ni:book}$, to prove that the spacetime $\cal K$ does exist and follows from the requirement that ${M}/{R_0}$ has to be small. As discussed in the introduction, see eq.\ref{1.3}, the requirement of the smallness of the ratio $J/M$ angular momentum can be weakened increasing $R_0$.}
\smallskip

\NI To prove the results described in i).....vii) we have to prove inequality \ref{ass4} and the first inequality of \ref{ass5a}. This is the content of the following 
\begin{thm}\label{T3.1}
In the spacetime ${\cal K}$, see Theorem \ref{Ktheorem2}, the following inequalities hold
\bea
&&\ML\sup_{(u,\ub)\in{\cal K}}|r^3|u|^{3-\de}(\overline{\ro({\tilde R})},\overline{\si({\tilde R})})|^2\leq 
\sup_{\Si_0\cap{{\cal K}}}|r^3|u|^{3-\de}(\overline{\ro({\tilde R})},\overline{\si({\tilde R})})|^2\eql{aass5b}\\
&&\ML+c\left[\sup_{(u,\ub)\in{\cal K}}\left(\QQ(u,\ub)+\QQb(u,\ub)\right)
+\varepsilon_0{\ \!\!^2}\bigg(1+\sup_{(u,\ub)\in{\cal K}}\big(|r^3|u|^{3-\de}\overline{\ro({\tilde R})}|^2+|r^3|u|^{3-\de}\overline{\si({\tilde R})}|^2\big)\bigg)\right]\nn\\
&&\nn\\
&&\nn\\
&&\ML\sup_{(u,\ub)\in{\cal K}}\widehat{\cal E}(u,\ub)\eql{aass4b}\\
&&\ML\leq \varepsilon_0f(R_0)\!\!\left(\sup_{(u,\ub)\in{\cal K}}\left(\QQ(u,\ub)+\QQb(u,\ub)\right)
+\sup_{(u,\ub)\in{\cal K}}\big(|r^3|u|^{3-\de}\overline{\ro({\tilde R})}|^2+|r^3|u|^{3-\de}\overline{\si({\tilde R})}|^2\big)\right)\ .\nn
\eea
\end{thm}
\NI{\bf Proof of \ref{aass5b}:} We prove the estimate for the $\ro$ component, the one for $\si$ goes exactly in the same way. 
\smallskip

\NI The evolution equation of $\ro$ along the incoming cones is:
\bea
\frac{d}{du}\ro+\frac{3}{2}\oom tr\chib\ro=\oom\left(-\divv\bb-2\eta\cdot\bb-\frac{1}{2}\chih\cdot\aa+\zeta\cdot\bb\right)\ .
\eea
Applying Lemma 4.1.5 of \cite{Kl-Ni:book} it follows:
\bea
|r^{(3-\frac{2}{p})}\ro({\tilde R})|_{p,S}(u,\ub)\leq c_0\left(|r^{(3-\frac{2}{p})}\ro({\tilde R})|_{p,S}(u_0,\ub)
+\int_{u_0}^u|r^{(3-\frac{2}{p})}{\tilde F}|_{p,S}(\la,\ub)d\la\right)\ \ \ \ \ \eql{roeqq}
\eea
where
\bea
{\tilde F}:=\oom\!\left(\!-\divv\bb({\tilde R})-2\eta\cdot\bb({\tilde R})-\frac{1}{2}\chih\cdot\aa({\tilde R})+\zeta\cdot\bb({\tilde R})\!\right)\ 
\eea
and
\bea
S(u_0,\ub)=\Cb(\ub)\cap\Si_0\ .
\eea
As in $\cal K$, $|u|\leq |u_0|$, multiplying the left hand side of \ref{roeqq} with $|u|^{3-\de}$ we obtain the inequality
\bea
&&|r^{(3-\frac{2}{p})}|u|^{3-\de}\ro({\tilde R})|_{p,S}(u,\ub)\eql{roeqq1}\\
&&\leq c_0\left(|r^{(3-\frac{2}{p})}|u_0|^{3-\de}\ro({\tilde R})|_{p,S}(u_0,\ub)
+\int_{u_0}^u|r^{(3-\frac{2}{p})}|\la|^{3-\de}{\tilde F}|_{p,S}(\la,\ub)d\la\right)\ .\nn
\eea
We consider this inequality for $p=4$ as this is what we need to get $\sup$ estimates.

\NI We have to prove that the estimate of $|r^{(3-\frac{2}{p})}|\la|^{3-\de}{\tilde F}|_{p=4,S}(\la,\ub)$
in terms of the $\tilde{\cal Q}$ norms makes the integral in \ref{roeqq1} uniformily bounded. In the inequality
\bea
|r^{(3-\frac{2}{p})}|\la|^{3-\de}{\tilde F}|_{p=4,S}\!&\leq&\! c\left(|r^{(3-\frac{2}{4})}|\la|^{3-\de}\nabb{\bb}|_{p=4,S}
+2|r^{(3-\frac{2}{4})}|\la|^{3-\de}\eta{\bb}|_{p=4,S}\right.\ \ \ \ \ \ \ \ \ \ \ \ \ \eql{2.50}\\
\!&+&\!\left.\frac{1}{2}|r^{(3-\frac{2}{4})}|\la|^{3-\de}\chih\aa|_{p=4,S}+|r^{(3-\frac{2}{4})}|\la|^{3-\de}\ze{\bb}|_{p=4,S}\right)\nn
\eea
\medskip

\NI Next Lemma estimates  the term $|r^{(3-\frac{2}{4})}|\la|^{3-\de}\nabb{\bb}|_{p=4,S}(\la,\ub)$ and the term $|r^{(3-\frac{2}{4})}|\la|^{{(3-\de)}}\eta\!\cdot\!{\bb}|_{p=4,S}$, the other ones have a better decay and can be treated in a similar way.
\begin{Le}\label{L2.2new}
The following estimate holds
\bea
&&\ML|r^{(3-\frac{2}{4})}|\la|^{3-\de}\nabb{\bb}|_{p=4,S}(\la,\ub)\eql{esta}\\
&&\leq \frac{c}{|\la|^{1+\frac{(\ep+2\de)}{2}}}\left({\QQb}^{\frac{1}{2}}(\la,\ub)
+{\varepsilon_0}\bigg(1+\sup_{(\la,\ub)\in{\cal K}_2}\!\big(|r^3|\la|^{3-\de}\overline{\ro({\tilde R})}|+|r^3|\la|^{3-\de}\overline{\si({\tilde R})}|\big)\bigg)\right)\nn\\
&&\nn\\
&&\ML|r^{(3-\frac{2}{4})}|\la|^{{(3-\de)}}\eta\!\cdot\!{\bb}|_{p=4,S}\eql{estb}\\
&&\leq \frac{1}{|\la|^{2-\frac{\de}{2}}}\left(\sup_{(\la,\ub)\in{\cal K}_2}\QQ(\la,\ub)^{\frac{1}{2}}
+{\varepsilon_0}\bigg(1+\sup_{(\la,\ub)\in{\cal K}_2}\!\big(|r^3|\la|^{3-\de}\overline{\ro({\tilde R})}|+|r^3|\la|^{3-\de}\overline{\si({\tilde R})\big)}|\bigg)\right)\nn
\eea
\end{Le}
\medskip

\NI{\bf Proof:} See the appendix. This implies that  
\bea
&&\ML\int_{u_0}^u|r^{(3-\frac{2}{4})}|\la|^{3-\de}\nabb{\bb(\tilde R)}|_{p=4,S}d\la=
O\!\left(\frac{1}{|u|^{\frac{({\tilde\ep}+2\de)}{2}}}\right)\left(\varepsilon_0+\!\!\sup_{(u,\ub)\in{\cal K}}\QQb(u,\ub)^{\frac{1}{2}}\right.\nn\\
&&\ML\left.\ \ \ \ \ \ \ \ \ \ \ \ \ \ \ \ \ \ \ \ 
+\frac{\varepsilon_0}{|\la|^{1-\frac{2\de+\ep}{2}}}\sup_{(u,\ub)\in{\cal K}}(|r^3|\la|^{3-\de}\overline{\ro({\tilde R})}|+|r^3|\la|^{3-\de}\overline{\si({\tilde R})}|)\right)\ .\nn
\eea
and
\bea
&&\int_{u_0}^u|r^{(3-\frac{2}{4})}|\la|^{{(3-\de)}}\eta\cdot{\bb}|_{p=4,S}d\la\\
&&\leq c\frac{1}{|u|^{1-\frac{\de}{2}}}\left(\varepsilon_0+\sup_{(u,\ub)\in{\cal K}}\QQ(u,\ub)^{\frac{1}{2}}+{\varepsilon_0}\!\!\sup_{(u,\ub)\in{\cal K}}\!\big(|r^3|\la|^{3-\de}
\overline{\ro({\tilde R})}|+|r^3|\la|^{3-\de}\overline{\si({\tilde R})}|\big)\right)\ .\nn
\eea
The other terms in \ref{2.50} can be treated in the same way, therefore we obtain finally
\bea
&&\ML|r^{(3-\frac{2}{p})}|u|^{3-\de}\ro({\tilde R})|_{p,S}(u,\ub)\leq c_0\left.\bigg[|r^{(3-\frac{2}{p})}|u|^{3-\de}\ro({\tilde R})|_{p,S}(u_0,\ub)\right.\nn\\
&&\ML\left.+c\frac{1}{|u|^{1-\frac{\de}{2}}}\left(\sup_{(u,\ub)\in{\cal K}}\!\!\big(\QQ(u,\ub)^{\frac{1}{2}}+\QQb(u,\ub)^{\frac{1}{2}}\big)
+\varepsilon_0\bigg(1+\sup_{(u,\ub)\in{\cal K}}\!\big(|r^3|\la|^{3-\de}\overline{\ro({\tilde R})}|+|r^3|\la|^{3-\de}\overline{\si({\tilde R})}|\big)\bigg)\right)\right]\ .\nn
\eea
Deriving tangentially the transport equation for $\ro$ and proceeding as before, we obtain analogous estimates for 
$|r^{(4-\frac{2}{p})}|u|^{3-\de}\nabb\ro({\tilde R})|_{p,S}(\la,\ub)$, then applying Lemma 4.1.3 of \cite{Kl-Ni:book} and taking the $\sup$ norms
we obtain the desired inequality \ref{aass5b} proving the first part of the theorem.
\smallskip

\NI{\bf Remarks:}{\em We do not prove here this estimate for $\nabb\ro({\tilde R})$ as we will prove a better estimate in Lemma \ref{L2.2nabbro}. This improved estimate will be needed in the sequel.}
\smallskip

\NI{\bf Proof of \ref{aass4b}:}
\smallskip

\NI In the $\tilde{\cal Q}$ norms, there are some differences with respect to those defined in \cite{Kl-Ni:book}, the more important being, obviously, the presence in these norms of the extra weight factor $|u|^{5+\ep}$; the control of the norms with an extra factor $|u|^{\si}$ has been already examined  in \cite{Kl-Ni:peeling}.

\NI The important property which has been proved there and which is one of the central features of the whole approach is that the ``Error", $\widehat{\cal E}(u,\ub)$, which has to be controlled to prove the boundedness of the $\tilde{\cal Q}$ norms, has exactly the same structure as the one relative to the $\cal Q$ norms. This is because, in the external region (or more in general, in the region where $u$ is negative), the extra terms in $\widehat{\cal E}(u,\ub)$ produced by the weight factor $|u|^{\si}$ give a negative contribution and, therefore, can be neglected.\footnote{See eq. (3.15) of \cite{Kl-Ni:peeling}.}

\NI We start examining  the general structure of $\widehat{\cal E}(u,\ub)$. Proceeding exactly as in \cite{Kl-Ni:peeling}, see equations (3.15), (3.16), (3.17), we know that the expression
of $\widehat{\cal E}(u,\ub)$ for a Weyl tensor $W$, (which in the present case is $\lie_O{\tilde R}\ ,\ \lie_T{\tilde R}$ or terms with more (modified) Lie derivatives) is, neglecting the negative terms of the error,\footnote{The final expression of the error will be the sum of the
$\widehat{\cal E}(W)(u,\ub)$ with respect to all the possible $W$ appearing in the $\tilde{\cal Q}$ norms.}
\bea
&&\widehat{\cal E}(W)(X,Y,Z)=\nn\\
&&-\int_{V_{(u,\ub)}}|u|^{\si}\bigg[DivQ(W)_{\b\ga\de}X^{\b}Y^{\ga}Z^{\de}+\frac{1}{2}Q^{\a\b\ga\de}(W)\!\left(\pii{X}_{\a\b}Y_{\ga}Z_{\de}\right.\nn\\
&&\left.\ \ \ \ \ \ \ \ \ \ \ \ \ \ \ \ \ +\pii{Y}_{\a\b}Z_{\ga}X_{\de}+\pii{Z}_{\a\b}X_{\ga}Y_{\de}\right)\bigg]
\eea
where $\pii{X}_{\a\b}$ is the deformation tensor relative to the vector field $X$.

\NI Symbolically we can write, denoting with $\Pi$ a generic term at the connection coefficients level:\footnote{With this we mean terms proportional to connection coefficients or linear combinations of them.}

\bea
\widehat{\cal E}(W)=\sum\int_{\cal K}|u|^{5+\ep}\tau_{-}^{a}\tau_{+}^{b}(D^{\a}{\tilde R})\c(D^{\b}{\tilde R})\c(D^{\ga}\Pi)\eql{3.20}
\eea
where $\a+\b+\ga\geq 2$, the factor $\tau_{-}^{a}\tau_{+}^{b}$ represents, with appropriate $a$ and $b$,
the weights already present in the analogous terms of \cite{Kl-Ni:book} and $W$ denote the term $\lie^{\de}_X{\tilde R}$
with $\de\in\{1,2,...,q\!-\!1\}$, see \ref{2.2},..,\ref{2.4}. The factor $|u|^{5+\ep}$ is the extra weight factor which we have to cope with.

\NI{\bf Remark:} {\em That all the terms of the ``Error" have the structure \ref{3.20} can be seen explicitely in Chapter 6 of \cite{Kl-Ni:book} where  all the terms are explicitely written and analyzed.}

\NI At first glance we might expect that each of the terms $D^\a{\tilde R}, D^\b{\tilde R}, D^\ga \Pi$ appearing on the right hand side of
\ref{3.20},\footnote{$D^{\ga}\Pi$ denotes the covariant derivative of an arbitrary deformation tensor.
Although the term $\Pi$ is at the level of the connection coefficient and therefore has ``a derivative less" with respect to the $R$ terms this fact is already
present at the level of decay in \cite{Kl-Ni:book} and, therefore, does not change the possible presence of the extra decay factor.}
can absorb a factor $|u|^{\frac{5+\ep}{2}}$ for the reason that the $L^2$ norms of the various Riemann components have these weight factors as these norms are estimated in terms of these $\tilde{\cal Q}$ norms (defined with these weight factors).
This will more than compensate the presence of the weight $|u|^{5+\ep}$ in \ref{3.20}.\footnote{To make this statement clearer recall that, apart from a $\ro$ term the estimate we need to prove for the ``Error" is ${\widehat{\cal E}}\leq c\varepsilon_0{\tilde{\cal Q}}$.}

\NI Unfortunately this is not quite true, indeed the symbolic expression $D^\a{\tilde R}\!\c\!D^\b{\tilde R}\c D^\ga \Pi$ hides the
presence of terms such as $\rho({\tilde R}),\si({\tilde R})$  or $\tr\chi$, $\tr\chib $, $\om$, $\omb$. To prove that the term $\rho({\tilde R}),\si({\tilde R})$ can absorb in its $L^2$ norms some additional weight requires more care and has been proved when we estimate separately, via a bootstrap mechanism, $\sup_{{\cal K}}|r^3|u|^{3-\de}{\ro({\tilde R})}|$ and $\sup_{{\cal K}}|r^3|u|^{3-\de}\si({\tilde R})|$.\footnote{The additional weight which $\ro({\tilde R}),\si({\tilde R})$ can tolerate is smaller than the one for the remaining Riemann components, but still sufficient to prove the result. Observe, moreover, that the proof we present here of inequality \ref{aass4b} has to be used ``before" to prove inequality \ref{ass4} and to complete the bootstrap mechanism.} 

\NI Moreover the norms of the connection coefficients $\tr\chi$, $\tr\chib $, $\om$, $\omb$
cannot absorb  any additional weight factor as they are different from zero even in the {\Sch} spacetime. Nevertheless, with the present definition of the vector field $T$ \!\footnote{The same as in \cite{Kl-Ni:peeling}, different from \cite{Kl-Ni:book}.} and of the canonical foliation these connection coefficients  only appear in the ``Error", when together with $\ro$, in the linear combinations $\tr\chi+\tr\chib$ and $\om+\omb$ which are identically zero in Kerr spacetime and have  a better and sufficient asymptotic behaviour as it has been show in  Theorem \ref{Ktheorem2}.

\NI Most of the error terms, nevertheless, do not have this problem and they can be estimated as in \cite{Kl-Ni:book}; the control of the extra weight can be done as said before and has been carefully proved in a kind of linearized version discussed in a paper by G.Caciotta and T.Raparelli to appear. Here we do not repeat their proof and we examine only some of the more problematic error terms.

\subsubsection{Some error terms}
As we said the more delicate terms are those containing $\ro({\tilde R})$ (and  also $\si({\tilde R})$, but the terms with $\si$ have always an extra $r^{-1}$ decay and therefore are easier). These terms are of the following kind:
\bea
\int_{\cal K}|u|^{5+\ep}\tau_{-}^{a}\tau_{+}^{b}{\ro({\tilde R})}(\lie^{\a}_X{\tilde R})(D^{\ga}\Pi)\eql{err1}
\eea
where $(\lie^{\a}_X{\tilde R})$ is, symbolically, a null component of $\lie^{\a}_X{\tilde R}$, with $\a\!\geq\!1$. This last term multiplied by its weight including the factor $|u|^{\frac{5+\ep}{2}}$ can be estimated by a $\tilde{\cal Q}$ norm. The  factor $\ro({\tilde R})$  
has been estimated before, using the evolution equation for $\ro$ along the incoming cones, in terms of the initial data term, $\sup_{\Si_0\cap{\cal K}}|r^3|u|^{3-\de}{\ro({\tilde R})}|$. 

\NI To show how to control the error term \ref{err1} we estimate it following \cite{Kl-Ni:book}, Chapter 6. More precisely we estimate one of the terms with this structure, namely a term analogous to the one in the first line of eq.(6.2.1).
\smallskip 

\NI The term we are considering is one of the many  terms which compose the part of the error ${\cal E}_1$, see \cite{Kl-Ni:book} page 242, relative to the spacetime region $V(\overline{u},\overline{\ub})\subset{\cal K}$, the part over $\Si_0$ of $J^{(-)}(S(\overline{u},\overline{\ub}))$. The boundaries of this region are 
\[\Si_0\cap J^{(-)}(S(\overline{u},\overline{\ub}))\ \ ,\ \ C(\overline{u},[\ub_0(\overline{u}),\overline{\ub}])\ \ ,
\ \ \Cb(\overline{\ub},[u_0(\overline{\ub}),\overline{u}])\ ,\]
where
\[\ub_0(\overline{u})=|\overline{u}|\ \ ,\ \ u_0(\overline{\ub})=-\overline{\ub}\ .\]

\NI Therefore we consider 
\bea
\int_{V(\overline{u},\overline{\ub})}|u|^{5+\ep}\tau_{+}^{6}\a(\lie_{T}{\tilde R})\cdot\Theta(T,{\tilde R})\eql{err2}
\eea
a term of
$\int_{V(\overline{u},\overline{\ub})}|u|^{5+\ep}DivQ(\lie_{T}{\tilde R})_{\b\ga\de}\acc^{\b}\acc^{\ga}\acc^{\de}$,
and, more specifically, between the various terms which compose $\Theta(T,{\tilde R})$ we choose, see equation (6.1.14) of \cite{Kl-Ni:book}, the term proportional to ${^{(T)}}p_4(\ro({\tilde R}),\si({\tilde R}))$.
In conclusion the term we want to control is
\bea
\int_{V(\overline{u},\overline{\ub})}|u|^{5+\ep}\tau_{+}^{6}\a(\lie_{T}{\tilde R})\cdot\left({^{(T)}}p_4(\ro({\tilde R}),\si({\tilde R}))\right)\eql{err2a}\ .
\eea
Proceeding as in \cite{Kl-Ni:book} Chapter 6 page 260, we write\footnote{We consider only the $\ro$ part as the easier $\si$ part can be treated in a similar way.}
\bea
&&\ML\left|\int_{V(\overline{u},\overline{\ub})}|u|^{5+\ep}\tau_{+}^{6}\a(\lie_{T}{\tilde R})\cdot\left({^{(T)}}p_4(\ro({\tilde R}),\si({\tilde R}))\right)\right|\\
&&\ML\leq c\int_{u_0(\overline{\ub})}^{\overline u}du'\left(\int_{C(u')\cap{V(\overline{u},\overline{\ub})}}|u'|^{5+\ep}\tau_{+}^{6}|\a(\lie_{T}{\tilde R})|^2\right)^{\frac{1}{2}}
\!\left(\int_{C(u')\cap{V(\overline{u},\overline{\ub})}}|{^{(T)}}p_4|^2|u'|^{5+\ep}\tau_{+}^{6}|\ro({\tilde R})|^2\right)^{\frac{1}{2}}\nn\\
&&\ML\leq c{\tilde{\cal Q}_1}^{\frac{1}{2}}\int_{u_0(\overline{\ub})}^{\overline u}du'\left(\int_{C(u')\cap{V(\overline{u},\overline{\ub})}}|{^{(T)}}p_4|^2||u'|^{\frac{5+\ep}{2}}\tau_{+}^{3}\ro({\tilde R})|^2\right)^{\frac{1}{2}}\ .\nn
\eea
The estimate of the last integral is:
\bea
&&\ML\left[\!\!\left(\int_{C(u')\cap{V(\overline{u},\overline{\ub})}}|{^{(T)}}p_4|^2\left||u'|^{\frac{5+\ep}{2}}\tau_{+}^{3}\ro({\tilde R})\right|^2\right)^{\frac{1}{2}}\right]\!\!\nn\\
&&\ML\leq\!\left(\sup_{{V(\overline{u},\overline{\ub})}}\left||u'|^{3-\de}\tau_{+}^{3}\ro({\tilde R})\right|\right)\!\frac{1}{|u|^{\frac{1}{2}-(\frac{\ep}{2}+\de)}}\!\left(\int_{C(u')\cap{V(\overline{u},\overline{\ub})}}|{^{(T)}}p_4|^2\!\right)^{\!\!\frac{1}{2}}\nn\\
&&\ML\leq\left(\sup_{{V(\overline{u},\overline{\ub})}}\left||u'|^{3-\de}\tau_{+}^{3}\ro({\tilde R})\right|\right)\frac{1}{|u'|^{\frac{3}{2}-(\frac{\ep}{2}+\de)}}\left(\int_{\ub_0(u')}^{\infty}\frac{1}{\tau_+^2}|\tau_+\tau_-{^{(T)}}p_4|_{p=2,S}^2\right)^{\frac{1}{2}}\nn\\
&&\ML\leq\left(\sup_{{V(\overline{u},\overline{\ub})}}\left||u'|^{3-\de}\tau_{+}^{3}\ro({\tilde R})\right|\right)\frac{1}{|u'|^{2-(\frac{\ep}{2}+\de)}}\left(\sup_{V(\overline{u},\overline{\ub})}|\tau_+\tau_-{^{(T)}}p_4|_{p=2,S}^2\right)^{\frac{1}{2}}\nn\\
&&\ML\leq c\frac{\varepsilon_0}{|u'|^{2-(\frac{\ep}{2}+\de)}}\left(\sup_{{V(\overline{u},\overline{\ub})}}\left||u'|^{3-\de}\tau_{+}^{3}\ro({\tilde R})\right|\right)\eql{err11a}
\eea
recalling that $\ub_0(u')=|u'|$. The last two inequalities use the estimate for $p_4$ proved in \cite{Kl-Ni:book}, Proposition 6.1.4, which could be improved. 
Therefore as from (the first part of) Theorem \ref{T3.1} we have
\[\left||u'|^{3-\de}\tau_{+}^{3}\ro({\tilde R})\right|\leq \big({\varepsilon_0}+c\ \!{\tilde Q}_{\cal K}^{\frac{1}{2}}\big)\]
we conclude that
\bea
&&\ML \int_{\cal K}|u|^{5+\ep}\tau_{+}^{6}\a(\lie_{T}{\tilde R})\cdot\Theta(T,{\tilde R})
\leq c{\tilde{\cal Q}_1}^{\frac{1}{2}}\int_{u_0}^{\overline u}du'
\left(\int_{C(u')\cap{\cal K}}|{^{(T)}}p_4|^2||u'|^{\frac{5+\ep}{2}}\tau_{+}^{3}\ro({\tilde R})|^2\right)^{\frac{1}{2}}\nn\\
&&\ML\leq c{\tilde{\cal Q}_1}^{\frac{1}{2}}\left(\int_{u_0}^{\overline u}\frac{1}{|u'|^{\frac{5}{2}-(\frac{\ep}{2}+\de)}}du'\right)
\left(\sup_{{\cal K}}\left||u'|^{3-\de}\tau_{+}^{3}\ro({\tilde R})\right|\right)\left(\sup_{\cal K}|\tau_+\tau_-{^{(T)}}p_4|_{p=2,S}^2\right)^{\frac{1}{2}}\nn\\
&&\ML\leq c\frac{\varepsilon_0}{R_0^{\frac{3}{2}-(\frac{\ep}{2}+\de)}}\QQ_{\cal{K}}^{\frac{1}{2}}\big({\varepsilon_0}+c\ \!{\tilde Q}_{\cal K}^{\frac{1}{2}}\big)
\leq c\frac{\varepsilon_0}{R_0}\big({\varepsilon_0}^2+\ \!{\tilde Q}_{\cal K}\big)\ .\ \eql{err3}
\eea
in agreement with \ref{aass4b}.
\smallskip

\NI{\bf Conclusions:}\ {\em The previous discussion and estimate, although done for a specific term of the ``Error" should convince the reader that the $\tilde{\cal Q}$ norms are bounded. The remaining part of the paper is devoted to obtain decay informations from them. }

\subsection{The decay estimates for the $\lie_TR$ null components} Form the boundedness of the $\tilde{\cal Q}$ norms we obtain, proceeding as in \cite{Kl-Ni:book} and in the first part of \cite{Kl-Ni:peeling}, the following $\sup$ estimates:\footnote{Recall that the decay of the initial data has an $\ep'>\ep$.  Hereafter all the capital $C$ constants, $C,C_0,C_1,...$ are proportional to $\varepsilon_0$.}
\begin{eqnarray}
&&\sup_{\mathcal{K}}r^{\frac{7}{2}}|u|^{\frac{5}{2}+\frac{\e}{2}}|\a(\lie_{T}R)| \leq C_0,
\quad\sup_{\mathcal{K}}r^{\frac{7}{2}}|u|^{\frac{5}{2}+\frac{\e}{2}} |\b(\lie_{T}R)|\leq C_0\nn\\
&&\sup_{\mathcal{K}}r^3|u|^{3+\frac{\e}{2}}|\ro(\lie_{T}R)-\overline{\ro(\lie_{T}R)}| \leq C_0,\quad \sup r^3|u|^{3+\frac{\e}{2}}|\si(\lie_{T}R)
-\overline{\si(\lie_{T}R)}|\leq C_0\nn\\
&& \sup_{\mathcal{K}}r^2|u|^{4+\frac{\e}{2}}\bb(\lie_{T}R)\leq C_0,\quad\sup_{\mathcal{K}}r|u|^{5+\frac{\e}{2}}|\aa(\lie_{T}R)|\leq C_0\ .\eql{LTRdec1}
\end{eqnarray}
As in \cite{Kl-Ni:peeling} we use the transport equations for $\b(\lie_{T}R)$ and $\a(\lie_{T}R)$ along the incoming cones to obtain a better decay in $r$ for these components. More precisely we prove the following theorem:
\begin{thm}\label{T4.1}
In the spacetime $\cal K$, from the estimates \ref{LTRdec1} we derive the following ones
\begin{eqnarray}
&&\sup_{\mathcal{K}}r^{4}|u|^{2+\frac{\e''}{2}} |\b(\lie_TR)|\leq C_1\eql{bbest}\\
&&\sup_{\mathcal{K}}r^{5}|u|^{1+\frac{\e''}{2}}|\a(\lie_{T}R)| \leq C_1\eql{aaest}
\end{eqnarray} 
where $\e''<\e$.
\end{thm}
\NI{\bf Proof of \ref{bbest}:}
From the Bianchi equations, see (3.2.8) of \cite{Kl-Ni:book}, it follows that
$\b=\b(\lie_TR)$ satisfies, along the incoming null hypersurface $\Cb(\nu)$, the evolution equation
\bea
\dddd_3\b+\tr\chib\b=\nabb\ro+\left[2\omb\b+\dual\nabb\si+2\hat{\chi}\c\bb+
3(\eta\ro+\dual\eta\si)\right]\eql{1.20a}
\eea
which can be rewritten as
\bea
\frac{\partial\b_a}{\partial\la}+\oom\tr\chib\b_a=2\oom\omb\b_a+\oom\left[\nabb_a\ro+\dual\nabb_a\si+2(\hat{\chi}\c\bb)_a+
3(\eta\ro+\dual\eta\si)_a\right]\ \ \ \eql{1.8}
\eea
where\! \footnote{All the notations used in
this paper without an explicit definition are those already introduced in \cite{Kl-Ni:book}. The movig frame compatible with equation \ref{1.8} is the Fermi transported, see detailed discussion in \cite{Kl-Ni:book}.} $\b_a=\b(e_a)(\lie_TR)$. From this equation, see Chapter 4 of \cite{Kl-Ni:book}, we obtain the following inequality, 
\bea
\frac{d}{d\la}|r^{(2-\frac{2}{p})}\b|_{p,S}
&\leq&||2\oom\omb-{(1-1/p)}(\oom\tr\chib-\overline{\oom\tr\chib})||_{\infty}|r^{(2-\frac{2}{p})}\b|_{p,S}\eql{4.7}\\
&+&\|\oom\|_{\infty}\!\left(|r^{(2-\frac{2}{p})}\nabb\ro|_{p,S}
+3|r^{(2-\frac{2}{p})}\eta\ro|_{p,S}+|r^{(2-\frac{2}{p})}\tilde F|_{p,S}\right)\ ,\nn
\eea
where ${\tilde F}(\c)=2\hat{\chi}\c\bb+(\dual\nabb\si+3\dual\eta\si)$.\footnote{The term $\dual\nabb\si+3\dual\eta\si$ behaves as
the term $\nabb_a\ro+3\eta_a\ro$ and, therefore, we will not consider it explicitely.} Integrating along $\Cb(\nu)$, with $\la_1\!=\!u|_{\Cb(\nu)\cap\Si_0}$, we obtain
\bea
&&|r^{(2-\frac{2}{p})}\b|_{p,S}(\la,\nu)\leq|r^{(2-\frac{2}{p})}\b|_{p,S}(\la_1)\nn\\
&&\ \  +\int_{\la_1}^{\la}||2\oom\omb-{(1-1/p)}(\oom\tr\chib-\overline{\oom\tr\chib})||_{\infty}
|r^{(2-\frac{2}{p})}\b|_{p,S}(\la',\nu)\nn\\
&&\ \ +\  \|\oom\|_{\infty}\!\left(\int_{\la_1}^{\la}|r^{(2-\frac{2}{p})}\nabb\ro|_{p,S}\!
+\!3\!\int_{\la_1}^{\la}|r^{(2-\frac{2}{p})}\eta\ro|_{p,S}
\!+\!\frac{1}{2}\!\int_{\la_1}^{\la}|r^{(2-\frac{2}{p})}\tilde F|_{p,S}\right)\  .\nn
\eea 
In $\cal K$ we have at least the following behaviour \footnote{Consistent with the decay properties proved in
\cite{Kl-Ni:book}. In fact in ${\cal K}$ the estimates for the second term are better with respect to the $|\la|$ decay.}
\[\|\oom\omb\|_{\infty}=O(r^{-1}|\la|^{-1})\ \  \mbox{and}\ \ 
\|\oom\tr\chib-\overline{\oom\tr\chib}\|_{\infty}=O(r^{-2}|\la|^{-\frac{1}{2}})\ .\]
Therefore we can apply the Gronwall's Lemma obtaining:
\bea
|r^{2-\frac{2}{p}}\b|_{p,S}(\la,\nu)&\leq&
c\left[|r^{2-\frac{2}{p}}\b|_{p,S}(\la_1)+\|\oom\|_{\infty}\!\left(\!\int_{\la_1}^{\la}|r^{2-\frac{2}{p}}\nabb\ro|_{p,S}
\right.\right.\nn\\
&+&\left.\left.3\!\int_{\la_1}^{\la}|r^{2-\frac{2}{p}}\eta\ro|_{p,S}
+\frac{1}{2}\!\int_{\la_1}^{\la}|r^{2-\frac{2}{p}}\tilde{F}|_{p,S}\right)\right]
\eea
and recalling that $\|\oom\|_{\infty}\leq C$, we will inglobe,  hereafter, this factor in the constant $c$.  Multiplying both sides by
$r^2|\la|^{2+\frac{\e''}{2}}$, with $\ep>{\ep'}>\ep''>0$, remembering that $r(\la,\nu)<r(\la_1,\nu)$ and $|\la|<|\la_1|$, we obtain
\bea
&&\ML|r^{4-\frac{2}{p}}|\la|^{2+\frac{\e''}{2}}\b|_{p,S}(\la,\nu)
\leq c\left(|r^{4-\frac{2}{p}}|\la|^{2+\frac{\e''}{2}}\b|_{p,S}(\la_1)\right.\eql{6.17a}\\
&&\ML+\!\int_{\la_1}^{\la}\!|r^{4-\frac{2}{p}}|\la'|^{2+\frac{\e''}{2}}\nabb\ro|_{p,S}
\!+\!\left.3\!\int_{\la_1}^{\la}|r^{4-\frac{2}{p}}|\la'|^{2+\frac{\e''}{2}}\eta\ro|_{p,S}
\!+\!\frac{1}{2}\!\int_{\la_1}^{\la}|r^{4-\frac{2}{p}}|\la'|^{2+\frac{\e''}{2}}\tilde{F}|_{p,S}\!\right)\ .\nn
\eea
We examine the integrals in \ref{6.17a} and prove that they are uniformily bounded  in $\nu$.\footnote{Recall that all the null components $\a,\b,\ro,\si,...$ refer to $\lie_TR$.}
\medskip

\NI a) $\int_{\la_1}^{\la}|r^{4-\frac{2}{p}}|\la'|^{2+\frac{\e''}{2}}\nabb\ro(\lie_TR)|_{p,S}$:
\smallskip

\NI To prove the boundedness of this integral we use the following lemma:
\begin{Le}\label{L2.2nabbro}
In the spacetime $\cal K$ from the boundedness of the $\tilde{\cal Q}$ norms the following estimate holds:
\bea
\sup_{{\cal K}}\big|r^{4-\frac{2}{p}}|\la|^{\frac{6+\ep'}{2}}\nabb\ro({\tilde R})\big|_{p,S}\leq C\ .\eql{estq}
\eea
\end{Le}
\NI{\bf Proof:} See the appendix.
\medskip

\NI b) $\int_{\la_1}^{\la}|r^{4-\frac{2}{p}}|\la'|^{2+\frac{\e''}{2}}\eta\ro|_{p,S}$\ :
Recalling that $\eta$ satisfies in \cite{Kl-Ni:book},\bea
|r^{2-2/p}|\la|^{\frac{1}{2}}\eta|_{p,S}(\la,\nu)\leq c\ \ ,\ p\in [2,\infty]\ .\eql{6.19q}
\eea
Using the previous estimate for $\ro(\tilde R)$ in ${\cal K}$, 
\beaa
\sup_{{\cal K}}\left||u|^{(3-\de)}\tau_{+}^{3}\ro({\tilde R})\right| \leq \frac{c}{R_0}(\varepsilon_0+\QQ_{\cal{K}}^{\frac{1}{2}})\leq C\ ,
\eeaa
it follows immediately
\bea
\int_{\la_1}^{\la}|r^{4-\frac{2}{p}}|\la'|^{2+\frac{\e''}{2}}\eta\ro|_{p,S}\leq c
\int_{\la_1}^{\la}\frac{1}{r|\la'|^{\frac{3}{2}-\frac{2\de+\ep''}{2}}}\leq  c\ .
\eea
\medskip

\NI c) $\int_{\la_1}^{\la}|r^{4-\frac{2}{p}}|\la'|^{2+\frac{\e''}{2}}\tilde{F}|_{S,p}$\ :
\medskip

\NI From the expression ${\tilde F}(\c)=\dual\nabb\si+3\dual\eta\si+2\hat{\chi}\c\bb$ and the previous remark concerning
$\dual\nabb\si+3\dual\eta\si$, we have only to prove that
\bea
\int_{\la_1}^{\la}|r^{4-\frac{2}{p}}|\la'|^{2+\frac{\e''}{2}}\chih\bb|_{p,S}\leq c
\eea
This is easy, as, from the estimates \ref{LTRdec1}, we have
\beaa
\sup_{\cal K}|r^{2}|\la|^{4+\frac{\ep}{2}}\bb|\leq C_0\ ,\ \sup_{\cal K}||\la|^{\frac{1}{2}}r^{2-2/p}\chih|_{p,S}\leq C_0\ .
\eeaa
Therefore
\bea
\ML\int_{\la_1}^{\la}|r^{4-\frac{2}{p}}|\la'|^{2+\frac{\e''}{2}}\tilde{F}|_{p,S}\leq c
\int_{\la_1}^{\la}\frac{1}{|\la'|^{\frac{5}{2}+\frac{\ep-\ep''}{2}}}\leq \int_{\la_1}^{\la_0}\frac{1}{|\la'|^{\frac{5}{2}+\frac{\ep-\ep''}{2}}}
\leq \frac{c}{R_0^{\ \!\frac{3}{2}+\frac{\ep-\ep''}{2}}}\ ,\ \ \ \ 
\eea
where $\la_0=R_0$.
Collecting all these estimates for the integrals in \ref{6.17a} we infer that
\bea
|r^{4-\frac{2}{p}}|\la|^{2+\frac{\e''}{2}}\b|_{p,S}(\la,\nu)\leq
c\left(|r^{4-\frac{2}{p}}|\la|^{2+\frac{\e'}{2}}\b|_{p,S}(\la_1)+1\right)
\eea
and using the initial data assumptions we conclude that, 
\bea
|r^{4-\frac{2}{p}}|\la|^{2+\frac{\e''}{2}}\b|_{p,S}(\la,\nu)\leq c\ ,\ \  p\in [2,4]\ .
\eea
To prove the $\sup$ estimate \ref{bbest} we have to repeat for $\nabb\b$ the previous estimate for $\b$. This requires again the transport equation for $\nabb\b$ along the $\Cb$ ``cones" which at its turn requires the control of an extra derivative for $\ro$ and $\si$. This is the reason why we need a greater regularity in the initial data which translates in the introduction of $\cal Q$ norms with more Lie derivatives than in \cite{Kl-Ni:book}. We do not write the proof here as it goes, with the obvious changes, exactly as for the $\b$ estimate.

\smallskip

\NI{\bf Proof of \ref{aaest}:}
We look at the transport equation for $|r^{(1-\frac{2}{p})}\a(\lie_TR)|_{p,S}$\ .
From the evolution equation satified by $\a$, see \cite{Kl-Ni:book}, Chapter 3, equation (3.2.8),
\beaa
\frac{\partial\a}{\partial\la}+\frac{1}{2}\oom\tr\chib\a=4\oom\omb\a+\oom\left[\nabb\hot\b+\left(-3(\hat{\chi}\ro+
\dual\hat{\chi}\si)+(\zeta+4\eta)\hot\b\right)\right]\ ,\ \ \ \ \ 
\eeaa
it follows that
\bea
\frac{\partial |\a|^p}{\partial\la}&=&p|\a|^{p-1}\frac{\partial|\a|}{\partial\la}=
p|\a|^{p-2}\a\c\frac{\partial \a}{\partial\la}\nn\\
&=&p|\a|^{p-2}\a\c\left((-\frac{1}{2}\oom\tr\chib+4\oom\omb)\a+\oom\nabb\hot\b+F(\c)\right)\\
&=&p(-\frac{1}{2}\oom\tr\chib+4\oom\omb)|\a|^{p}+p|\a|^{p-2}\oom\a\c(\nabb\hot\b)+p|\a|^{p-2}\a\c F(\c)\ \nn
\eea
where
\[F=\oom\left(-3(\hat{\chi}\ro+\dual\hat{\chi}\si)+(\zeta+4\eta)\hot\b\right)\ .\]
Immediately,
\bea
\frac{\partial |\a|^p}{\partial\la}+\frac{p}{2}\oom\tr\chib|\a|^{p}=4p\oom\omb|\a|^{p}
+p|\a|^{p-2}\oom\a\c(\nabb\hot\b)+p|\a|^{p-2}\a\c F(\c)\ .\ \ \ \ \ \ 
\eea
As, see \cite{Kl-Ni:book}, Chapter 4,
\bea
\frac{d}{d\la}\left(\int_{S(\la,\nu)}r^{\si}|\a|^pd\mu_{\ga}\right)&=&
\int_{S(\la,\nu)}r^{\si}\left(\frac{\partial|\a|^p}{\partial\la}+(1+\frac{\si}{2})\oom\tr\chib|\a|^p\right)\nn\\
&-&\frac{\si}{2}\int_{S(\la,\nu)}r^{\si}|\a|^p(\oom\tr\chib-\overline{\oom\tr\chib})\ ,
\eea
choosing $\si=p(1-\frac{2}{p})$ we obtain
\bea
&&\frac{d}{d\la}\left(\int_{S(\la,\nu)}|r^{(1-\frac{2}{p})}\a|^pd\mu_{\ga}\right)=
\int_{S(\la,\nu)}r^{p(1-\frac{2}{p})}\left(\frac{\partial|\a|^p}{\partial\la}+\frac{p}{2}\oom\tr\chib|\a|^p\right)\nn\\
&&-\frac{p(1-\frac{2}{p})}{2}\int_{S(\la,\nu)}|r^{(1-\frac{2}{p})}\a|^p(\oom\tr\chib-\overline{\oom\tr\chib})\nn\\
&&=\int_{S(\la,\nu)}r^{p(1-\frac{2}{p})}\left(4p\oom\omb|\a|^{p}
+p|\a|^{p-2}\oom\a\c(\nabb\hot\b)+p|\a|^{p-2}\a\c F(\c)\right)\nn\\
&&-\left(\frac{p}{2}-1\right)\int_{S(\la,\nu)}|r^{(1-\frac{2}{p})}\a|^p(\oom\tr\chib-\overline{\oom\tr\chib})\ .
\eea
Therefore with $p\in [2,4]$ we have
\bea
\ML\frac{d}{d\la}|r^{(1-\frac{2}{p})}\a|_{p,S}^p\!&=&\!\int_{S(\la,\nu)}
\left(4p\oom\omb-\left(\frac{p}{2}-1\right)\!(\oom\tr\chib-\overline{\oom\tr\chib})\right)|r^{(1-\frac{2}{p})}\a|^p\nn\\
\ML\!&+&\!\int_{S(\la,\nu)}pr^{p(1-\frac{2}{p})}|\a|^{p-2}\oom\a\c(\nabb\hot\b)+\int_{S(\la,\nu)}pr^{p(1-\frac{2}{p})}|\a|^{p-2}\a\c F(\c)\nn
\eea
and, therefore,
\bea
&&p|r^{(1-\frac{2}{p})}\a|_{p,S}^{p-1}\frac{d}{d\la}|r^{(1-\frac{2}{p})}\a|_{p,S}\leq
\|\oom\|_{\infty}\!\left(4p|\omb|_{\infty}+\left(\frac{p}{2}-1\right)\!|\tr\chib-\overline{\tr\chib}|_{\infty}\right)\!|r^{(1-\frac{2}{p})}\a|_{p,S}^p\nn\\
&&+p\|\oom\|_{\infty}\|r^{5}|\la|^{2+\frac{\ep''}{2}}\nabb\b\|_{\infty}\frac{1}{r^{(4+\frac{2}{p})}|\la|^{2+\frac{\ep''}{2}}}\int_{S(\la,\nu)}|r^{(1-\frac{2}{p})}\a|^{p-1}+p\|r^{1-\frac{2}{p}}F\|_{\infty}\int_{S(\la,\nu)}|r^{(1-\frac{2}{p})}\a|^{p-1}\nn\\
&&\nn\\
&&\leq \|\oom\|_{\infty}\left(4p|\omb|_{\infty}+\big(\frac{p}{2}-1\big)\!|\tr\chib-\overline{\tr\chib}|_{\infty}\right)|r^{(1-\frac{2}{p})}\a|_{p,S}^{p}\nn\\
&&+p\|\oom\|_{\infty}\|r^{5}|\la|^{2+\frac{\ep''}{2}}\nabb\b\|_{\infty}\frac{1}{r^{4}|\la|^{2+\frac{\ep''}{2}}}|r^{(1-\frac{2}{p})}\a|_{p,s}^{p-1}
+p\|rF\|_{\infty}|r^{(1-\frac{2}{p})}\a|_{p,s}^{p-1}\nn
\eea
where in the last inequality we applied the Holder inequality
\beaa
&&\int_{S(\la,\nu)}r^{(p-1)(1-\frac{2}{p})}|\a|^{p-1}\leq\left(\int_{S(\la,\nu)}1^p\right)^\frac{1}{p}
\left(\int_{S(\la,\nu)}[r^{(p-1)(1-\frac{2}{p})}|\a|^{p-1}]^{\frac{p}{p-1}}\right)^\frac{p-1}{p}\nn\\
&&\leq cr^{\frac{2}{p}}\left(\int_{S(\la,\nu)}r^{p(1-\frac{2}{p})}|\a|^p\right)^\frac{p-1}{p}\leq cr^{\frac{2}{p}}|r^{(1-\frac{2}{p})}\a|_{p,s}^{p-1}\ .
\eeaa
Therefore the following inequality holds:
\bea
\frac{d}{d\la}|r^{(1-\frac{2}{p})}\a|_{p,S}
\!&\leq&\!\|\oom\|_{\infty}\left(4p|\omb|_{\infty}+\big(\frac{p}{2}-1\big)\!|\tr\chib-\overline{\tr\chib}|_{\infty}\right)|r^{(1-\frac{2}{p})}\a|_{p,S}
\eql{atras}\nn\\
&+&p\|\oom\|_{\infty}\frac{1}{r^{4}|\la|^{2+\frac{\ep''}{2}}}\|r^{5}|\la|^{2+\frac{\ep''}{2}}\nabb\b\|_{\infty}+p\|rF\|_{\infty}\ .
\eea
$F=\oom\left(-3(\hat{\chi}\ro+\dual\hat{\chi}\si)+(\zeta+4\eta)\hot\b\right)$ satisfies the following bound: 
\bea
&&\|F\|_{\infty}\leq c\|\oom|_{\infty}\|\hat{\chi}\|_{\infty}\left(\|\ro\|_{\infty}+\|\si\|_{\infty}\right)+(\|\zeta\|_{\infty}+\|\eta\|_{\infty})\|\b\|_{\infty}\nn\\
&&\leq c\!\left(\frac{1}{r^2|\la|^{\frac{1}{2}}}\|r^2|\la|^{\frac{1}{2}}\hat{\chi}\|_{\infty}\frac{1}{r^3|\la|^{3-\de}}\|r^3|\la|^{3-\de}(\ro,\si)\|
+\frac{1}{r^2|\la|^{\frac{1}{2}}}\bigg(\|r^2|\la|^{\frac{1}{2}}\zeta\|_{\infty}\right.\nn\\
&&\left.+\|r^2|\la|^{\frac{1}{2}}\eta\|_{\infty}\bigg)\frac{1}{r^4|\la|^{2+\ep''}}\|r^4|\la|^{2+\ep''}\!\b\|_{\infty}\right)\nn\\
&&\leq c\!\frac{1}{r^5|\la|^{\frac{7-2\de}{2}}}\left(\|r^3|\la|^{3-\de}(\ro,\si)\|_{\infty}+\|r^4|\la|^{2+\ep''}\!\b\|_{\infty}\right)\leq cC_1\!\frac{1}{r^5|\la|^{\frac{7-2\de}{2}}}\ \ \ \ \ \ \ \ \ \ \ \ 
\eea
and the previous inequality becomes
\bea
&&\ML\frac{d}{d\la}|r^{(1-\frac{2}{p})}\a|_{p,S}\leq\|\oom\|_{\infty}\left(4p|\omb|_{\infty}+\big(\frac{p}{2}-1\big)\!|\tr\chib-\overline{\tr\chib}|_{\infty}\right)|r^{(1-\frac{2}{p})}\a|_{p,S}\nn\\
&&\ML+p\|\oom\|_{\infty}\frac{1}{r^{4}|\la|^{2+\frac{\ep''}{2}}}\|r^{5}|\la|^{2+\frac{\ep''}{2}}\nabb\b\|_{\infty}
+c\!\frac{1}{r^5|\la|^{\frac{7-2\de}{2}}}\left(\|r^3|\la|^{3-\de}(\ro,\si)\|_{\infty}+\|r^4|\la|^{2+\ep''}\b\|_{\infty}\right)\eql{atras2}\nn\\
&\leq&\frac{c}{r|\la|}|r^{(1-\frac{2}{p})}\a|_{p,S}+\frac{c}{r^{4}|\la|^{2+\frac{\ep''}{2}}}
\leq\frac{c}{|\la|^2}|r^{(1-\frac{2}{p})}\a|_{p,S}+\frac{c}{r^{4}|\la|^{2+\frac{\ep''}{2}}}\ .\nn
\eea
Integrating along $\Cb(\nu)$ we obtain
\bea
|r^{(1-\frac{2}{p})}\a|_{p,S}(\la,\nu)\leq c\left(|r^{(1-\frac{2}{p})}\a|_{p,S}(\la_1,\nu)+\frac{1}{r^{4}|\la|^{1+\frac{\ep''}{2}}}\right)
\eea
where $\la_1=u_{\Si_0\cap\Cb(\nu)}$\ , and multiplying the inequality by $r^4|\la|^{1+\frac{\ep''}{2}}$,
we obtain
\bea
|r^{(5-\frac{2}{p})}|\la|^{1+\frac{\ep''}{2}}\a(\lie_TR)|_{p,S}(\la,\nu)\leq c\!\left(|r^{(5-\frac{2}{p})}|\la|^{1+\frac{\ep''}{2}}\a(\lie_TR)|_{p,S}(\la_1,\nu)+1\right)\ \ \ \ \ \ \ 
\eea
which completes the proof of Theorem \ref{T4.1}.

\subsection{The decay estimates for the $\Lie_TR$ null components}
In the final section \ref{S.s2.7} we use the estimates of the null components of $\Lie_TR$ to get estimates for the asymptotic behaviour of $R$. Observe that from the boundedness of the $\tilde{\cal Q}$ norms we have, instead, a control of the  null components of the $\lie_TR$ tensor. Therefore we have to get from these estimates those for $\Lie_TR$. The two tensors are connected by the following equation,
\bea
\Lie_TR= \lie_{T}R+\frac 1 2 {^{(T)}}[R]-\frac 3 8(\tr{^{(T)}}\pi)R\ ,\eql{piT1}
\eea
where
\bea
{^{(T)}}[R]_{\a\b\gamma\delta}={^{(T)}}\pi_\a^\mu R_{\mu\b\gamma\delta}
+{^{(T)}}\pi_\b^\mu R_{\a\mu\gamma\delta}+{^{(T)}}\pi_\gamma^\mu R_{\a\b\mu\delta}
+{^{(T)}}\pi_\delta^\mu R_{\a\b\gamma\mu}\ .\ \ \ \ \ \ \ \eql{piT2a}
\eea
In the previous section, see Theorem \ref{T4.1}, we have proved the following decay for the $\lie_TR$  null components, 
\beaa
&&\sup_{\mathcal{K}}r^{5}|u|^{1+\frac{\e''}{2}}|\a(\lie_{T}R)| \leq C_1,\quad
\sup_{\mathcal{K}}r^{4}|u|^{2+\frac{\e''}{2}} |\b(\lie_TR)|\leq C_1\nn\\
&&\sup_{\mathcal{K}}r^3|u|^{3+\frac{\e'}{2}}|\ro(\lie_TR)-{\overline{\ro(\lie_TR)}}| \leq C_1,\quad 
\sup r^3|u|^{3+\frac{\e'}{2}}|\si(\lie_TR)-{\overline{\si(\lie_TR)}}|\leq C_1\ \ \ \ \eql{LTWdec11bb}\\
&& \sup_{\mathcal{K}}r^2|u|^{4+\frac{\e}{2}}\bb(\lie_TR)\leq C_1,\quad\sup_{\mathcal{K}}r|u|^{5+\frac{\e}{2}}|\aa(\lie_TR)|\leq C_1\nn\ \ ,
\eeaa
Therefore from \ref{piT1} and \ref{piT2a} it is clear that, due to the fact that for the null components of $R$ the asymptotic estimates at our disposal are those proved in ${\cal K}$, see \ref{LTRdec2}, which are weaker than those suggested from the ``Peeling Theorem" by a factor $r^{-\frac{3}{2}}$ for $\a$ and a factor $r^{-\frac{1}{2}}$ for $\b$, the decay in $r$ for the $\Lie_TR$ components are analogous to those for the $\lie_TR$ ones if the various components of ${^{(T)}\!}\pi$ decay at least as $r^{-2}$.\footnote{The connection coefficients decay in $r$ with integer powers.} Nevertheless we will see in a moment that a $r^{-2}$ decay for ${^{(T)}\!}\pi$ is not yet enough to get the decay of the $R$ null components as to get it we have to perform an extra integration bringing from the decay of $\Lie_TR$ to the decay of $R$. 

\NI Intuitively we expect that the various components of ${^{(T)}\!}\pi$ have a sufficiently strong decay for the following argument: The $T$ vector field is Killing in the Kerr spacetime, therefore the only reason why here ${^{(T)}\!}\pi$ is not exactly zero is due to the extra correction of the initial data metric. As this correction is chosen to decay very fast it follows that the decay of the connection coefficients (which are zero in Kerr) should also decay as faster as we need, satisfying our goal. This argument, although correct, has to be proven and is the content of the next subsection.

\subsection{The ${^{(T)}\!}\pi$ estimates}\label{s.s2.6}
Defining the vector field $T$  as in \cite{Kl-Ni:peeling},
\bea
T=\oom(e_3+e_4)
\eea
we have the following expression for the various components of ${^{(T)}\!}\pi$:
\bea
&&\li{T}_{ab}=2\oom\left(\chih_{ab}+\chibh_{ab}+\de_{ab}(\om+\omb)\right)\nn\\
&&\lj{T}=\oom(\tr\chi+\tr\chib)+4\oom(\om+\omb)\eql{5.20}\\
&&\lm{T}_{a}=-4\oom\zeta_a\ ,\ \lmm{T}_{a}=4\oom\zeta_a\nn\\
&&\lnn{T}=0\ ,\ \lnnn{T}=0\ .\nn
\eea
From Theorem \ref{Ktheorem2} we have:
\bea
&&|r^{2-\frac{2}{p}}|u|^{1-\de}\li{T}|_{p,S}\leq c\varepsilon_0\nn\\
&&|r^{2-\frac{2}{p}}|u|^{1-\de}\lj{T}|_{p,S}\leq c\varepsilon_0\eql{piK2}\\
&&|r^{2-\frac{2}{p}}|u|^{1-\de}\tr{^{(T)}\!}\pi|_{p,S}\leq c\varepsilon_0\nn\\
&&|r^{2-\frac{2}{p}}|u|^{1-\de}{(\lm{T},\lmm{T})}|_{p,S}\leq c\varepsilon_0\nn
\eea
with arbitrary $\de>0$.
Using estimates \ref{piK2} we can obtain the decay of the $\Lie_TR$ null components and prove the following result:
\begin{thm}\label{T5.1}
In the spacetimes ${\cal K}$, whose initial conditions are given in Theorem \ref{spaceK}, the following inequalities hold
\beaa
&&\ML\sup_{\mathcal{K}}r^{5}|u|^{1+\frac{\e'''}{2}}|\a(\Lie_{T}R)| \leq C_2,\quad
\sup_{\mathcal{K}}r^{4}|u|^{2+\frac{\e''}{2}} |\b(\Lie_TR)|\leq C_2\nn\\
&&\ML\sup_{\mathcal{K}}r^3|u|^{3+\frac{\e'}{2}}|\ro(\Lie_TR)-{\overline{\ro(\Lie_TR)}}| \leq C_2,\quad
\sup_{\mathcal{K}} r^3|u|^{3+\frac{\e'}{2}}|\si(\Lie_TR)-{\overline{\si(\Lie_TR)}}|\leq C_2\ \ \ \ \eql{T51}\\
&&\ML\sup_{\mathcal{K}}r^2|u|^{4+\frac{\e}{2}}\bb(\Lie_TR)\leq C_2,\quad\sup_{\mathcal{K}}r|u|^{5+\frac{\e}{2}}|\aa(\Lie_TR)|\leq C_2\nn\ \ .
\eeaa
\end{thm}
{\bf Proof:}
From equations \ref{piT1} and \ref{piT2a} it follows:
\bea
&&\a(\Lie_TR)_{ab}=\a(\lie_TR)_{ab}+\frac 1 2 {^{(T)}}[R]_{a4b4}-\frac 3 8(\tr{^{(T)}}\pi)\a(R)_{ab}\nn\\
&&\b(\Lie_TR)_{a}=\b(\lie_TR)_{a}+\frac 1 2 {^{(T)}}[R]_{a434}-\frac 3 8(\tr{^{(T)}}\pi)\b(R)_a\nn\\
&&\ro(\Lie_TR)=\ro(\lie_TR)+\frac 1 2 {^{(T)}}[R]_{3434}-\frac 3 8(\tr{^{(T)}}\pi)\ro(R)\nn\\
&&\si(\Lie_TR)=\si(\lie_TR)+\frac 1 2 {^{(T)}}[{^*\!}R]_{3434}-\frac 3 8(\tr{^{(T)}}\pi)\si(R)\eql{LieTlieT}\\
&&\bb(\Lie_TR)_{a}=\bb(\lie_TR)_{a}+\frac 1 2 {^{(T)}}[R]_{a343}-\frac 3 8(\tr{^{(T)}}\pi)\bb(R)_a\nn\\
&&\aa(\Lie_TR)_{ab}=\aa(\lie_TR)_{ab}+\frac 1 2 {^{(T)}}[R]_{a3b3}-\frac 3 8(\tr{^{(T)}}\pi)\aa(R)_{ab}\nn
\eea
and, observing that ${^{(T)}}\pi_{44}=0$,
\bea
&&{^{(T)}}[R]_{a4b4}={^{(T)}}\pi_a^\mu R_{\mu4b4}+{^{(T)}}\pi_4^\mu R_{a\mu b4}+{^{(T)}}\pi_b^\mu R_{a4\mu4}
+{^{(T)}}\pi_4^\mu R_{a4b\mu}\nn\\
&&=-\frac{1}{2}{^{(T)}}\pi_{a4}R_{34b4}+{^{(T)}}\pi_{ac}R_{c4b4}
-\frac{1}{2}{^{(T)}}\pi_{43}R_{a4b4}+{^{(T)}}\pi_{4c} R_{acb4}\nn\\
&&\ \ \ -\frac{1}{2}{^{(T)}}\pi_{b4}R_{a434}+{^{(T)}}\pi_{bc}R_{a4c4}
-\frac{1}{2}{^{(T)}}\pi_{43}R_{a4b4}+{^{(T)}}\pi_{4c}R_{a4bc}\eql{piT3a}\\
&&=-\frac{1}{2}{^{(T)}}{\hat \pi}_{a4}R_{34b4}+\big({^{(T)}}{\hat \pi}_{ac}R_{c4b4}+\frac{1}{4}({\tr{^{(T)}}\pi})R_{a4b4}\big)
-\frac{1}{2}\big({^{(T)}}{\hat\pi}_{43}R_{a4b4}-\frac{1}{2}({\tr{^{(T)}}\pi})R_{a4b4}\big)
\nn\\
&&\ \ \ +{^{(T)}}{\hat \pi}_{4c} R_{acb4}-\frac{1}{2}{^{(T)}}{\hat \pi}_{b4}R_{a434}
+\big({^{(T)}}{\hat\pi}_{bc}R_{a4c4}+\frac{1}{4}({\tr{^{(T)}}\pi})R_{a4b4}\big)\nn\\
&&\ \ \ -\frac{1}{2}\big({^{(T)}}{\hat \pi}_{43}R_{a4b4}-\frac{1}{2}({\tr{^{(T)}}\pi})R_{a4b4}\big) +{^{(T)}}{\hat \pi}_{4c} R_{a4bc}\nn\\
&&=-\frac{1}{2}{^{(T)}}{\hat \pi}_{a4}R_{34b4}+\big({^{(T)}}{\hat \pi}_{ac}R_{c4b4}
+{^{(T)}}{\hat\pi}_{bc}R_{a4c4}+\frac{3}{4}({\tr{^{(T)}}\pi})R_{a4b4}\big)\nn\\
&&\ \ \ -\frac{1}{2}{^{(T)}}{\hat\pi}_{43}R_{a4b4}+{^{(T)}}{\hat \pi}_{4c}(R_{acb4}+ R_{a4bc})-\frac{1}{2}{^{(T)}}{\hat \pi}_{b4}R_{a434}\ .\nn
\eea
Therefore estimating ${^{(T)}}[R]_{a4b4}$ we obtain
\bea
&&\ML\|{^{(T)}}[R]_{a4b4}\|_{\infty}\leq c\left((\||\li{T}|\|_{\infty}+\|\lj{T}\|_{\infty})\|\a(R)\|_{\infty}
+\|\lm{T}\|_{\infty}\|\b(R)\|_{\infty}\right)\nn\\
&&\ML\leq c\left(\frac{1}{r^{2+\frac{7}{2}}|\la|^{1-\de}}(\|r^2|\la|^{1-\de}\li{T}\|_{\infty}
+\|r^2|\la|^{1-\de}\lj{T}\|_{\infty})\|r^{\frac{7}{2}}\a(R)\|_{\infty}\right.\nn\\
&&\left.+\frac{1}{r^{2+\frac{7}{2}}|\la|^{1-\de}}\|r^2|\la|^{1-\de}\lm{T}\|_{\infty}\|r^{\frac{7}{2}}\b(R)\|_{\infty}\right)\nn\\
&&\ML\leq c\left(\frac{1}{r^{5+\frac{1}{2}}|\la|^{1-\de}}\|r^{\frac{7}{2}}\a(R)\|_{\infty}
+\frac{1}{r^{5+\frac{1}{2}}|\la|^{1-\de}}\|r^{\frac{7}{2}}\b(R)\|_{\infty}\right)\nn\\
&&\ML\leq c\left(\frac{1}{r^{5+\frac{1}{2}}|\la|^{\frac{3}{2}-2\de}}\|r^{\frac{7}{2}}|\la|^{\frac{1}{2}-\de}\a(R)\|_{\infty}
+\frac{1}{r^{5+\frac{1}{2}}|\la|^{\frac{3}{2}-2\de}}\|r^{\frac{7}{2}}|\la|^{\frac{1}{2}-\de}\b(R)\|_{\infty}\right)\nn\\
&&\ML\leq \frac{c}{r^{5+\frac{1}{2}}|\la|^{\frac{3}{2}-2\de}}\leq \frac{c}{r^5|\la|^{2-2\de}}\ \ \ \ \ 
\eea
so that, finally,
\bea
\|{r^5|\la|^{(2-2\de)}}{^{(T)}}[R]_{a4b4}\|_{\infty}\leq c\ .\eql{aest1}
\eea
Moreover
\bea
\|(\tr{^{(T)}}\pi)\a(R)\|_{\infty}\!&\leq&\!\frac{c}{r^{5+\frac{1}{2}}|\la|^{\frac{3}{2}-2\de}} |r^{2}|\la|^{1-\de}\tr{^{(T)}\!}\pi|_{\infty}\|r^{\frac{7}{2}}|\la|^{\frac{1}{2}-\de}\a(R)\|_{\infty}\nn\\
\!&\leq&\!\frac{c}{r^{5}|\la|^{2-2\de}} |r^{2}|\la|^{1-\de}\tr{^{(T)}\!}\pi|_{\infty}\|r^{\frac{7}{2}}|\la|^{\frac{1}{2}-\de}\a(R)\|_{\infty}
\ \ \ \ \ 
\eea
and finally
\bea
\|r^{5}|\la|^{2-2\de}(\tr{^{(T)}}\pi)\a(R)\|_{\infty}\leq c\ .\eql{aest2}
\eea
using estimates \ref{aest1} and \ref{aest2} we obtain choosing $\de$ such that
\[1+\frac{\ep'''}{2}\leq 2-2\de\]
\bea
\sup_{\mathcal{K}}|r^{5}|u|^{1+\frac{\e'''}{2}}\a(\Lie_{T}R)| \leq C_2\ .
\eea

\NI Let us repeat the previous estimate for the $\b$ term.
\bea
&&{^{(T)}}[R]_{a434}={^{(T)}}\pi_a^\mu R_{\mu434}+{^{(T)}}\pi_4^\mu R_{a\mu 34}+{^{(T)}}\pi_3^\mu R_{a4\mu4}
+{^{(T)}}\pi_4^\mu R_{a43\mu}\nn\\
&&=-\frac{1}{2}{^{(T)}}\pi_{a4}R_{3434}+{^{(T)}}\pi_{ac}R_{c434}-\frac{1}{2}{^{(T)}}\pi_{43}R_{a434}+{^{(T)}}\pi_{4c} R_{ac34}\nn\\
&&\ \ \ -\frac{1}{2}{^{(T)}}\pi_{34}R_{a434}+{^{(T)}}\pi_{3c}R_{a4c4}
-\frac{1}{2}{^{(T)}}\pi_{43}R_{a434}+{^{(T)}}\pi_{4c}R_{a43c}\eql{piT3aa}\\
&&=-\frac{1}{2}{^{(T)}}{\hat \pi}_{a4}R_{3434}+\big({^{(T)}}{\hat \pi}_{ac}R_{c434}+\frac{1}{4}({\tr{^{(T)}}\pi})R_{a434}\big)
-\frac{1}{2}\big({^{(T)}}{\hat\pi}_{43}R_{a434}-\frac{1}{2}({\tr{^{(T)}}\pi})R_{a434}\big)
\nn\\
&&\ \ \ +{^{(T)}}{\hat \pi}_{4c} R_{ac34}-\frac{1}{2}{^{(T)}}{\hat \pi}_{34}R_{a434}
+\big({^{(T)}}{\hat\pi}_{3c}R_{a4c4}+\frac{1}{4}({\tr{^{(T)}}\pi})R_{a434}\big)\nn\\
&&\ \ \ -\frac{1}{2}\big({^{(T)}}{\hat \pi}_{43}R_{a434}-\frac{1}{2}({\tr{^{(T)}}\pi})R_{a434}\big)\nn\\
&&=-\frac{1}{2}{^{(T)}}{\hat \pi}_{a4}R_{3434}+\big({^{(T)}}{\hat \pi}_{ac}R_{c434}
+{^{(T)}}{\hat\pi}_{3c}R_{a4c4}+\frac{3}{4}({\tr{^{(T)}}\pi})R_{a434}\big)\nn\\
&&\ \ \ -\frac{1}{2}{^{(T)}}{\hat\pi}_{43}R_{a434}+{^{(T)}}{\hat \pi}_{4c} R_{ac34}-{^{(T)}}{\hat \pi}_{34}R_{a434}
+{^{(T)}}{\hat \pi}_{4c}R_{a43c}\nn\\
&&=\left(-\frac{1}{2}{^{(T)}}{\hat \pi}_{a4}R_{3434}+{^{(T)}}{\hat\pi}_{3c}R_{a4c4}+{^{(T)}}{\hat \pi}_{4c} R_{ac34}
+{^{(T)}}{\hat \pi}_{4c}R_{a43c}\right)\nn\\
&&+\left({^{(T)}}{\hat \pi}_{ac}R_{c434}+\frac{3}{4}({\tr{^{(T)}}\pi})R_{a434}-\frac{1}{2}{^{(T)}}{\hat\pi}_{43}R_{a434}
-{^{(T)}}{\hat \pi}_{34}R_{a434}\right)\nn
\eea
Therefore estimating ${^{(T)}}[R]_{a434}$ we obtain 
\bea
&&\ML\|{^{(T)}}[R]_{a434}\|_{\infty}\leq c\left((\|\lm{T}\|_{\infty}+\|\lmm{T}\|_{\infty})\|\ro(R)\|_{\infty}
+\big(\|\li{T}\|_{\infty}+\|\lj{T}\|_{\infty}+(\tr{^{(T)}\!}\pi)\big)\|\b\|_{\infty}\right)\nn\\
&&\ML\leq c\left(\frac{1}{r^{4+1}|\la|^{1-\de}}(\|r^2|\la|^{1-\de}\lm{T}\|_{\infty}+\|r^2|\la|^{1-\de}\lmm{T}\|_{\infty})\|r^{3}\ro(R)\|_{\infty}\right.\nn\\
&&\left.+\frac{1}{r^{2+\frac{7}{2}}|\la|^{1-\de}}\big(\|r^2|\la|^{1-\de}\li{T}\|_{\infty}+\|r^2|\la|^{1-\de}\lj{T}\|_{\infty}
+(r^2|\la|^{1-\de}\tr{^{(T)}\!}\pi)\big)\|r^{\frac{7}{2}}\b(R)\|_{\infty}\right)\nn\\
&&\ML\leq c\!\left(\frac{1}{r^{4+1}|\la|^{1-\de}}\|r^{3}\ro(R)\|_{\infty}
+\frac{1}{r^{4+\frac{3}{2}}|\la|^{\frac{3}{2}-2\de}}\|r^{\frac{7}{2}}|\la|^{\frac{1}{2}-\de}\b(R)\|_{\infty}\right)\nn\\
&&\ML\leq c\!\left(\frac{1}{r^{4}|\la|^{2-\de}}\|r^{3}\ro(R)\|_{\infty}
+\frac{1}{r^{4}|\la|^{3-2\de}}\|r^{\frac{7}{2}}|\la|^{\frac{1}{2}-\de}\b(R)\|_{\infty}\right)\nn\\
&&\ML\leq \frac{c}{r^{4}|\la|^{2-\de}}+\frac{c}{r^{4}|\la|^{3-2\de}}\leq \frac{c}{r^{4}|\la|^{2-\de}}\ \ \ \ \ 
\eea
and finally
\bea
\|{r^4|\la|^{(2-\de)}}{^{(T)}}[R]_{a434}\|_{\infty}\leq c\ .\eql{aest11}
\eea
Moreover
\bea
\|(\tr{^{(T)}}\pi)\b(R)\|_{\infty}\!&\leq&\!\frac{c}{r^{5+\frac{1}{2}}|\la|^{\frac{3}{2}-2\de}} |r^{2}|\la|^{1-\de}\tr{^{(T)}\!}\pi|_{\infty}\|r^{\frac{7}{2}}|\la|^{\frac{1}{2}-\de}\b(R)\|_{\infty}\nn\\
\!&\leq&\!\frac{c}{r^{4}|\la|^{3-2\de}} |r^{2}|\la|^{1-\de}\tr{^{(T)}\!}\pi|_{\infty}\|r^{\frac{7}{2}}|\la|^{\frac{1}{2}-\de}\b(R)\|_{\infty}\ \ \ \ \ 
\ \ \ \ \ 
\eea
and finally
\bea
\|r^{4}|\la|^{3-2\de}(\tr{^{(T)}}\pi)\b(R)\|_{\infty}\leq c\ .\eql{aest22}
\eea
using estimates \ref{aest11} and \ref{aest22} we obtain choosing $\de$ such that
\[1+\frac{\ep''}{2}\leq \min\left(2-\de,3-2\de\right)\]
\bea
\sup_{\mathcal{K}}|r^4|u|^{1+\frac{\e''}{2}}\b(\Lie_{T}R)| \leq C_2\ .
\eea
For the other terms there is no need to repeat this computation. In fact the $R$ null components already satisfy the peeling and going from $\lie_T $ to $\Lie_T$  the $|u|^{\ga}$ weight factor cannot become worst. 
\medskip

\NI If we use only the estimates proved in \cite{Kl-Ni:book}, see the remark at page 13, we can prove a weaker result, namely
\begin{thm}\label{T5.1(K1)}
Using only the estimates proved in  \cite{Kl-Ni:book} the following inequalities hold
\beaa
&&\ML\sup_{\mathcal{K}}r^{5}|u||\a(\Lie_{T}R)| \leq C_3,\quad
\sup_{\mathcal{K}}r^{4}|u|^{1+\frac{\e''}{2}} |\b(\Lie_TR)|\leq C_3\nn\\
&&\ML\sup_{\mathcal{K}}r^3|u|^{3+\frac{\e'}{2}}|\ro(\Lie_TR)-{\overline{\ro(\Lie_TR)}}| \leq C_2,\quad
\sup_{\mathcal{K}} r^3|u|^{3+\frac{\e'}{2}}|\si(\Lie_TR)-{\overline{\si(\Lie_TR)}}|\leq C_2\ \ \ \ \\
&&\ML\sup_{\mathcal{K}}r^2|u|^{4+\frac{\e}{2}}\bb(\Lie_TR)\leq C_2,\quad\sup_{\mathcal{K}}r|u|^{5+\frac{\e}{2}}|\aa(\Lie_TR)|\leq C_2\nn\ \ .
\eeaa
\end{thm}

\NI To prove the peeling for both the components $\a(R)$ and $\b(R)$ we have to use the result of Theorem \ref{T5.1} as the decay proved in Theorem \ref{T5.1(K1)} is not sufficient for getting the ``peeling estimate" for $\a(R)$. In fact to obtain it, as we explain in detail in the next subsection, we integrate along the integral curves of $T$ for a fixed $r={\overline r}$ and the decay factor ${|u|^{-\ga}}$ makes, if $\ga>1$, the integral uniformily bounded. This implies that the estimates of the norms of the null components times their appropriate $r$ weights are bounded if the corresponding initial data are bounded. Using the estimates of Theorem \ref{T5.1(K1)}, the decay factor ${|u|^{-1}}$ of the $\a(\Lie_T)$ null component is not integrable and therefore it is impossible to obtain the expected  decay for $\a(R)$. Viceversa the result proved in Theorem \ref{T5.1} gives a decay factor  ${|u|^{-(1+\frac{\ep'''}{2})}}$ which makes the integration uniformily bounded and the decay of $\a$ tied to the that of the associated initial data and, therefore, satisfying the peeling estimate.

\subsection{The decay estimates for the $R$ null components}\label{S.s2.7}
\NI Once we have a control of the decay of the various null components of $\Lie_TR$ we can easily obtain analogous estimates for
the $\pr_T$ derivatives of the (pointwise) norms of the $R$ components. Let us consider the more delicate $\a$ term. From the relation
\bea
&&\a(\Lie_TR)(e_a,e_b)=(\Lie_TR)(e_a,e_4,e_b,e_4)\\
&&=\pr_T(\a(e_a,e_b))+\sum R([T,e_a],e_4,e_b,e_4)+\sum R(e_a,[T,e_4],e_b,e_4)\ ,\nn
\eea
observing that in the Kerr spacetime $[T,e_a]$ and $[T,e_4]$ decay sufficiently fast, it follows that these terms cannot have a slower decay in ${\cal K}$ due to the extra correction in the metric whose initial data have been chosen to decay faster than the Kerr metric's ones. In fact
\bea
&&[T,e_a]=[\oom(e_3+e_4),e_a]=\oom[e_3,e_a]-e_a(\oom)e_3+\oom[e_4,e_a]-e_a(\oom)e_4\nn\\
&&=\oom([e_3,e_a]+ [e_4,e_a])-e_a(\oom)(e_3+e_4)\nn\\
&&=\oom(\dddd_3e_a+\dddd_4e_a)-\oom(\chi+\chib)_{ab}e_b-e_a(\oom)(e_3+e_4)
\eea 
and
\bea
&&[T,e_4]=[\oom(e_3+e_4),e_4]=\oom[e_3,e_4]-e_4(\oom)e_3-e_4(\oom)e_4\nn\\
&&=\oom(\dd_4\log\oom)e_3-\oom(\dd_3\log\oom)e_4+4\oom\ze(e_a)e_a-\oom(\dd_4\log\oom)e_3-\oom(\dd_4\log\oom)e_4\nn\\
&&=-\oom(\dd_3\log\oom+\dd_4\log\oom)e_4+4\oom\ze(e_a)e_a\eql{Te4com}
\eea 
Observing that in Kerr spacetime 
\bea
\dd_3\log\oom+\dd_4\log\oom=0\ \ ,\ \ (\chi+\chib)=0\ \ ,\ \ e_a(\oom)=0
\eea
The only terms which survive in Kerr spacetime are $\oom(\dddd_3e_a+\dddd_4e_a)$ and $4\oom\ze(e_a)e_a$ which decay as \footnote{With $\dddd_3e_a$,
$\dddd_4e_a$ we denote the projection of the$\dd_4e_a$ and $\dd_3e_a$ derivatives on $TS$, see \cite{Kl-Ni:book}, Chapter 3.}
\bea
&&\oom(\dddd_3e_a+\dddd_4e_a)=\frac{1}{r^3}c_{ab}e_b\nn\\
&&4\oom\ze(e_a)e_a=\frac{1}{r^3}c_{ab}e_b\ .
\eea
The first line can be proved observing that as shown in the appendix, eq. \ref{3.208}, the following estimate holds in Kerr spacetime, see also eq. \ref{IPframe} for the definition of the vector field $e_{\la}$:
\bea
[T,e_{\la}]=-2R\sin\th e_{\la}(\om_B)e_{\phi}=O(ar^{-3})e_{\phi}\ \ ,\ \ [T,e_{\phi}]=0\ ,
\eea
the second line from an explicit computation of $\ze(e_{\phi})$.
Therefore as the corrections to the Kerr metric decay faster also these commutators satisfy in ${\cal K}$ a decay at least of the following type
\bea
|{\bf g}([T,e_a],e_\a)|\leq c\frac{1}{r^2|u|^{1-\de}}\ \ ,\ \ |{\bf g}([T,e_4],e_\a)|\leq c\frac{1}{r^2|u|^{1-\de}}
\eea
and observing that $[T,e_4]$ does not produce terms proportional to $e_3$,\footnote{This will produce a term proportional to $\ro(R)$ whose decay would not be sufficient for the conclusion.} it follows that $\pr_T(\a(e_a,e_b))$ has the same decay as $\a(\Lie_TR)(e_a,e_b)$. 
\medskip

\NI The same result holds for $\b(\Lie_TR)$ and more easily for the remaining components which we already know satisfy the ``Peeling decay". Therefore we have the following estimates 
\begin{eqnarray}
&&\sup_{\mathcal{K}}r^{5}|u|^{1+\frac{\e'''}{2}}|\pr_T(\a(R)(e_a,e_b))| \leq C_3\nn\\
&&\sup_{\mathcal{K}}r^{4}|u|^{2+\frac{\e''}{2}}|\pr_T(\b(R)(e_a))|\leq C_3\ .
\end{eqnarray}
The final step is to integrate along the integral curves of $T$ at fixed $r$. It is clear that the $|u|$ weight factors will allow to bound uniformily these integrals. Basically, the details are in the appendix, identifying $T$ with $\frac{\pr}{\pr t}$ and $u$ with $t-r_*$ we have integrating $|{\overline r}^{5}\a(R)|$ on the curve starting at $r={\overline r}$,
\bea
|{\overline r}^{5}\a(R)|({\overline t},{\overline r})\leq c\left(|{\overline r}^{5}\a(R)|(0,{\overline r})
+\int_0^{{\overline t}=\overline r-r_1}\frac{1}{({\overline r}-t')^{1+\frac{\e'''}{2}}}dt'\right)\ ,\eql{intcon1}
\eea
where, identifying $r$ with $r_*$,
\bea
&&{\overline r}=r(u,\ub)=\frac{\ub-u}{2}\ \ ,\ \ {\overline t}=t(u,\ub)=\frac{\ub+u}{2}\nn\\
&&u=u({\overline t},{\overline r})=u(0,r_1)=-r_1\ \ ,\ \ \ub=\ub({\overline t},{\overline r})=\ub(0,r_2)=r_2\nn\\
&&{\overline t}=t(u,\ub)=\frac{r_2-r_1}{2}=\frac{{\overline t}+{\overline r}-r_1}{2}\ .
\eea
As the exponent in the integrand is greater than one, the integrand is uniformily bounded in $\overline r$ and the previous inequality becomes:
\bea
|{\overline r}^{5}\a(R)|({\overline t},{\overline r})=|{\overline r}^{5}\a(R)|(u,\ub)\leq c\left(|{\overline r}^{5}\a(R)|(0,{\overline r})+1\right)\ .
\eea
As the initial data are chosen in such a way that 
\[|{\overline r}^{5}\a(R)|(0,{\overline r})\leq c\ ,\]
we conclude that 
\bea
\sup_{{\cal K}}|r^{5}\a(R)|\leq c\ ,
\eea
which is the desired result as it implies the expected asymptotic null decay, that is when $r$ goes to infinity, with $|u|$ constant.
\medskip

\NI{\bf Remark:} If we used only the estimates proved in \cite{Kl-Ni:book} we would prove the estimate
\[\sup_{{\cal K}_1}r^{5}|u||\pr_T(\a(R)(e_a,e_b))| \leq C_4\]
and from it we obtain a decay for $\a$ which does not satisfy the peeling for the presence of a $\log r$ factor, namely we would obtain
\bea
\sup_{{\cal K}_1}\left|\frac{{\overline r}^{5}}{\log{\overline r}}\a(R)\right|(0,{\overline r})\leq c\ .
\eea
\section{Conclusions}
 Let us recall some general aspects of the problem we are considering.

\NI The assumptions done at the beginning on the Kerr mass upper bound allows to treat the Kerr part of the metric as a perturbation and to apply the result of \cite{Kl-Ni:book} in the slightly modified version discussed in \cite{Kl-Ni:peeling} . On the other side if the metric correction, $\de{{\bf g}}$, of the initial Kerr data metric were absent it is clear that the (very external region of the) spacetime would coincide with the one of the corresponding Kerr spacetime. If the initial data metric differ from the Kerr metric by  some small corrections then the solutions are still near to \Sch, but also near to Kerr. Therefore, although we are not making any linearization in the metric correction, but treating a true non linear problem, nevertheless these considerations allow us to write the Riemann tensor as

\bea
R=R_{Kerr}+\de R\ \ .\eql{basic}
\eea
This is the reason why to prove the {\bf ``expected result"} is more difficult than 
 the {\bf ``weaker result"} we are proving here. In that case, in fact, we have to prove a non linear perturbation around a Kerr spacetime and we cannot fully use the machinery in \cite{Kl-Ni:book}. The difficulty is, of course, that the rotation generators vector fields cannot be considered in this case as ``nearly Killing" and this makes difficult to bound the ${\cal Q}(\lie_OR)$ norms.

\section{Appendix}
\subsection{The initial data $\tilde{\cal Q}$ norms.}\label{S.S2.8}
Looking at the initial data for one of the $\tilde{\cal Q}$ norms we have that the following quantity has to be bounded
\beaa
 \int_{\Si_0\cap{\cal K}}|u|^{5+\e}Q(\lie_O {\tilde R})(\bar{K},\bar{K},T,T)\ 
\eeaa
Recalling the expression of $Q(\lie_O {\tilde R})(\bar{K},\bar{K},T,T)$,
\beaa 
Q(\lie_O {\tilde R})(\bar{K},\bar{K},T,T)\!&=&\! \frac 1 8 \ub^4|\a|^2+\frac 1 8u^4|\aa|^2+\frac 1 2(\ub^4+\frac{1}{2}u^2\ub^2)|\b|^2\\
\!&+&\!\frac 1 2 (u^4+4\ub^2u^2+\ub^4)(\ro^2+\si^2)+\frac 1 2(u^4+\frac{1}{2}u^2\ub^2)|\bb|^2
\eeaa
where $\a=\a(\Lie_TR),\b=\b(\Lie_TR)....$. Observing that on $\Si_0\cap{\cal K}$ $u,\ub$ behave like $r$, we can write
\beaa
 &&\int_{\Si_0\cap{\cal K}}|u|^{5+\ep}Q(\lie_O R)(\bar{K},\bar{K},T,T)\ \cong\nn\\
 &&\int_{R_0}^{\infty}dr r^{2+(5+\ep)+4}\left[|\a|^2+|\aa|^2+|\b|^2+(\ro^2+\si^2)+|\bb|^2\right]
 \eeaa
 Therefore $\a(\lie_O\lie_TR)$ has to decay, as $r\rightarrow\infty$ as $o(r^{-(6+\tilde{\ep})})$ with $\tilde{\ep}>\ep$ and the same for all the other components. As $\lie_O$ does not increase the decay this means that all the null components of $\lie_TR$, with the exception of $\ro$ has to decay in the same way. This implies that, again with the exception of $\ro$, all the null components of $R$ has to decay toward spatial infinity as $r^{-(5+\tilde{\ep})}$. Therefore this requires that the metric correction $\de{\bf g}$ has to decay as $O(r^{-(3+\tilde{\ep})})$. To complete this analysis let us look at the more delicate term, namely $\ro(\lie_O\lie_TR)$. We have
 \bea
 \ro(\lie_O\lie_TR)=\Lie_O\ro(\lie_TR)-\frac{1}{8}{^{(O)}\!}\pi\ro(\lie_TR)+O(\frac{1}{r})(\b((\lie_TR)+\bb(\lie_TR))\ \ \ \ \ \ \ \ \ 
 \eea
 Observe that the term $O(\frac{1}{r})(\b(\lie_TR)+\bb(\lie_TR))$ has a very good decay as $\b(\lie_TR)$ and $\bb(\lie_TR)$ have. The term $\frac{1}{8}{^{(O)}\!}\pi\ro(\lie_TR)$ behaves as \[O(r^{-1})\ro(\lie_TR)=O(r^{-1})r^{-2}|u|^{-\ga}r^{-3}\] with $\ga=\frac{1}{2}\ \mbox{or}\ 1-\de$ depending if we are considering the spacetime ${\cal K}_1$ or ${\cal K}_2$.
 In both cases we have $O(r^{-1})\ro(\lie_TR)=r^{-3}|u|^{-\ga}$ which is what we need. We are left to consider the term $\Lie_O\ro(\lie_TR)=\pr_O\ro(\lie_TR)$. Observe that
 \bea
 \pr_O\ro(\lie_TR)=\pr_O\left(\ro(\Lie_TR)+\frac{3}{8}\tr{^{(T)}\!}\pi\ro(R)+\{corrections\}\right)
 \eea
 Observe that on $\Si_0$  $\tr{^{(T)}\!}\pi=o(r^{-3})$ so that $\frac{3}{8}\tr{^{(T)}\!}\pi\ro(R)=o(r^{-6})$ as we need. Finally 
 \bea
 \pr_O\ro(\Lie_TR)=\pr_O\left(\pr_T\ro(R)+O(r^{-(3+\tilde{\ep})}(\b(R)+\bb(R))\right)\ .
 \eea
 The last term is a correction decaying very fast and in the first term $\pr_T\pr_O\ro(R)$ the derivatives with respect to $O$ adds a factor $r$ in the decay as the part of $\ro$ which decays as $r^{-3}$ does not depend on the angular variables, the $\pr_T$ derivative does not give zero as ${^{(T)}\!}\pi$ is not zero due to the corrections to Kerr, but this term goes at least as $r^{-(3+\tilde{\ep})}$.\footnote{We can see it in a slightly different way: on $\Si_0$ we can write
 $\pr_OR=\pr_OR_{Kerr}+\pr_O\de R$ and $\pr_T=\pr_{T_{Kerr}}+\pr_{\de T}$ so that $\pr_T\pr_OR=\pr_{T_{Kerr}}\pr_O\de R+\pr_{\de T}\pr_OR_{Kerr}$. Therefore as ${\de T}=o(r^{-2})c_{\a}e_{\a}$ it is clear that $\pr_T\pr_OR=o(r^{-6})$.} Therefore all the terms decay sufficiently fast and we can conclude that the following estimate holds:
 \bea
 |\ro(\lie_O\lie_TR)|_{\Si_0}=O\left(\frac{1}{r^{6+\tilde{\ep}}}\right)\ \ \mbox{with}\ \ \tilde{\ep}>\ep>0\ .
 \eea
 Therefore the present result is a true extension of the result in \cite{Kl-Ni:peeling} for the situation where in the metric to the $\de{\bf g}$ correction it is added, as a correction, a time independent part which is small and decay as $r^{-2}$.

\subsection{Some proofs}

\NI{\bf Proof of the last part of Theorem \ref{Ktheorem2}:}

\NI{\em In the spacetime $\mathcal{K}$, whose existence is proved in Theorem \ref{Ktheorem2},  
 the following inequalities hold:
\bea
&&\sup_{\mathcal{K}}r^{2}|u|^{1-\de}|\mbox{\em$\tr$}\chi+\mbox{\em$\tr$}\chib|\leq c\varepsilon_0\nn\\
&&\sup_{\mathcal{K}}r^{2}|u|^{1-\de}|\om+\omb|\leq c\varepsilon_0\ .\eql{Otheorem22}
\eea}
\NI {\bf Proof:} To prove the first line of inequalities \ref{Otheorem22} we have to look at the transport equation for  $\tr\chi+\tr\chib$. From the structure equations, see \cite{Kl-Ni:book}, equations (3.1.46),.....,(3.1.48), we have
\bea
&&\ML\dd_3\tr\chi\!=\!-\frac{1}{2}\tr\chib{\tr\chi}\!-\!(\dd_3\log\oom)\tr\chi+2\ro+\left[\!-\!\chibh\c\chih+2|\eta|^2
\!+\!2(\lapp\log\oom+\divv\zeta)\right]\ \ \ \ \nn\\
&&\ML\dd_3\tr\chib\!=\!-\frac{1}{2}(\tr\chib)^2+(\dd_3\log\oom)\tr\chib+|\chibh|^2\eql{3.26}
\eea
and from them
\bea
&&\dd_3(\tr\chi+\tr\chib)=-\frac{1}{2}\tr\chib({\tr\chi}+\tr\chib)+(\dd_3\log\oom)(\tr\chi+\tr\chib)\eql{3.25a}\nn\\
&&+2\left(\ro\!-\!(\dd_3\log\oom)\tr\chi\right)+\left[\!-\!\chibh\c\chih+2|\eta|^2\!+\!2(\lapp\log\oom+\divv\zeta)+|\chibh|^2\right]\nn \ \ \ 
\eea
which we can rewrite 
\bea
&&\frac{\partial}{\partial\la}(\tr\chi+\tr\chib)+\frac{\overline{\oom\tr\chib}}{2}({\tr\chi}+\tr\chib)=
\left[2^{-1}\!\left({\overline{\oom\tr\chib}}\!-\!{\oom\tr\chib}\right)\!+\!\oom(\dd_3\log\oom)\!\right]\!(\tr\chi+\tr\chib)\nn\\
&&+2\oom\!\left(\ro\!-\!(\dd_3\log\oom)\tr\chi\right)+\oom\!\left[\!-\!\chibh\c\chih+2|\eta|^2\!+\!2(\lapp\log\oom+\divv\zeta)+|\chibh|^2\right]\ .\ \eql{3.27a}
\eea
Recalling Lemma 4.1.5 of \cite{Kl-Ni:book} from the previous equation, observing that $\left[2^{-1}\!\left({\overline{\oom\tr\chib}}\!-\!{\oom\tr\chib}\right)\!+\!\oom(\dd_3\log\oom)\!\right]$ is integrable in $\la$, we obtain
\bea
&&|r^{1-\frac{2}{p}}(\tr\chi+\tr\chib)|_{p,S}(\la,\nu)\leq |r^{1-\frac{2}{p}}(\tr\chi+\tr\chib)|_{p,\Cb(\nu)\cap\Si_0}\nn\\
&&+c\int_{\la_1}^{\la}d\la'|r^{1-\frac{2}{p}}\left(\ro\!-\!(\dd_3\log\oom)\tr\chi\right)|_{p,S}(\la',\nu)\eql{app1}\\
&&+c\int_{\la_1}^{\la}d\la'\left|r^{1-\frac{2}{p}}\left[\!-\!\chibh\c\chih+2|\eta|^2\!+\!2(\lapp\log\oom+\divv\zeta)+|\chibh|^2\right]\right|_{p,S}\ .\nn
\eea
Observe that the integrand of the last integral can be estimated in the following way, using the results of subsection 4.3.16 in \cite{Kl-Ni:book} and recalling that due to the better decay of the initial data we have an extra decay factor $|u|^{\frac{1}{2}-\de}$, with $\de>0$ and arbitrary, we obtain:
\bea
&&\left|r^{1-\frac{2}{p}}\left[\!-\!\chibh\c\chih+2|\eta|^2\!+\!2(\lapp\log\oom+\divv\zeta)+|\chibh|^2\right]\right|_{p,S}\\
&&\leq c{\varepsilon_0}r\!\left(\frac{1}{r^{3}|u|^{3-2\de}}+\frac{1}{r^4|u|^{2-2\de}}+\frac{1}{r^3|u|^{1-\de}}+\frac{1}{r^2|u|^{4-2\de}}\right)\leq c\frac{\varepsilon_0}{r|u|^{2-\de}}\ .\nn
\eea
Substituting this estimate in \ref{app1} we obtain
\bea
&&|r^{1-\frac{2}{p}}(\tr\chi+\tr\chib)|_{p,S}(\la,\nu)\leq |r^{1-\frac{2}{p}}(\tr\chi+\tr\chib)|_{p,\Cb(\nu)\cap\Si_0}\nn\\
&&+c\int_{\la_1}^{\la}d\la'|r^{1-\frac{2}{p}}\!\left(\ro\!-\!(\dd_3\log\oom)\tr\chi\right)|_{p,S}(\la',\nu)+c\frac{\varepsilon_0}{r|\la|^{1-\de}}\ \eql{app2}
\eea
and from it, using the following estimate we prove later on
\bea
\sup_{(\la,\nu)\in{\cal K}}|r^{3-\frac{2}{p}}|\la|^{1-\de}\!\left(\ro\!-\!(\dd_3\log\oom)\tr\chi\right)\!|_{p,S}(\la,\nu)\leq c\varepsilon_0\ ,\eql{app3}
\eea
\bea
&&|r^{1-\frac{2}{p}}(\tr\chi+\tr\chib)|_{p,S}(\la,\nu)\leq|r^{1-\frac{2}{p}}(\tr\chi+\tr\chib)|_{p,\Cb(\nu)\cap\Si_0}\eql{app2a}\\
&&+c\varepsilon_0\left(\int_{\la_1}^{\la}d\la'\frac{1}{r^2(\la',\nu)|\la'|^{1-\de}}\right)+c\frac{\varepsilon_0}{r|\la|^{1-\de}}\nn\\
&&\leq|r^{1-\frac{2}{p}}(\tr\chi+\tr\chib)|_{p,\Cb(\nu)\cap\Si_0}+
c\varepsilon_0\frac{1}{r(\la,\nu)|\la|^{1-\de}}+c\frac{\varepsilon_0}{r(\la,\nu)|\la|^{1-\de}}\nn
\eea
which implies
\bea
\ML|r^{2-\frac{2}{p}}|\la|^{1-\de}(\tr\chi+\tr\chib)|_{p,S}(\la,\nu)\leq |r^{2-\frac{2}{p}}|\la|^{1-\de}(\tr\chi+\tr\chib)|_{p,\Cb(\nu)\cap\Si_0}+c\varepsilon_0\ .\eql{app2b}
\eea
Therefore the first of inequalities \ref{Otheorem22} is proved if we can prove the estimate \ref{app3} \footnote{In fact the slightest weaker one: $\sup_{(\la,\nu)\in{\cal K}}|r^{2-\frac{2}{p}}|\la|^{2-\de}\left(\ro\!-\!(\dd_3\log\oom)\tr\chi\right)|_{p,S}(\la,\nu)\leq c\varepsilon_0$ would be sufficient.}
\beaa
\sup_{(\la,\nu)\in{\cal K}}|r^{3-\frac{2}{p}}|\la|^{1-\de}\!\left(\ro\!-\!(\dd_3\log\oom)\tr\chi\right)\!|_{p,S}(\la,\nu)\leq c\varepsilon_0\ .
\eeaa
The proof of this last estimate follows the analogous proof in \cite{Kl-Ni:peeling}, Proposition 3.1. We write the transport equation for 
$\left(\ro\!-\!(\dd_3\log\oom)\tr\chi\right)$ along an outgoing cone  $C(\la)$ of the $\cal K$ spacetime  using the structure
equations, see \cite{Kl-Ni:book} equations (3.1.46),...,(3.1.48) of \cite{Kl-Ni:book} and the null Bianchi equations.
We obtain
\bea
\frac{\partial}{\partial\nu}(\ro\!-\!\tr\chi\dd_3\log\oom)=\oom\dd_4\ro
\!-\!\left[\oom(\dd_3\log\oom)\dd_4\tr\chi\!+\!\oom\tr\chi\dd_4\dd_3\log\oom\right]\ .\ \ \ \eql{3.34aa}
\eea
where $\frac{\partial}{\partial\nu}=\oom\dd_4$. Recalling the explicit expression of the Bianchi equations, see
\cite{Kl-Ni:book}, equations (3.2.8)  we have
\bea
\oom\dd_4\ro&=&-\frac{3}{2}\oom\tr\chi\ro+\big[\divv\b-\big(-\frac{1}{2}\chibh\c\a-\ze\c\bb+2\eta\c\bb\big)\big]\eql{3.35aa}\\
&=&-\frac{3}{2}\oom\tr\chi\ro
+\varepsilon_0\left[O\left(\frac{1}{r^{\frac{9}{2}}|\la|^{\frac{1}{2}-\de}}\right)+O\left(\frac{1}{r^{\frac{9}{2}}|\la|^{\frac{5}{2}-2\de}}\right)+O\left(\frac{1}{r^4|\la|^{3-2\de}}\right)
+O\left(\frac{1}{r^4|\la|^{3-2\de}}\right)\right]\nn\\
&&= -\frac{3}{2}\oom\tr\chi\ro+O\left(\frac{\varepsilon_0}{r^4|\la|^{1-\de}}\right)
\eea
The second term in the right hand side of \ref{3.34aa} can be written, using the transport equation for $\tr\chi$, as:
\bea
&&-\left[\oom(\dd_3\log\oom)\dd_4\tr\chi\!+\!\oom\tr\chi\dd_4\dd_3\log\oom\right]\nn\\
&&=-\oom\tr\chi\left(-\ro-(\dd_3\log\oom)(\dd_4\log\oom)+\left[\etab\c\eta-2\ze^2-2\ze\c\nabb\log\oom\right]\right)\nn\\
&&\ -\oom\dd_3\log\oom\left(-\frac{1}{2}\tr\chi^2+\tr\chi(\dd_4\log\oom)+|\chih|^2\right)\nn\\
&&=-\frac{1}{2}\oom\tr\chi(-\tr\chi\dd_3\log\oom)+\oom\tr\chi\ro
-\oom\tr\chi\left[\etab\c\eta-2\ze^2-2\ze\c\nabb\log\oom+|\chih|^2\right]\nn\\
&&=-\frac{1}{2}\oom\tr\chi(-\tr\chi\dd_3\log\oom)+\oom\tr\chi\ro+O\left(\frac{\varepsilon_0}{r^4|\la| ^{2-2\de}}\right)\eql{3.36aa}
\eea
Collecting \ref{3.35aa} and \ref{3.36aa}, evolution equation \ref{3.34aa} can be written as
\bea
\frac{\partial}{\partial\nu}(\ro\!-\!\tr\chi\dd_3\log\oom)=-\frac{1}{2}\oom\tr\chi(\ro-\tr\chi\dd_3\log\oom)
+\varepsilon_0\left[O\left(\frac{1}{r^4|\la|^{1-\de}}\right)+O\left(\frac{1}{r^4|\la|^{2-2\de}}\right)\right]\ .\nn
\eea
Integrating along an outgoing cone from $(\la,\nu_*)$ to $(\la,\nu)$ with $\nu_*>\nu$ and applying Gronwall's lemma we obtain \footnote{Choosing $\de$ sufficiently small.}
\bea
&&\ML\ML|r^{3-\frac{2}{p}}|\la|^{1-\de}(\ro\!-\!\tr\chi\dd_3\log\oom)|_{p,S}(\la,\nu)\!\leq\!
c|r^{3-\frac{2}{p}}|\la|^{1-\de}(\ro\!-\!\tr\chi\dd_3\log\oom)|_{p,S}(\la,\nu_*)+c\varepsilon_0\ .\ \ \ \ \ \ \ \ \ \ \ \eql{app4}
\eea
As the spacetime ${\cal K}$ is already at our disposal, we can perform, in the previous inequality the limit $\nu_*\rightarrow\infty$. From \cite{Kl-Ni:book}, Chapter 8, Propositions 8.2.1, 8.3.1, we conclude that
\bea
&&\lim_{\nu_*\rightarrow\infty}r^3(\ro\!-\!\tr\chi\dd_3\log\oom)
=\lim_{\nu_*\rightarrow\infty}\left[r^3(\ro\!-\!\frac{2}{r}\dd_3\log\oom)+r^3O(\frac{1}{r^2}\omb)\right]\nn\\
&&=\lim_{\nu_*\rightarrow\infty}r^3(\ro\!-\!\frac{2}{r}\dd_3\log\oom)=
\lim_{\nu_*\rightarrow\infty}r^3(\ro\!+\!\frac{4\omb}{r})=0\eql{app5}
\eea
so that \ref{app4} becomes
\bea
|r^{3-\frac{2}{p}}|\la|^{1-\de}(\ro\!-\!\tr\chi\dd_3\log\oom)|_{p,S}(\la,\nu)\!\leq\! c\varepsilon_0\ ,
\eea
proving \ref{app3}. To prove the second inequality of \ref{Otheorem22} we proceed in a similar way.
Inequality \ref{app3} implies immediately the following relation:
\bea
\omb=-(2\tr\chi)^{-1}\ro+O\left(\frac{\varepsilon_0}{r^2|\la|^{1-\de}}\right)\ .
\eea
Using this relation we can estimate $\om+\omb$ from the estimate of $(\om-(2\tr\chi)^{-1}\!\ro)$ as
\bea
\om+\omb=\left(\om-(2\tr\chi)^{-1}\!\ro\right)+O\left(\frac{\varepsilon_0}{r^2|\la|^{1-\de}}\right)\ .\eql{3.47}
\eea

\NI To estimate of $\left(\om-(2\tr\chi)^{-1}\ro\right)$
we write the evolution equation for this quantity along $\Cb(\nu)$. We observe that, from equation {(4.3.59)} in
\cite{Kl-Ni:book},
\bea
\dd_3\om&=&-\frac{1}{2}\dd_3\dd_4\log\oom=-\frac{1}{2\oom}\dd_3\oom\dd_4\log\oom+\frac{1}{2}(\dd_3\log\oom)\dd_4\log\oom\nn\\
&=&-(\dd_3\log\oom)\om+\frac{1}{2}\ro-\frac{1}{2\oom}{\hat
F}=-(\dd_3\log\oom)\om-\tr\chi\omb+\left[O\left(\frac{\varepsilon_0}{r^3|\la|^{1-\de}}-\frac{1}{2\oom}{\hat F}\right)\right]\nn\\
&=&-(\dd_3\log\oom)\om+\tr\chib\omb-(\tr\chi+\tr\chib)\omb+\varepsilon_0\left[O\left(\frac{1}{r^3|\la|^{1-\de}}\right)+O\left(\frac{1}{r^4|\la|^{2-2\de}}\right)\right]\nn\\
&=&-(\dd_3\log\oom)\om+\tr\chib\omb+\left[O\left(\frac{\varepsilon_0}{r^3|\la|^{1-\de}}\right)\right]\eql{3.48}
\eea
where ${\hat F}\equiv 2\oom\zeta\cdot\nabb\log\oom+\oom(\etab\cdot\eta-2\zeta^2)$.
\smallskip

\NI From the structure equations and Bianchi equations, \cite{Kl-Ni:book}, Chapter 3,
\bea
&&\dd_3\left(-(2\tr\chi)^{-1}\ro\right)=\frac{1}{2}\frac{\ro}{\tr\chi^2}(\dd_3\tr\chi)-\frac{1}{2\tr\chi}\dd_3\ro\nn\\
&&=\left(-\omb+O\big(\frac{\varepsilon_0}{r^2|\la|^{1-\de}}\big)\right)\frac{1}{\tr\chi}\dd_3\tr\chi
-\frac{1}{2\tr\chi}\left(-\frac{3}{2}\tr\chib\ro+\left[-\divv\bb+O\left(\frac{\varepsilon_0}{r^3|\la|^{4-2\de}}\right)\right]\right)\nn\\
&&=-\omb\frac{1}{\tr\chi}\dd_3\tr\chi+\frac{1}{2\tr\chi}\frac{3}{2}\tr\chib\ro
+\left[O\big(\frac{\varepsilon_0}{r^2|\la|^{1-\de}}\big)\frac{1}{\tr\chi}\dd_3\tr\chi+\frac{1}{2\tr\chi}O\left(\frac{\varepsilon_0}{r^3|\la|^{4-2\de}}\right)
\right]\nn\\
&&=-\omb\frac{1}{\tr\chi}\left(-\frac{1}{2}\tr\chib\tr\chi-(\dd_3\log\oom)\tr\chi+2\ro
-\left[\chibh\c\chih-2\divv\ze-2\lapp\log\oom-2|\eta|^2\right]\right)\nn\\
&&+\frac{1}{2\tr\chi}\frac{3}{2}\tr\chib\ro
+\varepsilon_0\left[O\left(\frac{1}{r^2|\la|^{1-\de}}\right)\frac{1}{\tr\chi}\dd_3\tr\chi+\frac{1}{2\tr\chi}O\left(\frac{1}{r^3|\la|^{4-2\de}}\right)
\right]\nn\\
&&=\frac{1}{2}\tr\chib\omb+(\dd_3\log\oom)\omb-\omb\frac{1}{\tr\chi}2\ro+\frac{1}{2\tr\chi}\frac{3}{2}\tr\chib\ro+\varepsilon_0\left[O\!\left(\frac{1}{r^2|\la|^{2-\de}}\right)\right]\nn\\
&&=\frac{1}{2}\tr\chib\omb+(\dd_3\log\oom)\omb-\frac{1}{2}(\dd_3\log\oom)4\omb-\frac{3}{2}\tr\chib\omb+\left[O\!\left(\frac{\varepsilon_0}{r^2|\la|^{2-\de}}\right)\right]
\eea
Therefore
\bea
\dd_3\left(-(2\tr\chi)^{-1}\ro\right)&=&\frac{1}{2}\tr\chib\omb-(\dd_3\log\oom)\omb
-\frac{3}{2}\tr\chib\omb+\left[O\!\left(\frac{\varepsilon_0}{r^2|\la|^{2-\de}}\right)\right]\ \ \ \ \ \ \ \ \ \ \ \ \ \ \ \eql{3.49}\\
&=&-\tr\chib\omb-(\dd_3\log\oom)\!\left(-(2\tr\chi)^{-1}\ro\right)+\left[O\!\left(\frac{\varepsilon_0}{r^2|\la|^{2-\de}}\right)\right]\nn
\eea
which finally implies, together with \ref{3.48},
\bea
\dd_3\!\left(\om-(2\tr\chi)^{-1}\ro\right)=-(\dd_3\log\oom)\!\left(\om-(2\tr\chi)^{-1}\ro\right)
+\left[O\!\left(\frac{\varepsilon_0}{r^2|\la|^{2-\de}}\right)\right]\ \ \ \ \ \ \ 
\eea
Repeating the same calculation for the first tangential derivatives,
applying again Gronwall's lemma and using the initial data assumptions we conclude that
\bea
\sup_{{\cal K}_1}\left|r^2|\la|^{1-\de}\!\left(\om-(2\tr\chi)^{-1}\ro\right)\right|\leq c\varepsilon_0
\eea
so that, together with \ref{3.47}, we obtain
\bea
\sup_{{\cal K}_1}r^2|\la|^{1-\de}\left|\left(\om+\omb\right)\right|\leq c\varepsilon_0\ .\eql{3.53}
\eea

\NI{\bf A simplified proof of Theorem \ref{Ktheorem2}.}
As said in Section \ref{S.2} the proof of Theorem \ref{Ktheorem2} is a simple adaptation of Theorem 2.1 of \cite{Kl-Ni:peeling}. A direct proof based only on the results in \cite{Kl-Ni:book} can be, nevertheless, obtained. The proof of Theorem \ref{Ktheorem2} used there requires a different definition of the last slice foliation. This can be avoided proving the result in the following steps:
\smallskip

i) We introduce as in \cite{Kl-Ni:peeling}  some $\hat{\cal Q}$ norms relative to the Riemann tensor $R$, but with an extra weight factor $|u|^{1-2\de}$ and we prove that using the estimates \ref{Otheorem22} proved before, these $\hat{\cal Q}$ norms are bounded, provided the initial data decay at least as $r^{-2}$. This implies that all the null components of the Riemann tensor $\a(R),\b(R),(\ro-\overline{\ro})(R), (\si-\overline{\si})(R),\bb(R),\aa(R)$ have an extra decay factor $|u|^{\frac{1}{2}-\de}$.
\smallskip

ii) With these improved decay factors for the Riemann components using the transport equations for the connection coefficients one obtains that all the connection coefficients which are identically zero in Kerr spacetime have also the extra decay factor $|u|^{\frac{1}{2}-\de}$.\footnote{There is no need of defining a last slice foliation as the spacetime $\cal K$ is already at our disposal and the transport equations can be used starting from Scri where we know that the relation $\om-(2\tr\chi)^{-1}\ro=0$ holds.}
\smallskip

\NI Therefore the only thing which has to be proved is that the ``Error" for the $\hat{\cal Q}$ norms is bounded assuming the \cite{Kl-Ni:book} estimates for the connection coefficients plus inequalities \ref{Otheorem22}. Again the problem is similar to the estimate of the error for $\tilde{\cal Q}$ norms relative to $\lie_TR$, the only terms which have to be controlled are those involving $\ro(R)$ as differently from $(\ro-\overline{\ro})(R)$ the decay of $\overline{\ro}(R)$ cannot be improved.
\medskip

\NI{\bf Proof of Lemma \ref{L2.1a}:}
As the Kerr spacetime is of type D, following the Petrov classification, in the null frame where the null vector fields are the principal null directions, the only null Riemann component are $\ro$ and $\si$ or, in the Newman-Penrose classification, $\Psi_2$ and their explicit expression is 
\bea
\ro(R)+i\si(R)= \frac{1}{(r-i\cos\theta)^3}=\frac{1}{r^3}+i\frac{3a\cos\theta}{r^4}+O\left(\frac{1}{r^5}\right)
\eea
so that
\bea
&&\ro(R)=\frac{1}{r^3}+O\left(\frac{1}{r^5}\right)\nn\\
&&\si(R)=\frac{3a\cos\theta}{r^4}+O\left(\frac{1}{r^6}\right)\ .
\eea

The principal null directions $l,n$ defined, for instance, in Chandrasekar book, \cite{Ch}, are
\bea
&&l=\frac{r^2+a^2}{\Delta}\frac{\partial}{\partial t}+\frac{a}{\Delta}\frac{\partial}{\partial\phi}+\frac{\partial}{\partial r}\nn\\
&&n=\frac{\De}{2\Si}\left(\frac{r^2+a^2}{\Delta}\frac{\partial}{\partial t}+\frac{a}{\Delta}\frac{\partial}{\partial\phi}-\frac{\partial}{\partial r}\right)\nn\\
&&{\tilde e}_{\theta}=\frac{1}{\sqrt{\Si}}\frac{\partial}{\partial\theta}\\
&&{\tilde e}_{\phi}=\frac{1}{\sqrt{\Si}}\left(a\sin\theta\frac{\partial}{\partial t}+\frac{1}{\sin\theta}\frac{\partial}{\partial\phi}\right)\nn
\eea
where
\bea
&&\De=r^2+a^2-2Mr\nn\\
&&\Si=r^2+a^2\cos^2\theta\nn\\
&&\Si R^2={(r^2+a^2)^2-\Delta a^2\sin^2\theta}\eql{3.148}
\eea
Besides the fact that our initial data are not exactly those of the Kerr spacetime due to corrections $\de{\overline g}$ and $\de{\overline k}$, one has to recognize that our null orthonormal frame is not the one associated to the principal null directions and, therefore, the Kerr spacetime gives a contribution to all the Riemann components. In fact the frame which, apart from extra small corrections, is adapted to the double null foliation of $\cal K$ is the one proposed by Israel and Pretorius, see \cite{Is-Pr},
\bea
&&e_4=2\oom L=\sqrt{\frac{R^2}{\De}}\left(\frac{\partial}{\partial\ub}+\om_B\frac{\partial}{\partial\phi}\right)\nn\\
&&e_3=2\oom\Lb=\sqrt{\frac{R^2}{\De}}\left(-\frac{\partial}{\partial u}+\om_B\frac{\partial}{\partial\phi}\right)\eql{IPframe}\\
&&e_{\la}=\frac{R}{{\mathcal L}}\frac{\partial}{\partial\la}=\frac{1}{\Si R}\left(-\De P\frac{\partial}{\partial r}+Q\frac{\partial}{\partial\theta}\right)
=\frac{Q}{\Si R}\left(-P\frac{\partial}{\partial r_*}+\frac{\partial}{\partial\theta}\right)\nn\\
&&e_{\phi}=\frac{1}{R\sin\theta}\frac{\partial}{\partial\phi}\ .
\eea
Its relation with the previous one are
\bea
&&l=\frac{1}{2{\sqrt\De}R}\left[(r^2+a^2+Q)e_4+(r^2+a^2-Q)e_3+2{\sqrt\De}a\sin\th e_\phi\right]\nn\\
&&n=\frac{{\sqrt\De}}{4\Si R}\left[(r^2+a^2+Q)e_3+(r^2+a^2-Q)e_4+2 {\sqrt\De}a\sin\th e_\phi\right]\nn\\
&&{\tilde e}_{\th}=\frac{{\sqrt\Si}R}{Q}e_{\la}+\sqrt{\frac{\De}{\Si}}\frac{P}{2R}(e_4-e_3)\nn\\
&&{\tilde e}_{\phi}=\frac{1}{{\sqrt\Si}R}\left(r^2+a^2\right)e_{\phi}+\frac{{\sqrt\De}a\sin\th}{2{\sqrt\Si}R}(e_3+e_4)
\eea
and the inverse relation
\bea
&&e_4
=\frac{\sqrt\De}{2\Si R}\left[\left(r^2+a^2+\frac{R\Si r}{Q}\right)l+\left(r^2+a^2-\frac{R\Si r}{Q}\right)\frac{2\Si}{\De}n-2{\sqrt\Si}a\sin\theta{\tilde e}_{\phi}\right]\nn\\
&&e_3=\frac{\sqrt\De}{2\Si R}\left[\left(r^2+a^2-\frac{R\Si r}{Q}\right)l+\left(r^2+a^2+\frac{R\Si r}{Q}\right)\frac{2\Si}{\De}n-2{\sqrt\Si}a\sin\theta{\tilde e}_{\phi}\right]\nn\\
&&e_\la=\frac{Q}{{\sqrt\Si}R}{\tilde e}_\theta-\frac{\De P}{2\Si R}\left(l-\frac{2\Si}{\De}n\right)\nn\\
&&e_{\phi}=\frac{r^2+a^2}{{\sqrt\Si}R}{\tilde e}_{\phi}-\frac{\De}{2R\Si}a\sin\theta\left(l+\frac{2\Si}{\De}n\right)\ .
\eea
Under the assumption that $a$ and $M$ are small we have approximately
\bea
&&l=e_4+O\left(\frac{a^2}{r^2}\right)e_3+O\left(\frac{a\sin\th}{r}\right)e_\phi\nn\\
&&n=e_3+O\left(\frac{a^2}{r^2}\right)e_4+O\left(\frac{a\sin\th}{r}\right)e_\phi\nn\\
&&{\tilde e}_{\th}=e_{\la}+O\left(\frac{a^2}{r^2}\right)(e_4-e_3)\nn\\
&&{\tilde e}_{\phi}=e_{\phi}+O\left(\frac{a\sin\th}{r}\right)(e_3+e_4)\ .
\eea
These relations show that the ``Kerr part contribution" to the  null Riemann components different from $\ro$ is negligeable and decay always sufficiently fast when we are considering the null components of $\lie_TR$.  

\NI Therefore we can neglect these extra contributions and observe that from the expression of ${\tilde R}=\lie_TR$ and definition \ref{piT1a} it follows that the more delicate contribution to $\ro({\tilde R})$ is: ${^{(T)}\!}\pi\ro(R)$ and the explicit expression of ${^{(T)}\!}\pi$ is given in subsection \ref{s.s2.6}. In the spacetime ${\cal K}$, $|{^{(T)}\!}\pi|$ behaves as $r^{-2}|u|^{1-\de}$ as proved in subsection \ref{s.s2.6}. Therefore from the initial data we have 
\bea
|r^5|u|^{1-\de}\ro({\lie_TR})|_{\Si_0}\leq c\ \ \mbox{equivalent to}\ \ |r^{6-\de}\ro(\lie_TR)|_{\Si_0}\leq c\ .\eql{IIest}
\eea

\subsubsection{Proof of Lemma \ref{L2.2new}:}

\NI{\bf Proof of \ref{esta}:}
Using Corollary 4.1.1 of \cite{Kl-Ni:book} the following relation holds:
\bea
&&\ML|r^{(3-\frac{2}{4})}|\la|^{3-\de}\nabb{\bb}|_{p=4,S}(\la,\ub)
=\left(\int_{S(\la,\ub)}r^2|\la|^2(r^8 |\la|^{10-4\de})|\nabb{\bb}|^4\right)^{\frac{1}{4}}\nn\\
&&\ML\leq \left(\int_{S(u_0,\ub)}r^4|r|^{4({3-\de})}|r^{\frac{3}{2}}\nabb{\bb}|^4\right)^{\frac{1}{4}}\nn\\
&&\ML+c\left(\int_{\Cb(\ub;[u_0,\la])}\bigg[r^4|\la'|^{5-2\de}|\nabb{\bb}|^2+r^6|\la'|^{5-2\de}|\nabb^2{\bb}|^2+r^4|\la'|^{7-2\de}|\ddb_3\nabb{\bb}|^2\bigg]\right)^{\frac{1}{2}}\nn\\
&&\ML\leq\left(\int_{S(u_0,\ub)}r^4|r|^{4({3-\de})}|r^{\frac{3}{2}}\nabb{\bb}|^4\right)^{\frac{1}{4}}\eql{delicate}\\
&&\ML+c\frac{1}{|\la|^{1+\frac{(\ep+2\de)}{2}}}\left(\int_{\Cb(\ub;[u_0,\la])}|\la'|^{(5+\ep)}\bigg[r^4|\la'|^2|\nabb{\bb}|^2+r^6|\la'|^{2}|\nabb^2{\bb}|^2+r^4|\la'|^4|\ddb_3\nabb{\bb}|^2\bigg]\right)^{\frac{1}{2}}\nn\ .
\eea
We estimate the first of the three integrals in the last inequality of \ref{delicate}
 as the other ones are treated exactly in the same way; from the inequality, see \cite{Kl-Ni:book}, Lemma 5.1.1:
\bea
\int_{S(\la',\ub)}r^2|\nabb\bb|^2\leq c\int_{S(\la',\ub)}|\Lie_O\bb|^2
\eea
it follows
\bea
\int_{\Cb(\ub;[u_0,\la])}|\la'|^{(5+\ep)}r^4|\la'|^2|\nabb{\bb}|^2\leq c\int_{\Cb(\ub;[u_0,\la])}|\la'|^{(5+\ep)}r^2|\la'|^2|\Lie_O{\bb}|^2\ .
\eea
From Proposition 5.1.1 of \cite{Kl-Ni:book} and the estimates on the deformation tensors and connection coefficients proved there we have
\bea
\ML|\Lie_O{\bb}({\tilde R})|\!&\leq&\! c\left(|\bb(\lie_O{\tilde R})|+\frac{O(\varepsilon_0)}{r}\left(|\aa({\tilde R})|+|\bb({\tilde R})|)
+|\ro({\tilde R})-\overline{\ro({\tilde R})}|+|\si({\tilde R})-\overline{\si({\tilde R})}|\right)\right)\nn\\
\ML\!&+&\!\frac{O(\varepsilon_0)}{r}(|\overline{\ro({\tilde R})}|+|\overline{\si({\tilde R})}|)
\eea
This implies that
\bea
&&\ML\ML\int_{\Cb(\ub;[u_0,\la])}|\la'|^{(5+\ep)}r^4|\la'|^2|\nabb{\bb}({\tilde R})|^2\leq c\left(\int_{\Cb(\ub;[u_0,\la])}|\la'|^{(5+\ep)}r^2|\la'|^2|{\bb}(\lie_O{\tilde R})|^2\right.\nn\\
&&\ML\ML \left.+\frac{\varepsilon_0^2}{|\la|^2}\int_{\Cb(\ub;[u_0,\la])}{|\la'|^{(5+\ep)}}\bigg[r^2|\la'|^2|{\bb}({\tilde R})|^2+|\la'|^4|{\aa}({\tilde R}))|^2
+r^4[|\ro({\tilde R})-\overline{\ro({\tilde R})}|^2+r^4|\si({\tilde R})-\overline{\si({\tilde R})}|^2\bigg]\right.\nn\\
&&\ML\ML+{\varepsilon_0^2}\int_{\Cb(\ub;[u_0,\la])}\frac{|\la'|^{(5+\ep)}}{|\la'|^2}r^4[\overline{\ro({\tilde R})}|^2+\overline{\si({\tilde R})}|^2]\nn\\
&&\ML\ML\leq c\int_{\Cb(\ub)\cap V(\la,\ub)}|\la'|^{5+\ep}Q(\lie_O{\tilde R})(\bar{K},\bar{K},T,e_3)
+{\varepsilon_0^2}\int_{\Cb(\ub;[u_0,\la])}{|\la'|^{(3+\ep)}}r^4[\overline{\ro({\tilde R})}|^2+\overline{\si({\tilde R})}|^2]\nn\\
&&\ML\ML\leq c\tilde{\cal Q}(\la,\ub)
+c{\varepsilon_0^2}\sup_{(\la,\ub)\in{\cal K}_2}(|r^3|\la|^{3-\de}\overline{\ro({\tilde R})}|^2+|r^3|\la|^{3-\de}\overline{\si({\tilde R})}|^2)\int_{\Cb(\ub;[u_0,\la])}|\la'|^{(3+\ep)}\frac{r^4}{r^6|\la'|^{6-2\de}}\nn\\
&&\ML\ML\leq c\tilde{\cal Q}(\la,\ub)
+c\frac{\varepsilon_0^2}{|\la|^{2-2\de-\ep}}\sup_{(\la,\ub)\in{\cal K}_2}(|r^3|\la|^{3-\de}\overline{\ro({\tilde R})}|^2+|r^3|\la|^{3-\de}\overline{\si({\tilde R})}|^2)\ .
\eea
Therefore going back to \ref{delicate} we have
\bea
&&\ML|r^{(3-\frac{2}{4})}|\la|^{3-\de}\nabb{\bb}|_{p=4,S}(\la,\ub)\leq \left(\int_{S(u_0,\ub)}r^4|r|^{4({3-\de})}|r^{\frac{3}{2}}\nabb{\bb}|^4\right)^{\frac{1}{4}}\nn\\
&&\ML+c\frac{1}{|\la|^{1+\frac{(\ep+2\de)}{2}}}\left({\QQb}^{\frac{1}{2}}(\la,\ub)
+\frac{\varepsilon_0}{|\la|^{1-\de-\frac{\ep}{2}}}\sup_{(\la,\ub)\in{\cal K}_2}(|r^3|\la|^{3-\de}\overline{\ro({\tilde R})}|+|r^3|\la|^{3-\de}\overline{\si({\tilde R})}|)\right)\nn
\eea
which completes the proof of Lemma \ref{L2.2new}.
\smallskip

\NI The first term has also a decay $\frac{c}{|\la|^{1+\frac{({\tilde\ep}+2\de)}{2}}}$ in $|\la|$. In fact as it is made
by initial data we have $u_0=u_0(\ub)=-\ub =r(u_0,\ub)$ and $|\nabb{\bb}({\tilde R})|=O(r^{-(7+\frac{\tilde\ep}{2})})$ with ${\tilde\ep}>\ep$.\footnote{Recall that $\bb$ is the component of $\lie_TR$ and this implies an extra decay factor $r^{-1}$.} Then
\[|r^{1+({3-\de})}r^{\frac{3}{2}}\nabb{\bb}|^4=O(r^{\frac{5}{2}+({3-\de})-(7+\frac{\tilde\ep}{2})})^4
=O(r^{-\frac{3}{2}-\frac{({\tilde\ep}+2\de)}{2}})^4=O(r^{-6-2({\tilde\ep}+2\de)})\ \ \ \mbox{and}\]
\bea
\bigg(\int_{S(u_0,\ub)}|r^{1+({3-\de})}r^{\frac{3}{2}}\nabb{\bb}|^4\bigg)^{\!\frac{1}{4}}=O(r^{-4-2({\tilde\ep}+2\de)})^{\frac{1}{4}}
=O\!\left(r^{-1-\frac{({\tilde\ep}+2\de)}{2}}\right)\leq \frac{c\varepsilon_0}{|\la|^{1+\frac{({\tilde\ep}+2\de)}{2}}}\ . \ \ \ \ \ 
\eea
Therefore
\bea
&&\ML|r^{(3-\frac{2}{4})}|\la|^{3-\de}\nabb{\bb}|_{p=4,S}(\la,\ub)\eql{2.60new}\\
&&\ML\leq c\frac{1}{|\la|^{1+\frac{(\ep+2\de)}{2}}}\left(\varepsilon_0+{\QQb}^{\frac{1}{2}}(\la,\ub)
+\frac{\varepsilon_0}{|\la|^{1-\de-\frac{\ep}{2}}}\sup_{(\la,\ub)\in{\cal K}_2}(|r^3|\la|^{3-\de}\overline{\ro({\tilde R})}|+|r^3|\la|^{3-\de}\overline{\si({\tilde R})}|)\right)\ .\nn
\eea
\NI{\bf Proof of \ref{estb}:} The following inequality holds:
 \bea
&&\ML|r^{(3-\frac{2}{4})}|\la|^{{(3-\de)}}\eta\!\cdot\!{\bb}|_{p=4,S}(\la,\ub)
=\left(\int_{S(\la,\ub)}r^{2}|\la|^{{4(3-\de)}}|{\bb}|^4|r^2\eta|^4\right)^{\frac{1}{4}}\\
&&\ML\leq |r^2\eta|_{\infty}\!\left(\int_{S(\la,\ub)} r^2|\la|^{{4(3-\de)}}|{\bb}|^4\right)^{\frac{1}{4}}
\leq \frac{1}{|\la|^{1-\de}}|r^2|\la|^{1-\de}\eta|_{\infty}\!\left(\int_{S(\la,\ub)}r^2|\la|^{12-4\de}|\bb|^4\right)^{\!\frac{1}{4}}\ .\nn 
\eea
Applying again Corollary 4.1.1 of \cite{Kl-Ni:book}, we have
\bea
&&\left(\int_{S(\la,\ub)}r^2|\la|^2||\la|^{\frac{(5-2\de)}{2}}{\bb}|^4\right)^{\frac{1}{4}}\leq
\left(\int_{S(\la,\ub_0)}r^2|\la|^2||\la|^{\frac{(5-2\de)}{2}}{\bb}|^4\right)^{\frac{1}{4}}\\
&&+\!\frac{1}{|\la|^{2+\frac{\de+\ep}{2}}}\left(\int_{C(\la)}|\la|^{(5+\ep)}\bigg[||\la|^2{\bb}|^2+r^2||\la|^2\nabb{\bb}|^2+|\la|^2|\dd_4|\la|^2{\bb}|^2\bigg]\right)^{\frac{1}{2}}\ .\nn
\eea
Proceeding as in the previous case we have
\bea
&&\left(\int_{S(\la,\ub)}r^2|\la|^2||\la|^{\frac{(5-2\de)}{2}}{\bb}|^4\right)^{\frac{1}{4}}\leq
\left(\int_{S(\la,\ub_0)}r^2|\la|^2||\la|^{\frac{(5-2\de)}{2}}{\bb}|^4\right)^{\frac{1}{4}}\\
&&+\!\frac{1}{|\la|^{2+\frac{\de+\ep}{2}}}\left(\sup_{(\la,\ub)\in{\cal K}_2}\QQ(\la,\ub)^{\frac{1}{2}}+\frac{\varepsilon_0}{|\la|^{-\de-\frac{\ep}{2}}}\sup_{(\la,\ub)\in{\cal K}_2}\!\big(|r^3|\la|^{3-\de}\overline{\ro({\tilde R})}|+|r^3|\la|^{3-\de}\overline{\si({\tilde R})}|\big)\right)\nn\\
&&\leq\left(\int_{S(\la,\ub_0)}r^2|\la|^2||\la|^{\frac{(5-2\de)}{2}}{\bb}|^4\right)^{\frac{1}{4}}\\
&&+\!\frac{1}{|\la|^{2-\frac{\de}{2}}}\left(\sup_{(\la,\ub)\in{\cal K}_2}\QQ(\la,\ub)^{\frac{1}{2}}+{\varepsilon_0}\!\!\sup_{(\la,\ub)\in{\cal K}_2}\!\big(|r^3|\la|^{3-\de}\overline{\ro({\tilde R})}|+|r^3|\la|^{3-\de}\overline{\si({\tilde R})}|\big)\right)\ .\nn
\eea
The estimate of the first term goes in the following way
\bea
&&\left(\int_{S(\la,\ub_0)}r^2\tau_-^2||\la|^{\frac{(5-2\de)}{2}}{\bb}|^4\right)^{\frac{1}{4}}\leq
c\left(\int_{S_0(\la,\ub_0)}r^{2+2+10-4\de}|{\bb}|^4\right)^{\frac{1}{4}}\\
&&=O\!\left(r^{\frac{2+4+10-4\de}{4}}\right)|\bb(\lie_TR)|=O\!\left(r^{\frac{4+2+10-4\de}{4}-1-5-{\tilde\ep}}\right)
\leq O\!\left(\frac{\varepsilon_0}{|\la|^{2+{(\tilde\ep+\de)}}}\right)\ .\nn
\eea

\subsubsection{Proof of Lemma \ref{L2.2nabbro}}
Observe that from Lemma 5.1.1 in \cite{Kl-Ni:book} we have
\bea
&&\ML\int_{S(\la,\nu)}|\la|^{5+\ep}r^2|\la|^2r^2|\nabb\ro(\tilde R)|^2\leq \int_{S(\la,\nu)}|\la|^{5+\ep}r^2|\la|^2|\Lie_O\ro(\tilde R)|^2\\
&&\ML\leq\int_{S(\la,\nu)}|\la|^{5+\ep}r^2|\la|^2|\ro(\lie_O\tilde R)|^2+\int_{S(\la,\nu)}|\la|^{5+\ep}r^2|\la|^2|\tr{^{(O)}\!}\pi|^2|\ro(\tilde R)|^2+{\{good\ corrections\}}\ .\nn
\eea
and also
\bea
&&\ML\int_{S(\la,\nu)}|\la|^{5+\ep}r^4r^2|\nabb\ro(\tilde R)|^2\leq \int_{S(\la,\nu)}|\la|^{5+\ep}r^4|\Lie_O\ro(\tilde R)|^2\eql{CB1}\\
&&\ML\leq\int_{S(\la,\nu)}|\la|^{5+\ep}r^4|\ro(\lie_O\tilde R)|^2+\int_{S(\la,\nu)}|\la|^{5+\ep}r^4|\tr{^{(O)}\!}\pi|^2|\ro(\tilde R)|^2+{\{good\ corrections\}}\ .\nn
\eea
From the estimate for $\ro({\tilde R})$ in Theorem \ref{T3.1} 
\[\sup_{{\cal K}}|r^3|\la|^{3-\de}\ro({\tilde R})|\leq c\]
and from the estimate proved in \cite{Kl-Ni:book}, Chapter 6,
\[\sup_{{\cal K}}|r{^{(O)}\!}\pi|\leq c\ ,\]
the second integral in \ref{CB1} can be estimated by
\bea
&&\int_{S(\la,\nu)}|\la|^{5+\ep}r^4|\tr{^{(O)}\!}\pi|^2|\ro(\tilde R)|^2\leq
\int_{S(\la,\nu)}|\la|^{5+\ep}r^2\frac{1}{r^6|\la|^{6-2\de}}\nn\\
&&\leq cr^2|\la|^{5+\ep}r^2\frac{1}{r^6|\la|^{6-2\de}}\leq c\frac{1}{r^2|\la|^{1-2\de-\ep}}
\eea
This implies 
\bea
&&\int_{\Cb(\nu)}|\la|^{5+\ep}r^4r^2|\nabb\ro({\tilde R})|^2\leq \int_{\Cb(\nu)}|\la|^{5+\ep}r^4|\ro(\lie_O{\tilde R})|^2+c\eql{2.85ff}\\
&&\leq c{\tilde{\cal Q}}_{\Si_0}+c\int_{\Cb(\nu)}|\la|^{5+\ep}r^2\frac{1}{r^6|\la|^{6-2\de}}+c
\leq c{\tilde{\cal Q}}_{\Si_0}+c\frac{1}{r|\la|^{1-2\de-\ep}}+c\leq c_1\nn
\eea
Using Corollary 4.1.1 of \cite{Kl-Ni:book} we have
\bea
\left(\int_{S(\la,\nu)}r^2|\la|^2|F|^4\right)^{\frac{1}{4}}\leq \left(\int_{S(\la_1,\nu)=\Cb(\nu)\cap\Si_0}r^2|\la|^2|F|^4\right)^{\frac{1}{4}}
+\left(\int_{\Cb(\nu)} \left[|F|^2+\c\c\c\c \right]\right)^{\frac{1}{2}}\ \ \ \ \ \ \eql{bbqq}
\eea
Choosing \ $F=|\la|^{\frac{5+\ep}{2}}r^3\nabb\ro({\tilde R})$ we obtain
\bea
&&\ML|r^{4-\frac{2}{4}}|\la|^{3+\frac{\ep}{2}}\nabb\ro({\tilde R})|_{p=4,S}
=\left(\int_{S(\la,\nu)}||\la|^{3+\frac{\ep}{2}}r^{4-\frac{2}{4}}\nabb\ro({\tilde R})|^4\right)^{\!\frac{1}{4}}\eql{newest}\\
&&\ML\leq \left(\int_{S(\la_1,\nu)=\Cb(\nu)\cap\Si_0}||\la|^{3+\frac{\ep}{2}}r^{4-\frac{2}{4}}\nabb\ro({\tilde R})|^4\right)^{\!\frac{1}{4}}
\!+\left(\int_{\Cb(\nu)}\left[||\la|^{\frac{5+\ep}{2}}r^{3}\nabb\ro({\tilde R})|^2+\c\c\c\c\right]\right)^{\!\frac{1}{2}}\nn\\
&&\ML\leq \left(\int_{S(\la_1,\nu)=\Cb(\nu)\cap\Si_0}||\la|^{3+\frac{\ep}{2}}r^{4-\frac{2}{4}}\nabb\ro({\tilde R})|^4\right)^{\!\frac{1}{4}}
\!+\left(\int_{\Cb(\nu)}|\la|^{5+\ep}\left[r^{4}|\ro(\lie_O{\tilde R})|^2+\c\c\c\c\right]\right)^{\!\frac{1}{2}}\nn\\
&&\ML\leq \left(\int_{S(\la_1,\nu)=\Cb(\nu)\cap\Si_0}||\la|^{3+\frac{\ep}{2}}r^{4-\frac{2}{4}}\nabb\ro({\tilde R})|^4\right)^{\!\frac{1}{4}}
\!+{\tilde{\cal Q}}_{\Si_O}^{\frac{1}{2}}\leq c\ ,\nn
\eea

\NI where the last inequality uses the boundedness of the $\tilde{\cal Q}$ norms. Observe now that inequality \ref{2.85ff} 
\bea
&&\int_{\Cb(\nu)}|\la|^{5+\ep}r^4r^2|\nabb\ro({\tilde R})|^2=\int_{\la_1}^{\la}||\la|^{\frac{5+\ep}{2}}r^3\nabb\ro({\tilde R})|^2_{p=2,S}\leq c_1\nn
\eea
imply, with $\tilde{\ep}>0$, 
\bea
||\la|^{\frac{5+\ep}{2}}r^3\nabb\ro({\tilde R})|^2_{p=2,S}\leq c\frac{1}{|\la|^{1+\tilde{\ep}}}
\eea
and from it
\bea
||\la|^{3+\tilde{\ep}+\frac{\ep}{2}}r^{4-\frac{2}{p}}\nabb\ro({\tilde R})|_{p=2,S}\leq c\ .\eql{2.93ff}
\eea
Interpolating between \ref{newest} and \ref{2.93ff} we obtain for $p\in[2,4]$,
\bea
|r^{4-\frac{2}{p}}|\la|^{3+\ep}\nabb\ro({\tilde R})|_{p,S}\leq c \eql{2.94ff}
\eea
proving the lemma. 
\subsection{Some technical details}

\NI{\bf 1.} The Kerr metric in the Boyer-Linquist coordinates $\{t,r,\theta,\phi\}$ is:
\begin{eqnarray}\label{BL}
ds^2 =-\frac{\De-a^2\sin^2\theta}{\Si}dt^2+\frac{\Sigma}{\Delta}dr^2+\Sigma
d\theta^2-\frac{4Mar\sin^2\theta}{\Sigma}d\phi dt+R^2\sin^2\theta d\phi^2 \nn
\end{eqnarray}
and its restriction to $\Si_0$:
\bea
ds^2\!&=&\!\frac{\Sigma}{\Delta}dr^2+\Si d\theta^2+R^2\sin^2\theta d\phi^2\\
\!&=&\!g_S+\left(\frac{a^2\sin^2\theta}{r^2}+O\left(\frac{a^2M}{r^3}\right)+O\left(\frac{a^4}{r^4}\right)\right)\!dr^2+\left(\frac{a^2\cos^2\theta}{r^2}\right)\!r^2d\theta^2\nn\\
\!&+&\!\left(\frac{a^2}{r^2}+O\left(\frac{a^2M}{r^3}\right)+O\left(\frac{a^4}{r^4}\right)\right)\!r^2\sin^2\theta d\phi^2\ .\nn
\eea
It follows that the components of the correction to the $g_S$ metric start with terms  of order $O({a^2}/{r^2})$ and if ${a^2}/{R_0^2}$ is $O(\varepsilon_0)$ then it is clear that these terms are corrections compatible with the assumptions in \cite{Kl-Ni:book}. 

\NI We expect, therefore, that we can connect to double null foliation of \cite{Is-Pr} to the one introduced in \cite{Kl-Ni:book} when $a$ is small.

\NI In principle once we have proved that the Riemann components can be written as those of Kerr plus corrections small and decaying fast we could look again at the transport equations for the connection coefficients and show that we can subtract the Kerr part and prove that what remains is small and faster decaying. This is, nevertheless, somewhat delicate as the null frame adapted to the Israel-Pretorius foliation does not coincide exactly with the one used in \cite{Kl-Ni:book}. 

\NI Before discussing this point let us recall the main properties of the Israel-Pretorius foliation. 

Let us define $u,\ub$ as two functions solutions of the eikonal equation
\[g^{\mu\nu}\pr_{\mu}w\pr_{\nu}w=0\ ,\]
\bea
&&u=\frac{t-r_*(\theta,r)}{2}\nn\\
&&\ub=\frac{t+r_*(\theta,r)}{2}\ .
\eea
The Kerr metric in the Boyer-Lindquist coordinates $\{x^{\mu}\}=\{t,r,\theta,\phi\}$ is 
\bea
{\bf g}_{(Kerr)}(\c,\c)\!&=&\!-\frac{\left(\Delta-a^2\sin^2\theta\right)}{\Si}dt^2
-a\sin^2\theta\frac{(r^2+a^2-\Delta)}{\Si}\left(dt\!\otimes\! d\phi+d\phi\!\otimes\!dt\right)\nn\\ 
&+&\frac{\Si}{\Delta}dr^2+\Si d\theta^2+\frac{(r^2+a^2)^2-{\Delta}a^2\sin^2\theta}{\Si}\sin^2\theta d\phi^2\eql{2.24}
\eea
and
\bea
&&g^{00}=-\frac{g_{\phi\phi}}{\lap\sin^2\theta}\ ,\ g^{0\phi}=\frac{g_{0\phi}}{\lap\sin^2\theta}\ ,\ g^{\phi\phi}=-\frac{g_{00}}{\lap\sin^2\theta}\nn\\
&&g^{rr}=g_{rr}^{-1}\ ,\ g^{\theta\theta}=g_{\theta\theta}^{-1}\ .\nn
\eea
The null (radial) geodesic vector field $L,\Lb$
\[L=L^{\ro}\frac{\pr}{\pr x^{\ro}}\ \ ,\ \ \Lb=\Lb^{\ro}\frac{\pr}{\pr x^{\ro}}\]
have the components $\Lb^{\ro}=-g^{\ro\mu}\pr_{\mu}\frac{(t+r_*)}{2}\ \  ,\ \  L^{\ro}=-g^{\ro\mu}\pr_{\mu}\frac{(t-r_*)}{2}$, therefore
\bea
&&\Lb^{0}=-\frac{g^{00}}{2}\ \ \Lb^{r}=-g^{rr}\frac{(\pr_rr_*)}{2} \ \ \Lb^{\theta}=-g^{\theta\theta}\frac{(\pr_{\theta}r_*)}{2}\ \ \Lb^{\phi}=-\frac{g^{\phi0}}{2}\nn\\
&&L^{0}=-\frac{g^{00}}{2}\ \ L^{r}=g^{rr}\frac{(\pr_rr_*)}{2} \ \ L^{\theta}=g^{\theta\theta}\frac{(\pr_{\theta}r_*)}{2}\ \ L^{\phi}=-\frac{g^{\phi0}}{2}
\eea
From these definitions with an easy calculation we obtain 
\bea
&&\Lb=\frac{1}{2\oom^2}\left(\frac{\pr}{\pr u}+\om_B\frac{\pr}{\pr\phi}\right)\nn\\
&&L=\frac{1}{2\oom^2}\left(\frac{\pr}{\pr\ub}+\om_B\frac{\pr}{\pr\phi}\right)
\eea
where
\bea
\om_B=\frac{2Mar}{\Si R^2}\ \ ,\ \ \oom=\sqrt{\frac{\Delta}{R^2}}\ .
\eea
Let us define
\bea
&&e_3=\frac{1}{\oom}\left(\frac{\pr}{\pr u}+\om_B\frac{\pr}{\pr\phi}\right)\ \ ,\ \ 
e_4=\frac{1}{\oom}\left(\frac{\pr}{\pr\ub}+\om_B\frac{\pr}{\pr\phi}\right)\nn\\
&&\Nb=\left(\frac{\pr}{\pr u}+\om_B\frac{\pr}{\pr\phi}\right)\ \ ,\ \ 
N=\left(\frac{\pr}{\pr\ub}+\om_B\frac{\pr}{\pr\phi}\right)\ .
\eea
Therefore
\bea
T=\oom(e_3+e_4)=\left(\frac{\pr}{\pr u}+\frac{\pr}{\pr \ub}\right)+2\om_B\frac{\pr}{\pr\phi}=\frac{\pr}{\pr t}+2\om_B\frac{\pr}{\pr\phi}
\eea
and , in Kerr,
\bea
&&\ML[T,e_{\la}]=[\frac{\pr}{\pr t},e_{\la}]+2[\om_B\frac{\pr}{\pr\phi},e_{\la}]=-e_{\la}(\om_B)\frac{\pr}{\pr\phi}
=-2R\sin\th e_{\la}(\om_B)e_{\phi}\nn\\
&&\ML[T,e_{\phi}]=[\frac{\pr}{\pr t},e_{\phi}]+2[\om_B\frac{\pr}{\pr\phi},e_{\phi}]=0
\eea
Observe that $\om_B=O(r^{-3})$ then 
\bea
&&e_{\la}(\om_B)=\frac{\pr r}{\pr\la}\frac{\pr \om_B}{\pr r}+\frac{\pr \th}{\pr\la}\frac{\pr \om_B}{\pr \th}
=\frac{\pr r}{\pr\la}O(ar^{-4})+\frac{\pr \th}{\pr\la}O(a^2r^{-5})\nn\\
&&=O(1)\left(O(ar^{-4})+O(a^2r^{-5})\right)=O(ar^{-4})\eql{3.208}
\eea
as, see \cite{Is-Pr},
\bea
\frac{\partial r}{\partial\la}=-\frac{\mu P^2Q\Delta}{\Si R^2}=O(1)\ \ ,\ \ \frac{\partial \th}{\partial\la}=\frac{\mu PQ^2}{\Si R^2}=O(1)\ .
\eea
\bea
\mbox{where}\ \ \ P^2=a^2(\la-\sin^2\th)\ \ ,\ \  Q^2=(r^2+a^2)^2-a^2\la \De\ .\nn
\eea
Therefore
\bea
[T,e_{\la}]=-2R\sin\th e_{\la}(\om_B)e_{\phi}=O(ar^{-3})e_{\phi}\ .
\eea
It is immediate to recognize that $N$ and $\Nb$ are equivariant vector fields, that is they send $S(u,\ub)$ into $S(u+\ep,\ub)$ and $S(u,\ub+\ep)$ respectively.

\NI A simple computation gives the following expressions
\bea
&&[N,\Nb]=-4\oom^2\zeta(e_{\phi})e_{\phi}\\
&&\nn\\
&&\zeta(e_{\phi})=-\frac{R\sin\theta}{2\Si}Q\frac{\pr\om_B}{\pr r}\ \ ,\ \ \zeta(e_{\la})=0
\eea
where $e_{\la}$ is the unit vector which form a null orthonormal frame with $e_3,e_4,e_\phi$.

\NI In the previous sections we defined $T=\oom(e_3+e_4)$ and we expect that $[T,e_4]$ decays fast. Let us compute it in Kerr and check the decay.\footnote{If we hade chosen $T$ such that in Kerr becomes $\frac{\pr}{\pr t}$ then the commutator (in Kerr) would be zero.}  Clearly if it is good in this case it will work also in the perturbed Kerr.
\bea
&&[T,e_4]=[\oom(e_3+e_4),e_4]=[\oom e_3,e_4]+[\oom e_4,e_4]=[\Nb,\frac{1}{\oom}N]-e_4(\oom)e_4\nn\\
&&=\frac{1}{\oom}[\Nb,N]-\frac{1}{\oom^2}\Nb(\oom)N-e_4(\oom)e_4\nn\\
&&=\frac{1}{\oom}[\Nb,N]-e_3(\oom)e_4-e_4(\oom)e_4=-4\oom\zeta(e_{\phi})e_{\phi}-\oom\big(e_3(\log\oom)+e_4(\log\oom)\big)e_4\nn\\
&&-4\oom\zeta(e_{\phi})e_{\phi}+2(\om+\omb)e_4=-4\oom\zeta(e_{\phi})e_{\phi}
\eea
and we know that $\zeta(e_{\phi})=O(r^{-3})$ which satisfies our request.
\medskip

\NI{\bf 2.} Once we have the estimate $\sup_{\cal K}|r^5|u|^{1+\ep}\pr_T\a_{ab}|$ to perform the integral we proceed in the following way: let us consider the integral curves of the vector field $T=\oom(e_3+e_4)$, let us denote them $\ga(s)$ such that
\beaa
\frac{d\ga^{\mu}}{ds}=\oom(e_3+e_4)=(\Nb+N)^{\mu}
\eeaa
where \[\Nb=\oom e_3\ \ ,\ \ N=\oom e_4\]
are the equivariant vector fields associated to the foliation. Mimicking the situation of the exact Kerr case we can define
\bea
&&N=\frac{\partial}{\partial\ub}+X\ \ ;\ \ \Nb=\frac{\partial}{\partial u}+X\nn\\
&&e_4=\frac{1}{\oom}\!\left(\frac{\partial}{\partial\ub}+X\right)\ \ ;\ \ \e_3=\frac{1}{\oom}\!\left(\frac{\partial}{\partial u}+X\right)\eql{met1}
\eea
and we expect that
\bea
X=\om_B\frac{\pr}{\pr\phi}+\de X=X_{\phi}e_{\phi}+X_{\theta}e_{\theta}
\eea
where
\[X_{\phi}:={\bf g}(X,e_{\phi})\ \ ,\ \ X_{\theta}:={\bf g}(X,e_{\th})\ ,\]
where $X$ is a vector field tangent to $S$ and $\de X$ is the part which is identically zero in Kerr. As we know that
\[[N,\Nb]=-4\oom^2\zeta(e_a)e_a\] it follows that
\[(\frac{\pr}{\pr\ub}-\frac{\pr}{\pr u})X=-4\oom^2\zeta(e_a)e_a\ .\]
With these definitions
\bea
\frac{d}{ds}u(\ga(s))=\frac{d\ga^{\mu}}{ds}\frac{\pr u}{\pr x^{\mu}}=\frac{d\ga^{u}}{ds}=\oom(e_3+e_4)^u=1
\eea
so that
\bea
u(\ga(s;{\overline r}))=u(\ga(0;{\overline r}))+s\ .
\eea
In $\cal K$ it is easy to show that defining $e_3,e_4$ as in \ref{met1} and using $u,\ub$ as coordinates we can write the metric in the following way
\bea
{\bf g}(\c,\c)=-4\oom^2dud\ub+\ga_{ab}\left(d\om^a-X^a(du+d\ub)\right)\left(d\om^b-X^b(du+d\ub)\right)\ .\ \ \ \ \ \ 
\eea

\end{document}